\documentclass{bioinfo}
\usepackage{hhline}
\usepackage{soul}
\usepackage{colortbl}
\usepackage{tcolorbox}
\usepackage{multirow}
\copyrightyear{2019} \pubyear{2019}

\newcommand*{\tcent}[1]{\multicolumn{1}{c|}{\textbf{#1}}}
\newcommand*{\tcentd}[1]{\multicolumn{1}{c||}{\textbf{#1}}}
\newcommand*{\tcentb}[1]{\multicolumn{1}{|c|}{\textbf{#1}}}
\newcommand*{\tcentno}[1]{\multicolumn{1}{c}{\textbf{#1}}}
\newcommand*{\tdash}{\multicolumn{1}{c|}{\textemdash}}

\access{Advance Access Publication Date: Day Month Year}
\appnotes{Manuscript Category}

\begin{document}
\firstpage{1}

\subtitle{Sequence analysis}

\title[Apollo]{Apollo: A Sequencing-Technology-Independent, Scalable, and Accurate Assembly Polishing Algorithm}
\author[Firtina \textit{et~al}.]{
Can Firtina\,$^{\text{\sfb 1}}$,
Jeremie S. Kim\,$^{\text{\sfb 1,2}}$,
Mohammed Alser\,$^{\text{\sfb 1}}$,
Damla Senol Cali\,$^{\text{\sfb 2}}$,\\
A. Ercument Cicek\,$^{\text{\sfb 3}}$,
Can Alkan\,$^{\text{\sfb 3},\ast}$, and
Onur Mutlu\,$^{\text{\sfb 1}, \text{\sfb 2}, \text{\sfb 3},\ast}$}
\address{
$^{\text{\sf 1}}$Department of Computer Science, ETH Zurich, Zurich 8092, Switzerland\\
$^{\text{\sf 2}}$Department of Electrical and Computer Engineering, Carnegie Mellon University, Pittsburgh 15213, PA, USA \\
$^{\text{\sf 3}}$Department of Computer Engineering, Bilkent University, Ankara 06800, Turkey \\
}

\corresp{$^\ast$To whom correspondence should be addressed.}

\history{Received on XXXXX; revised on XXXXX; accepted on XXXXX}

\editor{Associate Editor: XXXXXXX}

\corresp{$^\ast$To whom correspondence should be addressed.}

\history{Received on XXXXX; revised on XXXXX; accepted on XXXXX}

\editor{Associate Editor: XXXXXXX}

\abstract{\textbf{Motivation:}
Third-generation sequencing technologies can sequence long reads that contain as many as 2 million base pairs (bp). These long reads are used to construct an assembly (i.e., the subject's genome), which is further used in downstream genome analysis. Unfortunately, third-generation sequencing technologies have \emph{high} sequencing error rates and a \emph{large} proportion of bps in these long reads are \emph{incorrectly} identified. These errors propagate to the assembly and affect the accuracy of genome analysis. \emph{Assembly polishing algorithms} minimize such error propagation by polishing or fixing errors in the assembly by using information from alignments between reads and the assembly (i.e., read-to-assembly alignment information). However, current assembly polishing algorithms can only polish an assembly using reads either from a certain sequencing technology or from a small assembly. Such technology-dependency and assembly-size dependency require researchers to 1) run multiple polishing algorithms and 2) use small chunks of a large genome to use all available read sets and polish large genomes, respectively.\\
\textbf{Results:}
We introduce Apollo, a \emph{universal} assembly polishing algorithm that scales well to polish an assembly of \emph{any} size (i.e., both large and small genomes) using reads from \emph{all} sequencing technologies (i.e., second- and third-generation). Our goal is to provide a single algorithm that uses read sets from all available sequencing technologies to improve the accuracy of assembly polishing and that can polish large genomes. Apollo 1) models an assembly as a profile hidden Markov model (pHMM), 2) uses read-to-assembly alignment to train the pHMM with the Forward-Backward algorithm, and 3) decodes the trained model with the Viterbi algorithm to produce a polished assembly. Our experiments with real read sets demonstrate that Apollo is the \emph{only} algorithm that 1) uses reads from any sequencing technology within a single run and 2) scales well to polish large assemblies without splitting the assembly into multiple parts.\\
\textbf{Contact Authors:} \href{onur.mutlu@inf.ethz.ch}{onur.mutlu@inf.ethz.ch}, \href{calkan@cs.bilkent.edu.tr}{calkan@cs.bilkent.edu.tr}\\
\textbf{Supplementary information:} Supplementary data is available at \textit{Bioinformatics} online.
online.\\
\textbf{Availability:} Source code is available at https://github.com/CMU-SAFARI/Apollo \vspace{-4mm}}

\maketitle

\section{Introduction} \label{sec:introduction}

High-Throughput Sequencing (HTS) technologies are being widely used in genomics due to their ability to produce a large amount of sequencing data at a relatively low cost compared to first-generation sequencing methods~\citep{Sanger1977}. Despite these advantages, HTS technologies have two significant limitations. The first limitation is that HTS technologies can only sequence fragments of the genome (i.e., \emph{reads}). This results in the need to reconstruct the original full sequence by either using 1) read alignment, the process of aligning the reads to a \emph{reference genome}, a genome representative of all individuals within a species, or 2) \emph{de novo genome assembly}, the process of aligning all reads against each other to construct larger fragments called \emph{contigs}, by identifying reads that overlap and combining them. The second limitation of HTS technologies is that they introduce non-negligible insertion, deletion, and substitution errors (i.e., ${\sim}$10 - 15\% error rate) into reads. Depending on the method for reconstructing the original sequence, HTS errors often cause either 1) reads aligned to an incorrect location in the reference genome, or 2) erroneously constructed assemblies. These two limitations of HTS technologies are partially mitigated with computationally expensive algorithms such as \emph{alignment} and \emph{assembly construction}. Despite the wide availability of these algorithms, imperfect sequencing technologies still affect the reliability of downstream analysis in the genome analysis pipeline (e.g., variant calling).

Based on the average read length and the error profile of their reads, HTS technologies are roughly categorized into two types: (1) second-generation and (2) third-generation sequencing technologies. Second-generation sequencing technologies (e.g., Illumina) generate the most accurate reads (${\sim}$99.9\% accuracy). However, the length of their reads are short (${\sim}$100-300bp)~\citep{Glenn2011}. This introduces challenges in both read alignment and de novo genome assembly. In read alignment, a short read can align to multiple candidate locations in a reference equally well~\citep{Xin2013, Alser2017a, Kim2018, Alser2019, Alser2019b}. Aligners must either deterministically select a matching location, which requires additional computation, or randomly select one of the candidate locations, which results in non-reproducible read alignments~\citep{Firtina2016}. In de novo genome assembly, high computational complexity is required to identify overlaps between reads. Even after completing de novo genome assembly, there are often multiple gaps in an assembly~\citep{MeltzSteinberg2017}. This means an assembly is composed of many smaller contigs rather than a few long contigs, or in the ideal case, a single genome-sized contig.

Third-generation sequencing technologies (i.e., PacBio's Single Molecule Real-Time (SMRT) and Oxford Nanopore Technologies (ONT)) are capable of producing long reads (${\sim}$10Kbps on average and up to 2Mbps) at the cost of high error rates (${\sim}$10 - 15\% error rate)~\citep{Huddleston2014, Jain2018, Payne2018}. Different third-generation sequencing technologies result in different error profiles. For example, PacBio reads tend to have more insertion errors than other error types whereas insertion errors are the least common errors for ONT reads~\citep{Weirather2017}. Long reads make it more likely to find longer overlaps between the reads in de novo genome assembly. As a result, there are usually fewer long contigs~\citep{Alkan2011, Chaisson2015, MeltzSteinberg2017}. Despite this, error-prone reads often result in a highly erroneous assembly, which may not be representative of the subject's actual genome. As a consequence, any analysis using the erroneous assembly (e.g., identifying variations/mutations in a subject's genome to determine proclivity for diseases) is often unreliable.

Existing solutions that try to overcome the problem of error-prone assemblies when using de novo genome assembly can be categorized into two types. First, a typical solution is to correct the errors of long reads. Errors are corrected by using high coverage reads (e.g., ${\sim}$100$\times$ coverage) from the same sequencing technology (i.e., self-correction) or additional reads from more reliable second-generation sequencing technologies (i.e., hybrid correction). There are several available \emph{error correction} algorithms that use additional reads to locate and correct errors in long reads (e.g., Hercules~\citep{Firtina2018}, LoRDEC~\citep{Salmela2014}, LSC~\citep{Au2012}, and LoRMA~\citep{Salmela2016}). The main disadvantage of error correction algorithms is that they require \emph{more} sequenced reads from either the same or different sequencing technologies. For example, LoRMA, a self-correction tool, uses reads to build a de Bruijn graph for error correction. The reads corrected using a de Bruijn graph method cannot span even half of the entire genome, if the coverage is lower than 100$\times$~\citep{Salmela2016}. When the coverage is low, the connections in a de Bruijn graph can be weak. These weak regions can be treated as bulges and tips, and can be removed from the graph~\citep{Chaisson2004}, which may fail to create a reliable consensus of the entire genome for error correction. Although hybrid correction tools (e.g., PBcR~\citep{Koren2012}) can use low coverage short reads (e.g., 25$\times$) to correct the long reads that can span 95\% of the genome after correction, these hybrid correction tools require \emph{additional} short reads. Therefore, in both cases (i.e., hybrid and self-correction), generating additional reads (i,e., either additional short reads or high coverage long reads) requires additional cost and time. While a higher-coverage dataset may lead to higher read accuracy~\citep{Berlin2015}, the cost of producing a high-coverage dataset for long reads is often prohibitively high~\citep{Rhoads2015}. For example, sequencing the human genome with ONT at only 30$\times$ coverage costs around \$36,000~\citep{Jain2018}. Unless there exist sufficient resources for multiple sequencing technologies or high-coverage, error correction algorithms may not be a viable option to generate accurate assemblies.

The second method for removing errors in an assembly is called \emph{assembly polishing}. An assembly polishing process attempts to correct the errors of the assembly using the alignments of \emph{either} long or short reads to the assembly. The \emph{read-to-assembly} alignment, which is the alignment of the reads to the assembly, allows an assembly polishing algorithm to decide whether the assembly should be \emph{polished} based on the similarity of the base pairs between the alignments of the reads and their corresponding locations in the assembly. If the assembly polishing algorithm finds a dissimilarity, the algorithm modifies the assembly to make it more similar to the aligned reads as it assumes that the alignment information is a more reliable source. In other words, the dissimilarity is attributed to errors in the assembly. Assembly polishing algorithms assume that such modifications correct, or polish, the errors of an assembly.

There are various assembly polishing algorithms that use various methods for discovering dissimilarities and modifying the assembly (e.g., Nanopolish~\citep{Loman2015}, Racon~\citep{Vaser2017}, Quiver~\citep{Chin2013}, and Pilon~\citep{Walker2014}). However, the primary limitation of many of these assembly polishing algorithms is that they work only with reads from a limited set of sequencing technologies. For example, Nanopolish can use \emph{only} ONT long reads~\citep{SenolCali2018}, while Quiver supports \emph{only} PacBio long reads. Thus, these assembly polishing algorithms are sequencing-technology-dependent. Even though Pilon can use long reads as it does not impose a hard restriction not to use them, Pilon does not suggest using long reads, and it is well tuned for using short reads. Therefore, we consider Pilon as only a partially-sequencing-technology-independent algorithm as it neither prevents nor truly supports using long reads. Even though Racon can use either short or long reads to polish an assembly, it can use only a single set of reads \emph{within a single run} (e.g., only a set of PacBio reads). This requires an assembly to be polished in multiple runs with Racon to use all the available set of reads from multiple sequencing technologies (i.e., a \emph{hybrid set of reads}). There is currently no single assembly polishing algorithm that can polish an assembly with an \emph{arbitrary} set of reads from various sequencing technologies (e.g., both ONT and PacBio reads) within a single run.

While the technology-dependency problem of such assembly polishing algorithms could be mitigated by consecutively using either different algorithms (e.g., Quiver and Pilon) or the same algorithm multiple times (e.g., running Racon twice to use both PacBio and Illumina reads), there are scalability problems associated with using polishing algorithms to polish a large genome and, therefore, running assembly polishing algorithms multiple times for two reasons. First, none of the polishing algorithms can scale well to polish large genomes within a single run as they require large computational resources (e.g., polishing a human genome requires more than 192GB of available memory) unless the coverage of a set of reads is low (e.g., less than 10$\times$). Therefore, these assembly polishing algorithms \emph{cannot} polish large genomes in a single run if the available computational resources are not tremendous, and they are restricted to polish smaller parts (e.g., contigs) of a large genome. Second, dividing a large genome into smaller contigs and running polishing algorithms multiple times requires extra effort to collect and merge the multiple results to produce the polished large genome assembly as a whole.

A \emph{universal technology-independent assembly polishing algorithm} that can use reads regardless of both 1) the sequencing technology used to produce them and 2) the size of the genome, enables the usage of all available reads for a more accurate assembly compared to using reads from a single sequencing technology. Such a universal assembly polishing algorithm would also not require running assembly polishing multiple times to take advantage of all available reads. Unfortunately, such an assembly polishing algorithm does not exist.

Our goal in this paper is to propose a \emph{technology-independent} assembly polishing algorithm that enables all available reads to contribute to assembly polishing and that scales well to polish an assembly of any size (e.g., both small and large genome assemblies) within a single run. To this end, we propose a machine learning-based \emph{universal technology-independent assembly polishing} algorithm, Apollo, that corrects errors in an assembly by using read-to-assembly alignment regardless of the sequencing technology used to generate reads. Apollo is the first \emph{universal technology-independent} assembly polishing algorithm. Apollo's machine learning algorithm is based on two key steps: (1) training and (2) decoding the profile hidden Markov model (pHMM) of an assembly. First, Apollo uses the Forward-Backward and Baum-Welch algorithms~\citep{Baum1972} to train the pHMM by calculating the probability of the errors based on aligned reads. Error probabilities in the pHMM reveal how reads and the assembly that the reads align to are similar to each other without making any assumptions on the sequencing technology used to produce the reads. This is the \emph{key} feature that makes Apollo sequencing-technology-independent. Second, Apollo uses the Viterbi algorithm~\citep{Viterbi1967} to decode the trained pHMM to correct the errors of an assembly. Apollo employs a recent pHMM design~\citep{Firtina2018}, as this design addresses the computational problems that make pHMMs otherwise impractical to use for training in machine learning. The design of the pHMM enables flexibility in adapting the pHMM based on the error profile of the underlying sequencing technology of an assembly. Therefore, Apollo can additionally apply the known error profile of a sequencing technology to improve upon its error probability calculations.

We compare Apollo with Nanopolish, Racon, Quiver, and Pilon using datasets that are sequenced with different technologies: Escherichia coli K-12 MG1655 (MinION and Illumina), Escherichia coli O157 (PacBio and Illumina), Escherichia coli O157:H7 (PacBio and Illumina), Yeast S288C (PacBio and Illumina), and the human Ashkenazim trio sample (HG002, PacBio and Illumina). We compare our polished assemblies against highly accurate and finished genome assemblies of the corresponding samples to determine the accuracy of the various assembly polishing algorithms.

Using the datasets from different sequencing technologies, we first show that Apollo scales better than other polishing algorithms in polishing assemblies of large genomes using moderate and high coverage reads. Second, Apollo is the \emph{only} algorithm that can use reads from \emph{multiple} sequencing technologies in a \emph{hybrid} manner (e.g., using both ONT and Illumina reads in a single run). Because of this, Apollo scales well to polish an assembly of any size within a \emph{single} run using \emph{any} set of reads, which makes Apollo a universal, sequencing-technology-independent assembly polishing algorithm. Third, we show that when Apollo uses a hybrid set of reads (i.e., both PacBio and Illumina reads), it polishes assemblies generated by Canu ~\citep{Koren2017} (i.e., Canu-generated assemblies) more accurately than any other polishing algorithm. Fourth, for all other remaining cases, when we compare Apollo to other competing algorithms, our experiments show that Apollo usually produces assemblies of similar accuracy to competing algorithms: Nanopolish, Pilon, Racon, and Quiver. However, when using long read sets to polish Miniasm-generated \textit{E. coli} O157:H7, \textit{E. coli} K-12, and Yeast S288C assemblies, Apollo produces assemblies with less accuracy than that of Racon and Quiver. These experiments are based on 1) a ground truth (i.e., reference-dependent comparison), 2) k-mer similarity calculation (i.e., Jaccard similarity~\citep{Niwattanakul2013}) between an Illumina set of reads and a polished assembly, and 3) the quality assessment of the assembly from mapped short reads (i.e., reference-independent comparison). These comparisons show that Apollo can polish an assembly using reads from any sequencing technology while still generating an assembly with accuracy usually comparable to the competing algorithms. Fifth, we use moderate long read coverage datasets (e.g., 30$\times$) and show that Apollo can produce accurate assemblies even with a moderate read coverage. We conclude that Apollo is the \emph{first} universal assembly polishing algorithm that 1) scales well to polish assemblies of both large and small genomes, and 2) can use both long and short reads as well as a hybrid set of reads from various sequencing technologies.

This paper makes the following contributions:
\begin{itemize} \vspace{-0.6mm}
    \item We introduce Apollo, a new assembly polishing algorithm that can make use of reads sequenced by \emph{any} sequencing technology (e.g., PacBio, ONT, Illumina reads). Apollo is the \emph{first} assembly polishing algorithm that 1) is scalable such that it can polish assemblies of both large and small genomes, and 2) can polish an assembly with a hybrid set of reads within a single run.
    \item We show that using both long and short reads in a hybrid manner to polish a Canu-generated assembly enables the construction of assemblies more accurate than those constructed by running other polishing tools multiple times.
    \item We show that four competing polishing algorithms cannot scale well to polish assemblies of large genomes within a single run due to large computational resources that they require.
    \item We provide an open source implementation of Apollo\\ (https://github.com/CMU-SAFARI/Apollo).
\end{itemize}

\begin{methods}
\vspace{-6mm} \section{Methods} \label{sec:methods}

\begin{figure*}
\centerline{\includegraphics[scale=0.20]{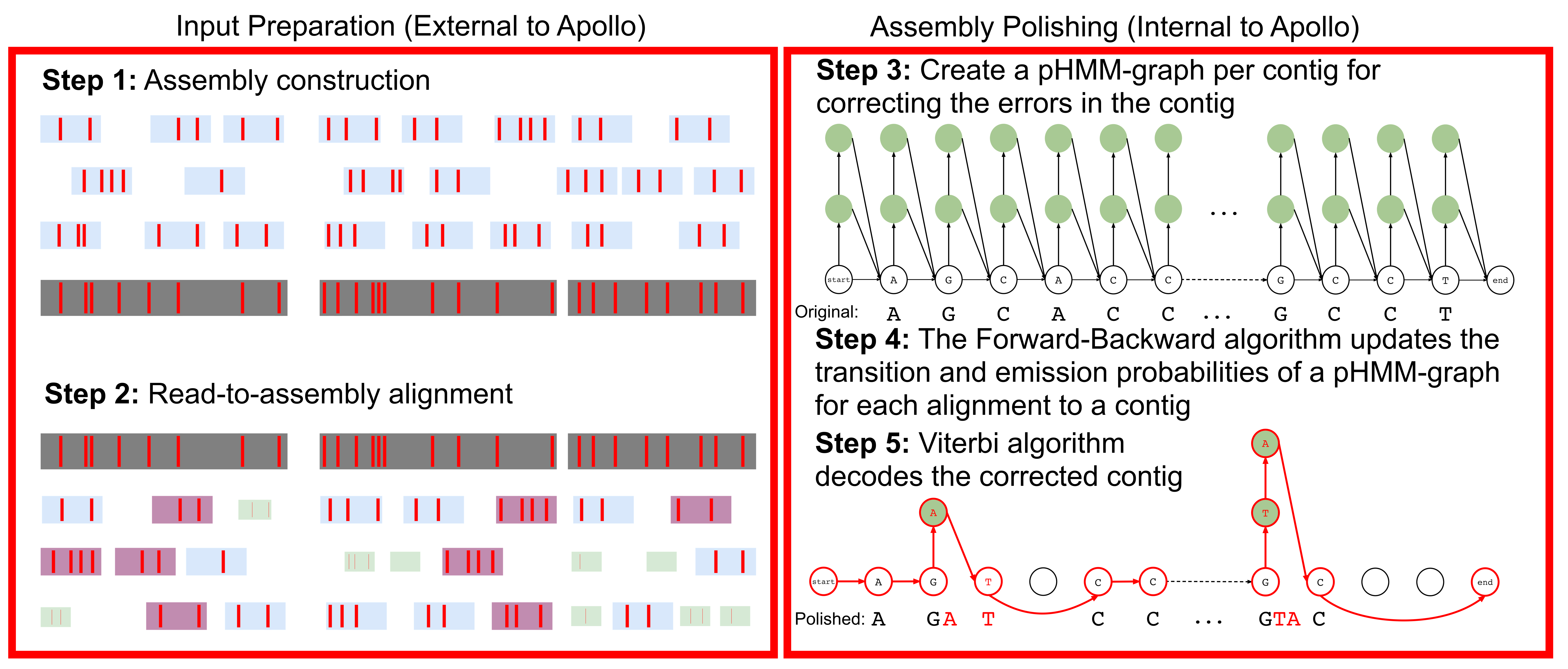}}
\vspace{-5mm}
\caption{Input preparation and the pipeline of Apollo algorithm in five steps. The first two steps refer to the use of external tools to generate the input for Apollo and are called \emph{input preparation steps} (left side). (Step 1) An assembler generates the assembly (dark gray, large rectangles) using erroneous reads (light blue rectangles). Here the errors are labeled with the red bars inside the rectangles. (Step 2) An aligner aligns the reads used in the first step as well as additional reads to the assembly. Here we show the reads sequenced using different sequencing technologies in different colors and sizes (e.g., a short rectangle indicates a short read) since it is possible to use any available read within a single run with Apollo. The rest of the three steps constitute the new Apollo algorithm and are called \emph{Internal to Apollo} (right side). (Step 3) Apollo creates a profile hidden Markov model graph (pHMM-graph) per assembly contig. Here, we show an example for the pHMM-graph generated for the contig that starts with "$AGCACC$" and ends with "$GCCT$" as we show the original sequence below the states labeled with a base pair. Each base pair in a contig is represented by a state labeled with the corresponding base pair (i.e., match state). A pHMM graph also consists of insertion states for each base pair labeled with green color as well as start and end states that do not correspond to any base pair in a contig. In this example, the maximum insertion that can be made between each base pair is two as we have two insertion states per match state. Each transition or emission of a base pair from a state has a probability associated with it. For simplicity, we omit deletion transitions from this graph. (Step 4) The Forward-Backward algorithm trains the pHMM-graph and updates the transition and emission probabilities based on read-to-assembly alignments. (Step 5) Using the updated probabilities, the Viterbi algorithm decodes the most likely path in the pHMM-graph and takes the path marked with the red transitions and states, which corresponds to the polished assembly. We also show the corresponding corrections in red text color below the states. For each contig, the output of Apollo is the sequence of base pairs associated with the states in the most likely path.}
\label{fig:overview}
\vspace{-5mm}
\end{figure*}

Apollo builds, trains, and decodes a profile hidden Markov model graph (pHMM-graph) to polish an assembly (i.e., to correct the errors of an assembly). Apollo performs assembly polishing using two input preparation steps that are external to Apollo (pre-processing) and three internal steps, as shown in Figure~\ref{fig:overview}. The first two pre-processing steps involve the use of external tools such as an \emph{assembler} and an \emph{aligner} to generate inputs for Apollo. First, an assembler uses reads (e.g., long reads) to generate assembly contigs (i.e., larger sequence fragments of the assembly). Second, an aligner aligns the reads used in the first step and any additional reads (e.g., short reads) of the same sample to the contigs to generate read-to-assembly alignment. Third, Apollo uses the assembly generated in the first step to construct a pHMM-graph per contig. A pHMM-graph is comprised of states, transitions between states, and probabilities that are associated with both states and transitions to account for all possible error types. Examples of errors that a sequencing technology can introduce into a read are insertion, deletion and substitution errors (which we handle in this work), and chimeric errors (which we do not handle). Therefore, correction of these errors can be accomplished by deleting, inserting, or substituting the corresponding base pair, respectively. Apollo identifies a path in the pHMM-graph such that the states that make the contig erroneous are excluded. Fourth, Apollo uses the read-to-assembly alignment to update, or train, the initial (\emph{prior}) probabilities of the pHMM-graph with the Forward-Backward and Baum-Welch algorithms. During training, the Forward-Backward algorithm uses each read alignment to change the prior probabilities of the graph based on the similarity between a read and the aligned region in the assembly. Fifth, Apollo implements the Viterbi algorithm to find the path in the pHMM-graph with the minimum error probability (i.e., decoding), which corresponds to the polished version of the corresponding contig.

\vspace{-3mm} \subsection{Assembly construction}
\label{sec:assembly}

An assembler takes a set of reads as input and identifies the overlaps between the reads in order to merge the overlapped regions into larger fragments called contigs. An assembler usually reports contigs in FASTA format~\citep{Pearson1988} where each element is comprised of an ID and the full sequence of the contig. The entire collection of contigs represents the whole assembly. Apollo requires the assembly to be constructed to correct the errors in each contig of the assembly. Thus, assembly generation is an external step to the assembly polishing pipeline of Apollo (Figure~\ref{fig:overview} Step 1). Apollo supports the use of any assembler that can produce the assembly in FASTA format~\citep{Pearson1988}, such as Canu~\citep{Koren2017} and Miniasm~\citep{Li2016a}.

\vspace{-3mm} \subsection{Read-to-assembly alignment}
\label{sec:alignment}

After assembly construction, the second external step is to generate the read-to-assembly alignment using 1) the reads that the assembler used to construct the assembly and 2) any additional reads sequenced from the same sample (Figure~\ref{fig:overview} Step 2). It is possible to use \emph{any} aligner that can produce the read-to-assembly alignment in SAM/BAM format~\citep{Li2009a} such as Minimap2~\citep{Li2018} or BWA-MEM~\citep{Li2009}. In the case where reads from multiple sequencing technologies are available for a given sample, an aligner aligns all reads to the assembly. Apollo assumes that the alignment file is coordinate sorted and indexed.

Apollo uses the assembly and the read-to-assembly alignment generated in the first two pre-processing steps in its assembly polishing steps. The next three steps (Steps 3-5) are the assembly polishing steps and implemented within Apollo.

\vspace{-3mm} \subsection{Creating a pHMM-graph per contig}
\label{sec:phmm}

The pHMM-graph that Apollo employs includes states that emit certain characters, directed transitions that connect a state to other states, and probabilities associated with character emissions and state transitions. The state transition probability represents the likelihood of following a path from a state to another state using the transitions connecting the states, and the character emission probability represents the likelihood for a state to emit a certain base pair when the state is visited. These pHMM-graph elements enable a pHMM-graph to provide the probability of generating a certain sequence when a certain path of states is followed using the directed transitions between the states.

This probabilistic behavior of pHMM-graphs makes them a good candidate to resolve errors of an assembly. Apollo represents each contig of an assembly as a pHMM-graph. The complete structure of a pHMM-graph allows Apollo to handle three major types of errors: substitution, deletion, and insertion errors. First, Apollo represents \emph{each} base pair of a contig as a state, called the \emph{match state}. The pHMM-graph preserves the sequence order of the contig by inserting a directed \emph{match transition} from the previous match state of a base pair to the next one. The match state of a certain base pair has a predefined (\emph{prior}) \emph{match emission probability} for the corresponding base pair, and \emph{mismatch emission probability} for the three remaining possible base pairs (i.e., a substitution error). A match state handles the cases when there is no error in the corresponding base pair (i.e., emitting the base pair that already exists in the certain position), or when there is a \emph{substitution error} (i.e., emitting a different base pair for the certain position). Second, there are $l$ many \emph{insertion states} for each base pair in the contig where $l$ is a parameter to Apollo, which defines the maximum number of additional base pairs that can be inserted between two base pairs (i.e., two match states). An insertion state inserts a single base pair in the location it corresponds to (e.g., visiting two subsequent insertion states after a match state inserts two base pairs between the two match states) in order to handle a \emph{deletion error}. Last, each match and insertion state has $k$ many \emph{deletion transitions} where $k$ is also a parameter to Apollo, which defines the maximum number of contiguous base pairs that can be deleted with a single transition. If there is an \emph{insertion error}, a deletion transition skips the match states between a state (e.g., an insertion or a match state) to a match state in order to delete the corresponding base pairs of the skipped match states. Further details of the pHMM-graph can be found in Supplementary Materials (Section 1).

The pHMM-graph structure that Apollo uses is identical to the one proposed in Hercules~\citep{Firtina2018}, a recently proposed error correction algorithm that uses pHMM-graphs. The key difference is that Apollo creates a graph \emph{for each contig} whereas Hercules creates a graph for each \emph{read}. As such, the pHMM-graph size in Apollo is usually larger than that in Hercules since contigs are typically longer than reads. Therefore, Apollo uses \emph{additional techniques} to handle large pHMM-graphs (e.g., dividing pHMM-graphs into smaller graphs without compromising correction accuracy) during both training and decoding steps, which has certain trade-offs with respect to implementation, as we explain in Sections~\ref{sec:training},~\ref{sec:decoding}, and~\ref{sec:experiment}.

\vspace{-3mm} \subsection{Training with the Forward-Backward algorithm}
\label{sec:training}

The training step of Apollo uses each read-to-assembly alignment to update transition and emission probabilities of a contig's pHMM-graph. The purpose of the training step is to make specific transitions and emissions more probable in a \emph{sub-graph} of the pHMM-graph such that it will be more likely to emit the entire read sequence for the region that the read aligns to. A sub-graph contains a subset of the states of a pHMM-graph and the transitions connecting these states. Each difference between a contig and the aligned read updates the probabilities so that it will be more likely to reflect the \emph{difference} observed in the read. The calculations during training do \emph{not} make assumptions about the sequencing technology of the read but only reflect the differences and similarities in the pHMM-graph. Thus, Apollo can update the sub-graph with \emph{any} read aligned to the contig. This makes Apollo a sequencing-technology-independent algorithm.

For each alignment to a contig, Apollo identifies the sub-graph that the read aligns to in the pHMM graph to update (train) the emission and transition probabilities in the sub-graph. Apollo locates the start and end states of the sub-graph to define its boundaries in the pHMM graph. First, Apollo identifies the start location of a read's alignment in the contig and marks the match state of the \emph{previous} base pair as the \emph{start state}. Second, Apollo estimates the location of the \emph{end state} such that the number of match states between the \emph{start state} and the \emph{end state} is longer than the length of the aligned read (i.e., up to $33.3$\% longer). This is to account for the case where there are more insertion errors than deletion errors. The Backward calculation uses the end state as the initial point to calculate the probabilities from backward as we explain later in this section. An accurate estimation of the end state is crucial as an inaccurate initial point for the Backward calculation may lead to inaccurate training. The insertion and the match states between the start and the end states as well as the transitions connecting these states constitute the sub-graph of the aligned region.

The sub-graphs that Apollo trains usually vary in size since the length of long reads (i.e., reads sequenced by the third-generation sequencing technologies) can fluctuate dramatically (e.g., from 15bps to 2Mbps) whereas the length of short reads is usually fixed (e.g., 100bps). As Apollo polishes the assembly using both short and long reads, the broad range of read lengths requires Apollo to be flexible in terms of defining the \emph{length of the sub-graph} (i.e., the number of match states that the sub-graph includes) to train. This is a key difference in requirements between Apollo and Hercules~\citep{Firtina2018}. Hercules defines the number of match states to include in a sub-graph with a \emph{fixed} ratio as the aligned reads are \emph{always} short reads. However, Apollo is more flexible in the selection of the region that a sub-graph covers since Apollo can use reads of any length. Apollo decides whether the aligned read is \emph{short} or \emph{long} based on the read length, of which we set the threshold at 500bps (i.e., if a read is longer than 500bps, it is considered as a long read). If the aligned read length is \emph{short} (i.e., shorter than 500bps), the sub-graph is $33.3$\% longer than the length of the short read. Otherwise, the sub-graph is $5$\% longer than the length of the aligned long read (empirically chosen).

Apollo uses the Forward-Backward and the Baum-Welch algorithms \citep{Baum1972} to train the sub-graph that a read aligns to. The Forward-Backward algorithm takes the aligned read as an observation and updates the emission and transition probabilities of the states in the sub-graph. There are three steps in the Forward-Backward algorithm: 1) Forward calculation, 2) Backward calculation, and 3) training by updating the probabilities (i.e., the expectation-maximization step using the Baum–Welch algorithm). First, Forward calculation visits each possible path from the start state up to but not including the end state until each visited state emits a single base pair from the read starting from the first (i.e., leftmost) base pair. Therefore, the number of visited states is equal to the length of the aligned read. Second, similar to Forward calculation, Backward calculation visits each possible path in a backward fashion (i.e., from the last base pair to the first base pair) starting with the state that the Forward calculation determines to be the most likely until the start state. Third, the Forward-Backward algorithm updates the transitions and emission probabilities based on how likely it is to take a certain transition or a state to emit a certain character. We refer to the updated probabilities as \emph{posterior probabilities}. In theory, the training step known as the Baum--Welch algorithm \citep{Baum1972} is separated from the Forward-Backward calculations, as described in Section 3 of Supplementary Materials. However, for the sake of simplicity, we assume that the Forward-Backward step includes both the Forward-Backward calculations and the training step when we refer to it in the remaining part of this paper. Apollo trains each sub-graph (i.e., each read alignment) independently even though the states and the transitions may overlap between the aligned reads. For overlaps, Apollo takes the average of the posterior transition and emission probabilities of the overlapping regions. Once Apollo trains each pHMM sub-graph using all the alignments to a contig, it completes the training phase for that contig. The trained pHMM-graph represents the polished version of the contig. Sections 2 and 3 in the Supplementary Materials describe in detail how Apollo locates a sub-graph per read alignment and the training phase of the Forward-Backward algorithm.

\vspace{-3mm} \subsection{Decoding with the Viterbi Algorithm}
\label{sec:decoding}

The last step in Apollo's assembly polishing mechanism is the decoding of the trained pHMM-graph in order to extract the path with the highest probability from the start of the graph to the end of the graph. Finding the path with the highest probability reveals the consensus of the aligned reads to correct the contig. To identify this path, Apollo uses the Viterbi algorithm~\citep{Viterbi1967} on the trained pHMM-graph (Figure~\ref{fig:overview} Step 5). The Viterbi algorithm is a dynamic programming algorithm that finds the most likely \emph{backtrace} from a certain state to the start state in a given graph. Each Viterbi value represents how likely it is to be in a certain state at a time $t$ (i.e., position in the contig) and is stored in the corresponding cell in a table called a dynamic programming table (DP table). Thus, a complete DP table reveals the most likely path of the entire pHMM-graph by backtracking the most likely path from the end state to the start state.

The Viterbi algorithm computes each entry of the dynamic programming table using the Viterbi values of the previously visited states. This data dependency makes the Viterbi algorithm less suitable for multi-threading support, as it prevents calculating the Viterbi values of the entire graph in parallel. Apollo overcomes this issue by dividing the pHMM-graph into sub-graphs (i.e., chunks), each of which includes a certain number of states. The Viterbi algorithm decodes each sub-graph (i.e., finds the optimal path in a graph) and merges the decoding results into one piece again. Since the Viterbi algorithm can decode each sub-graph independently, this allows Apollo to parallelize the Viterbi algorithm. We find that our parallelization greatly speeds up the Viterbi algorithm, by ${\sim}20\times$.

Apollo follows a slightly different approach than the actual Viterbi algorithm when decoding a graph. The actual Viterbi algorithm uses an observation provided as input (i.e., a sequence of base pairs) to calculate the Viterbi values of states in the graph. For Apollo, there is no observation provided as input. Apollo uses the base pair with the highest emission probability of a state as observation when calculating the Viterbi value of that state. For each state in the decoded path, Apollo outputs the base pair with the highest probability, which corresponds to the polished contig. Apollo reports each polished contig as a read in FASTA format. Details of the Viterbi algorithm are in Supplementary Materials (Section 4).

Note that Apollo can only polish contigs to which at least a single read aligns. Thus, Apollo reports an unpolished version of a contig, if there is no read aligned to it. In such cases, Apollo also reports the issue as output by informing that a certain contig cannot be polished because there is no read aligned to the contig. After raising the issue, Apollo continues polishing the remaining contigs, if any. We expect that such a case happens rarely. For example, a low coverage set of short reads may not be able to align to a too small and erroneous contig constructed using long reads, which would leave the contig with no read aligned to it. Another example would be having very similar regions (i.e., repetitive regions) in multiple contigs such that reads can be assigned to only one of the contigs sharing a similar region. Such a case may leave a contig without any read aligned to it since these reads may already be aligned to the similar regions in other contigs.

\end{methods}

\vspace{-3mm} \section{Results} \label{sec:results}
\subsection{Experimental Setup} \label{sec:experiment}
We implemented Apollo in C++ using the SeqAn library~\citep{Doring2008}. The source code is available at https://github.com/CMU-SAFARI/Apollo. Apollo supports multi-threading.

Our evaluation criteria include three different methods to assess the quality of the assemblies. First, we use the \emph{dnadiff} tool provided under MUMmer package~\citep{Kurtz2004} to calculate the accuracy of polished assemblies by comparing them with the highly-accurate reference genomes (i.e., ground truth genomes). We report the percentage of bases of an assembly that align to its reference (i.e., \emph{Aligned Bases}), the fraction of identical portions between the aligned bases of an assembly and the reference (i.e., \emph{Accuracy}), a score value that is the product of \emph{accuracy} and number of \emph{aligned bases} (as a fraction), which we call the \emph{Polishing Score}. \emph{Accuracy} value provides the accuracy of only the aligned portions of the polished assembly, not the entire assembly. However, \emph{polishing score} is a more comprehensive measure compared to \emph{accuracy}, as it normalizes the accuracy of the aligned portions of the polished assembly to the entire length of the assembly. Second, we use sourmash~\citep{TitusBrown2016} to calculate the k-mer similarity between filtered Illumina reads and an assembly. Third, we use QUAST~\citep{Gurevich2013} to report a further quality assessment of assemblies based on the mapping of filtered Illumina reads to assemblies. Both k-mer similarity and QUAST provide a reference-independent evaluation of assemblies.

Based on our evaluation criteria, we compare Apollo to four state-of-the-art assembly polishing algorithms: Nanopolish~\citep{Loman2015}, Racon~\citep{Vaser2017}, Quiver~\citep{Chin2013}, and Pilon~\citep{Walker2014}. If an assembly polishing algorithm does not support a certain dataset, we do not run the algorithm on that dataset. For example, we use Nanopolish only for the ONT dataset and Quiver only for PacBio datasets, and Pilon only for the Illumina dataset. We use Pilon with a PacBio dataset only once to show its capability to polish an assembly using long reads, albeit very inefficiently. We include Apollo and Racon in every comparison as they support a set of reads from any sequencing technology. For each dataset, we compare the algorithms that polish an assembly using the same set of reads. We run each assembly polishing algorithm with its default parameters.

We run all the tools (i.e., assemblers, read mappers, and assembly polishing algorithms) on a server with 24 cores (2 threads per core, Intel\textregistered Xeon\textregistered Gold 5118 CPU @ 2.30GHz), and 192GB of main memory. We assign 45 threads to all the tools we use and collect their runtime and memory usage using the \emph{time} command in Linux with the $-vp$ options. We report runtime and peak memory usage of the assembly polishing algorithms based on these configurations.

We use state-of-the-art tools to construct an assembly and to generate a read-to-assembly alignment before running Apollo, which correspond to the input preparation steps. We use Canu~\citep{Koren2017} and Miniasm~\citep{Li2016a} tools to construct assemblies of each set of long reads. For read-to-assembly alignment, we use Minimap2 and BWA-MEM to align long and short reads to an assembly. Quiver cannot work with alignment results that Minimap2 and BWA-MEM produce, but requires a certain type of aligner to align PacBio reads to an assembly. Thus, we use the \emph{pbalign} tool (https://github.com/PacificBiosciences/pbalign) that uses BLASR~\citep{Chaisson2012} to align PacBio reads to an assembly in order to generate a read-to-assembly alignment in the format that Quiver requires. We sort and index the resulting SAM/BAM read-to-assembly alignments using the SAMtools' sort and index commands~\citep{Li2009a}, respectively.

After assembly generation, we divide the long reads into smaller \emph{chunks} of size 1000bps (i.e., we perform \emph{chunking}). We do this because long reads cause high memory demand during the assembly polishing step, especially for large genomes (e.g., a human genome). This bottleneck exists not only for Apollo but also for other assembly polishing algorithms (e.g., Racon). For Apollo, dividing long reads into chunks prevents possible memory overflows due to the memory-demanding calculation of the Forward-Backward algorithm. Even though it is still possible to use long reads without chunking, we suggest using the resulting reads \emph{after chunking} if the available memory is not sufficient to run Apollo. We show that chunking results in producing more accurate assemblies (Supplementary Table S18).

Default parameters of Apollo are as follows: minimum mapping quality ($q=0$), maximum number of states that Forward-Backward ($f=100$) and the Viterbi algorithms ($v=5$) evaluate for the next time step, the number of insertion states per base pair ($i=3$), the number of base pairs decoded per sub-graph by Viterbi ($b=5000$), maximum deletions per transition ($d=10$), transition probability to a match state ($tm=0.85$), transition probability to an insertion state ($ti=0.1$), factor for the polynomial distribution to calculate each deletion transition ($df=2.5$), and match emission probability ($em=0.97$).

\begin{table*}
\begin{center}
\caption{Details of our datasets}
\label{tab:data}
\begin{tabular}{|l|l|l|}
\hline
\tcentb{Dataset} & \tcent{Accession Number}  & \tcent{Details}  \\\hhline{|=|=|=|}
\textit{E. coli} K-12 - ONT & Loman Lab$^*$ & 164,472 reads (avg. 9,010bps, 319$\times$ coverage) \\
\textit{E. coli} K-12 - Illumina & SRA SRR1030394 & 2,720,956 paired-end reads (avg. 243bps each, 285$\times$ coverage) \\
\textit{E. coli} K-12 - Ground Truth & GenBank NC\_000913 & Strain MG1655 (4,641Kbps) \\\hline
\textit{E. coli} O157 - PacBio & SRA SRR5413248 & 177,458 reads (avg. 4,724bps, 151$\times$ coverage) \\
\textit{E. coli} O157 - Illumina 	& SRA SRR5413247 & 11,856,506 paired-end reads (150bps each, 643$\times$ coverage) \\
\textit{E. coli} O157 - Ground Truth & GenBank NJEX02000001 & Strain FDAARGOS\_292 (5,566Kbps) \\\hline
\textit{E. coli} O157:H7 - PacBio & SRA SRR1509640 & 76,279 reads (avg. 8,270bps, 112$\times$ coverage) \\
\textit{E. coli} O157:H7 - Illumina  & SRA SRR1509643 & 2,978,835 paired-end reads (250bps each, 265$\times$ coverage) \\
\textit{E. coli} O157:H7 - Ground Truth & GCA\_000732965 & Strain EDL933 (5,639Kbps) \\\hline
Yeast S288C - PacBio & SRA ERR165511(8-9), ERR1655125 & 296,485 reads (avg. 5,735bps, 140$\times$ coverage) \\
Yeast S288C - Illumina & SRA ERR1938683 & 3,318,467 paired-end reads (150bps each, 82$\times$ coverage) \\
Yeast S288C - Ground Truth & GCA\_000146055.2 & Strain S288C (12,157Kbps) \\\hline
Human HG002 - PacBio & SRA SRR2036(394-471), SRR203665(4-9) & 15,892,517 reads (avg. 6,550bps, 35$\times$ coverage) \\
Human HG002 - Illumina  & SRA SRR17664(42-59) & 222,925,733 paired-end reads (148bps each, 22$\times$ coverage) \\
Human HG002 - Ground Truth & GCA\_001542345.1 & Ashkenazim trio - Son (2.99Gbps) \\\hline
\end{tabular}
\end{center}
{\small The datasets we use in our experiments. This data can be accessed through NCBI using the accession number. \\ $^*$The ONT datasets are available at http://lab.loman.net/2016/07/30/nanopore-r9-data-release/}
\vspace{-4mm}
\end{table*}

\vspace{-3mm} \subsection{Datasets} \label{sec:dataset}

In our experiments, we use DNA-seq datasets from five different samples sequenced by multiple sequencing technologies, as we show in Table~\ref{tab:data}.

We use a dataset from a large genome (i.e., a human genome) to demonstrate the scalability of polishing algorithms. For this purpose, we use the human genome sample from the Ashkenazim trio (HG002, Son) to compare the computational resources (i.e., time and maximum memory usage) that each polishing algorithm requires. We filter out the PacBio reads that have a length of less than 200 before calculating coverage and average read length.

We use the \textit{E. coli} O157 (Strain FDAARGOS\_292), \textit{E. coli} O157:H7, \textit{E. coli} K-12 MG1655, and Yeast S288C datasets to evaluate the polishing accuracy of Apollo and other state-of-the-art polishing algorithms in four ways. First, we evaluate whether using a hybrid set of reads with Apollo results in more accurate assemblies compared to polishing an assembly twice using a combination of other polishing tools (e.g., Racon + Pilon). Second, we measure the performance of the polishing algorithms when they polish the assemblies only once. Third, we subsample the \textit{E. coli} O157 and \textit{E. coli} K-12 datasets into 30$\times$ coverage to compare the performance of algorithms when long read coverage is moderate. Fourth, we additionally use the Human HG002 dataset to measure the k-mer distance and quality assessment of the assemblies using sourmash and QUAST, respectively.

\vspace{-3mm} \subsection{Applicability of Polishing Algorithms to Large Genomes} \label{sec:applicability}

We use the polishing algorithms to polish a large genome assembly (e.g., a human genome) to observe (1) whether the polishing algorithms can polish these large assemblies without exceeding the limitations of the computational resources we use to conduct our experiments and (2) the overall computational resources required to polish a large genome assembly (i.e., alignment and polishing). For this purpose, we use the PacBio and Illumina reads from the human genome sample of the Ashkenazim trio (HG002, Son) to polish a \emph{finished assembly} of the same Ashkenazim trio sample. The finished assembly was released by the Genome in a Bottle (GIAB) consortium (genomeinabottle.org). GIAB used 1) Celera Assembler with PbCR (v. 8.3rc2)~\citep{Koren2012} to assemble the PacBio reads from the HG002 sample and 2) Quiver to polish the assembly~\citep{Wenger2019}. Based on our experiments that we report in Table~\ref{tab:son}, we make four key observations. First, Pilon, Quiver, and Racon \emph{cannot} polish the assembly using the whole sets of PacBio (${\sim}35\times$ coverage) and Illumina (${\sim}22\times$ coverage) reads due to high computational resources that they require. Racon and Pilon exceed the memory limitations while using either the PacBio or Illumina reads to polish the human genome assembly. Quiver cannot start polishing the assembly as the required aligner (i.e., BLASR from the pbalign tool) cannot produce the alignment result due to memory limitations. Apollo can polish an assembly using \emph{both} PacBio and Illumina reads using at most nearly half of the available memory. Second, we reduce the coverage of the PacBio reads to 8.9$\times$ (SRA SRR2036394-SRR2036422) to observe whether Racon and Quiver can polish the large genome using a low coverage set of PacBio reads. We find that Racon is able to polish a human genome assembly using low coverage set of reads whereas BLASR \emph{cannot} produce the alignment results that Quiver requires due to memory limitations even when using a low coverage set of reads. Third, we split read-to-assembly alignment into multiple alignment files such that all reads mapped to each contig are represented in a separate alignment file (i.e., \emph{read-to-contig} alignment) to evaluate whether Pilon, Quiver, and Racon can polish the entire human genome using read-to-contig alignments. We observe that Pilon, Quiver, and Racon can polish contigs of a large genome, as Table~\ref{tab:son} shows. We note that when using pbalign, we align small batches of PacBio datasets (e.g., 1$\times$ coverage each) and later merge the alignments of these small batches. We also note that both the size of the longest contig (i.e., 35.2Mbp) and the number of short read alignments to the longest contig (i.e., 5,313,903) are $\sim85\times$ smaller than that of the entire assembly. When contigs longer than 35Mbp are available, we expect Pilon and Racon to require more memory for polishing longer contigs since these tools cannot scale well with contig size. Fourth, Apollo requires less memory than any polishing algorithm when polishing the human genome assembly contig by contig. We conclude that Apollo is the \emph{only algorithm} that scales well (i.e., memory requirements do not increase dramatically as the genome size increases) in polishing large genomes using a set of both PacBio and Illumina reads without reducing the coverage of the read set or splitting the read set or the alignment file into smaller batches. Pilon, Quiver, and Racon can polish a large genome assembly without reducing the coverage of a read set only if they polish the entire assembly contig-by-contig or split the readset into smaller batches before alignment.

\vspace{-4mm} \subsection{Polishing Accuracy} \label{sec:accuracy}
We first examine whether the use of a hybrid set of reads (e.g., long and short reads) within a single polishing run provides benefit over polishing an assembly twice using a set of reads from only a single sequencing technology (e.g., only PacBio reads) in each run. Second, we evaluate assembly polishing algorithms and compare them to each other given different options with respect to 1) the sequencing technology that produces long reads, 2) the assembler that constructs an assembly using long reads, 3) the aligner that generates read-to-assembly alignment, and 4) the set of reads that align to an assembly. We report the accuracy of unpolished assemblies as well as the performance of assembly polishing algorithms based on the evaluation criteria we explained in Section~\ref{sec:results}. We also compare the tools based on their performance given moderate (e.g., ${\sim}30\times$) and low (e.g., $2.6\times$) long read coverage.

\textbf{Apollo is either more accurate than or as accurate as running Pilon twice using a hybrid set of reads. Apollo also polishes Canu-generated assemblies more accurately for a species with PacBio reads than running other polishing tools multiple times.} In Table~\ref{tab:hybrid} (complete results in Supplementary Table S1) and Supplementary Table S2, we highlight the benefits of using a hybrid set of reads (e.g., PacBio + Illumina) within a single polishing run compared to polishing an assembly in multiple runs by using a set of reads from only a single sequencing technology (e.g., only PacBio or only Illumina) in each run. To this end, we compare the accuracy of polished assemblies using Apollo with that of the polished assemblies using other polishing tools (Nanopolish, Pilon, Quiver, and Racon) that we run multiple times. We use long (PacBio or ONT) and short (Illumina) reads from \textit{E. coli} O157, \textit{E. coli} O157:H7, \textit{E. coli} K-12 MG1655, and Yeast S288C datasets to polish Canu- and Miniasm-generated assemblies. For the \emph{first run}, we use the polishing algorithms to polish Canu- and Miniasm-generated assemblies. For the \emph{second run}, we provide Nanopolish, Pilon, Quiver, and Racon with the polished assembly from the first run and run these tools for the second time (i.e., \emph{Second Run}). Based on Supplementary Tables S1 and S2, we make three key observations. First, Apollo and Pilon are the only algorithms that \emph{always} polish a Canu-generated assembly with a \emph{polishing score} either equal to or better than that of the original Canu-generated assembly. Second, running other polishing tools multiple times to polish a Miniasm-generated assembly usually results in assemblies with higher polishing scores (e.g., by at most 3.79\% for PacBio and 7.57\% for ONT read sets) than using Apollo with a hybrid set of reads. Third, Apollo performs better when it uses PacBio reads in the hybrid set than using ONT reads. We conclude that the use of Apollo once with a hybrid set of reads that includes PacBio reads and a Canu-generated assembly is the best pipeline (i.e., one can construct the \emph{most} accurate assemblies for a species versus running other polishing tools multiple times).

\textbf{Apollo performs better than Pilon and comparable to Racon and Quiver when polishing a Canu-generated assembly using only a high coverage set of PacBio or Illumina reads}. In Supplementary Tables S3, S6, and S12, we use PacBio and Illumina datasets to compare the performance of Apollo with Racon~\citep{Vaser2017}, Quiver~\citep{Chin2013}, and Pilon~\citep{Walker2014}. Based on these datasets, we make five observations. First, Apollo usually outperforms Pilon (i.e., 4 out of 7, see the \emph{Polishing Score} column) using a set of short reads. Second, Apollo, Racon, and Quiver show significant improvements over the original Miniasm assembly in terms of accuracy. Third, Quiver and Racon polish the Miniasm-generated assembly more accurately than Apollo (see the \emph{Accuracy} and the \emph{Polishing Score} columns). Fourth, Apollo produces more accurate assemblies than the assemblies polished by Racon when we use moderate (${\sim}30\times$) and high coverage (151$\times$) PacBio read sets to polish Canu-generated assemblies. However, both algorithms generate assemblies with lower accuracy than the accuracy of the original Canu-generated assembly ($0.9998$ with the polishing score of $0.9992$) when we use high coverage read sets. Based on this observation, we suspect that the use of the original set of long reads (i.e., the set of reads that we use to construct an assembly) is not helpful as Canu corrects long reads before constructing an assembly. Thus, we also tried using the Canu-corrected long reads to polish a Canu-generated assembly. However, the use of corrected long reads did not consistently result in generating more accurate assemblies than the assemblies polished using the original set of long reads as we report in Supplementary Tables S3 and S9. We find that the alignment of Canu-corrected long reads to an erroneous assembly generates a smaller number of alignments than the alignment of the original long reads to the same erroneous assembly, as we show in Supplementary Table S17. We believe that the decrease in the number of alignments results in loss of information that assembly polishing algorithms use to polish an assembly, which subsequently leads to either similar or worse assembly polishing accuracy than using original set of long reads. Fifth, even though Pilon is not optimized to use long reads, we use Pilon to polish an assembly using long reads to observe if it polishes the assembly with comparable accuracy to the other polishing algorithms. We observe that Pilon significantly falls behind the other polishing algorithms in terms of our evaluation criteria. Thus, we do not use Pilon with long reads. We conclude that 1) Apollo usually performs better than Pilon when using short reads and 2) Apollo's performance is comparable to Racon and Quiver when using long PacBio reads to polish an assembly.

\textbf{Apollo performs better than Pilon and Nanopolish when polishing a Miniasm-generated assembly using only a set of Illumina and ONT reads, respectively.} We also investigate the performance of Apollo given the ONT dataset (\textit{E. coli} K-12 MG1655), compared to Nanopolish and Racon. We make two key observations based on the results we show in Supplementary Table S9. First, Racon provides the best performance in terms of the accuracy of contigs when the coverage is high (319$\times$) and the accuracy of the original assembly is low (e.g., a Miniasm-generated assembly). In the same setup, Apollo produces a more accurate assembly than Nanopolish. Second, even though Nanopolish produces the most accurate results with Canu using either high coverage (319$\times$) or moderate coverage (${\sim}30\times$) data, Apollo's polishing score differs only by at most ${\sim}1.21$\%. We conclude that Racon performs better than the competing state-of-the-art polishing algorithms if the coverage of a set of reads is high (e.g., 319$\times$). Apollo outperforms Nanopolish when polishing a Miniasm-generated assembly but Nanopolish outperforms Racon and Apollo when polishing a Canu-generated assembly. Thus, we also conclude that the accuracy of the original assembly dramatically affects the overall performance of Nanopolish as there is a significant performance difference between polishing Miniasm and polishing Canu assemblies. We suspect that the default parameter settings of Apollo may be a better fit for PacBio reads rather than ONT reads, which explains why Apollo performs worse with ONT datasets compared to PacBio datasets.

\textbf{Apollo is robust to different parameter choices}. In Supplementary Tables S19 - S21, we use the \textit{E. coli} O157 dataset to examine if Apollo is robust to using different parameter settings. To study the change in the performance of Apollo, we change the following parameters: maximum number of states that the Forward-Backward and the Viterbi algorithms evaluate for the next time step ($f$), number of insertion states per base pair ($i$), maximum deletion length per transition ($d$), transition probability to a match state ($tm$), transition probability to an insertion state ($ti$). We conclude that Apollo's performance is robust to different parameter choices as the accuracies of the Apollo-polished assemblies differ by at most 2\%.

\begin{table}
\begin{center}
\caption{Applicability, runtime, and memory requirements of four assembly polishing tools on a \emph{complete human genome assembly}}
\label{tab:son}
\vspace{-1mm}
{\renewcommand{\arraystretch}{0.8}
\resizebox{\columnwidth}{!}{
\begin{tabular}{|l|l|l||rr|}
\hline
\tcentb{Aligner}  & \tcent{Sequencing Tech.} & \tcentd{Polishing} & \tcentno{Runtime} & \tcent{Memory} \\
         		  & \tcent{of the Reads}     & \tcentd{Algorithm} &             	  & \tcent{(GB)}   \\\hhline{|=|=|=|==|}
Minimap2 & PacBio (35$\times$)   & Apollo     & 228h 43m 13s & 62.91  \\
BWA-MEM  & PacBio (35$\times$)   & Apollo     & 200h 13m 06s & 58.60  \\
Minimap2 & PacBio (35$\times$)   & Racon      & N/A          & N/A    \\
BWA-MEM  & PacBio (35$\times$)   & Racon      & N/A          & N/A    \\
pbalign  & PacBio (35$\times$)   & Quiver     & N/A          & N/A    \\\hhline{|=|=|=|==|}
Minimap2 & PacBio (8.9$\times$)  & Apollo     & 56h 21m 56s  & 44.99  \\
BWA-MEM  & PacBio (8.9$\times$)  & Apollo     & 42h 19m 09s  & 45.00  \\
Minimap2 & PacBio (8.9$\times$)  & Racon      & 3h 31m 37s   & 54.13  \\
BWA-MEM  & PacBio (8.9$\times$)  & Racon      & 2h 17m 21s   & 51.55  \\
pbalign  & PacBio (8.9$\times$)  & Quiver     & N/A          & N/A    \\\hhline{|=|=|=|==|}
Minimap2 & Illumina (22$\times$) & Apollo     & 98h 07m 05s  & 101.12 \\
BWA-MEM  & Illumina (22$\times$) & Apollo     & 105h 15m 05s & 107.06 \\
Minimap2 & Illumina (22$\times$) & Racon      & N/A          & N/A    \\
BWA-MEM  & Illumina (22$\times$) & Racon      & N/A          & N/A    \\
Minimap2 & Illumina (22$\times$) & Pilon      & N/A          & N/A    \\
Minimap2 & Illumina (22$\times$) & Pilon      & N/A          & N/A    \\\hhline{|=|=|=|==|}
Minimap2 & PacBio (35$\times$)   & Apollo$^*$ & 230h 37m 58s & 25.23 \\
pbalign  & PacBio (35$\times$)   & Quiver$^*$ & 104h 42m 35s & 29.92 \\
Minimap2 & PacBio (35$\times$)   & Racon$^*$  & 6h 48m 17s   & 132.51 \\
Minimap2 & Illumina (22$\times$) & Apollo$^*$ & 103h 27m 45s & 39.35 \\
BWA-MEM  & Illumina (22$\times$) & Apollo$^*$ & 111h 35m 15s & 39.35 \\
Minimap2 & Illumina (22$\times$) & Pilon$^*$  & 13h 59m 32s  & 66.67 \\
BWA-MEM  & Illumina (22$\times$) & Pilon$^*$  & 21h 15m 57s  & 49.93 \\\hline
\end{tabular}}}
\end{center}
{\small We polished the assembly of the Ashkenazim trio sample (HG002, Son) for different combinations of sequencing technology, aligner, and polishing algorithm. We report the runtime and the memory requirements of the assembly polishing tools (i.e., Aligner + Polishing). We report \emph{Runtime} and \emph{Memory} as N/A, if a polishing algorithm fails to polish the assembly. $^*$ denotes that we polish the assembly \emph{contig by contig} in these runs and collect the results once all of the contigs are polished separately.}
\vspace{-4mm}
\end{table}

\begin{table*}
\begin{center}
\caption{Comparison between using a hybrid set of reads with Apollo and running other polishing tools twice to polish a Canu-generated assembly}
\label{tab:hybrid}
\vspace{-1mm}
\begin{tabular}{|l|l|l|rrr||rr|}
\hline
\tcentb{Dataset} & \tcent{First Run} & \tcent{Second Run} & \tcentno{Aligned} 	 & \tcentno{Accuracy} & \tcentd{Polishing} & \tcentno{Runtime} & \tcent{Memory} \\
		   		 &       	   		 &  	  			  & \tcentno{Bases (\%)} &					  & \tcentd{Score}	   &             	   & \tcent{(GB)}   \\\hhline{|=|=|=|===|==|}
\textit{E. coli} O157	 & \tdash	    	& \tdash			& 99.94  & 0.9998 & 0.9992 & 43m 53s & 3.79 \\
\textit{E. coli} O157	 & Apollo (Hybrid) 	& \tdash	  		& 99.94  & 0.9999 & \textbf{0.9993}	 & 8h 16m 08s & 13.85 \\
\textit{E. coli} O157	 & Racon (PacBio)   & Racon (Illumina)	& 99.94  & 0.9994 & 0.9988 & 21m 44s & 22.65 \\
\textit{E. coli} O157	 & Pilon (Illumina) & Racon (PacBio)	& 99.94  & 0.9986 & 0.9980 & \textbf{4m 58s}	 & 11.40 \\
\textit{E. coli} O157	 & Quiver (PacBio)  & Pilon (Illumina)	& 99.94  & 0.9998 & 0.9992 & 5m 01s  & \textbf{7.50}  \\\hhline{|=|=|=|===|==|}
\textit{E. coli} O157:H7 & \tdash           & \tdash			& 100.00 & 0.9998 & 0.9998 & 43m 19s & 3.39  \\
\textit{E. coli} O157:H7 & Apollo (Hybrid)  & \tdash			& 100.00 & 0.9999 & \textbf{0.9999}  & 5h 58m 05s & 8.86  \\
\textit{E. coli} O157:H7 & Racon (PacBio)   & Racon (Illumina)	& 100.00 & 0.9995 & 0.9995 & 9m 43s  & \textbf{6.56}  \\
\textit{E. coli} O157:H7 & Pilon (Illumina) & Racon (PacBio)	& 100.00 & 0.9996 & 0.9996 & \textbf{6m 04s}  & 10.75 \\\hhline{|=|=|=|===|==|}
\textit{E. coli} K-12 	 & \tdash           & \tdash			& 99.98  & 0.9794 & 0.9792 & 34h 21m 46s & 5.06  \\
\textit{E. coli} K-12 	 & Apollo (Hybrid)  & \tdash 			& 99.99  & 0.9953 & 0.9952 & 9h 09m 50s & 9.35  \\
\textit{E. coli} K-12 	 & Racon (ONT)   	& Racon (Illumina)	& 100.00 & 0.9996 & \textbf{0.9996} & \textbf{11m 05s} & \textbf{5.10}  \\
\textit{E. coli} K-12 	 & Pilon (Illumina) & Racon (ONT)		& 99.99  & 0.9997 & \textbf{0.9996} & 15m 51s & 8.84 \\
\textit{E. coli} K-12	 & Nanopolish (ONT) & Pilon (Illumina)  & 99.99  & 0.9992 & 0.9991 & 9h 45m 01s & 18.10 \\\hhline{|=|=|=|===|==|}
Yeast S288C		 		 & \tdash        	& \tdash			& 99.89  & 0.9998 & 0.9987 & 1h 20m 39s & 6.24 \\
Yeast S288C		 		 & Apollo (Hybrid) 	& \tdash			& 99.89  & 0.9998 & \textbf{0.9987}  & 11h 08m 41s & \textbf{6.38} \\
Yeast S288C		 		 & Racon (PacBio)   & Racon (Illumina)	& 99.89  & 0.9994 & 0.9983 & 38m 21s & 6.93  \\
Yeast S288C		 		 & Pilon (Illumina) & Racon (PacBio)	& 99.89  & 0.9960 & 0.9949 & 21m 42s & 11.85 \\
Yeast S288C 			 & Quiver (PacBio)  & Pilon (Illumina)	& 98.95  & 0.9998 & 0.9893 & \textbf{12m 47s} & 13.28 \\\hline
\end{tabular}
\end{center}
{\small We use the long reads of \textit{E. coli} O157, \textit{E. coli} O157:H7, \textit{E. coli} K-12, and Yeast S288C datasets that are sequenced from PacBio and ONT (151$\times$, 112$\times$, 319$\times$, and 140$\times$ coverage, respectively) to generate their assemblies with \textbf{Canu}. Here, the polishing tools specified under \emph{First Run} and \emph{Second Run} polish the assembly using the set of reads specified in parentheses. The set of reads used in the second run is aligned to the assembly polished in the first run using Minimap2. PacBio and Illumina set of reads together constitute the hybrid set of reads (i.e., \emph{Hybrid}). We report the performance of the polishing tools in terms of the percentage of bases of an assembly that aligns to its reference (i.e., \emph{Aligned Bases}), the fraction of identical portions between the aligned bases of an assembly and the reference (i.e., \emph{Accuracy}) as calculated by dnadiff, and \emph{Polishing Score} value that is the product of \emph{Accuracy} and \emph{Aligned Bases} (as a fraction). We report the runtime and the memory requirements of the assembly polishing tools. We show the best result among \emph{assembly polishing algorithms} for each performance metric in \textbf{bold} text.}
\vspace{-4mm}
\end{table*}

\vspace{-3mm} \subsection{Reference-Independent Quality Assessment} \label{sec:kmerdistance}
We report both 1) the k-mer distance (i.e., Jaccard similarity~\citep{Niwattanakul2013} or \emph{k-mer similarity}) between \emph{filtered} Illumina reads and assemblies, and 2) quality assessment based on mapping these \emph{filtered} Illumina reads to assemblies to provide a reference-independent comparison between the polishing tools. We filter Illumina reads in three steps to get rid of erroneous short reads before using them. First, we remove the adapter sequences (i.e., adapter trimming). Second, we apply contaminant filtering for synthetic molecules. Third, we map the reads generated after the first three steps to the reference and filter out the reads that do not map to the reference. We use BBTools (sourceforge.net/projects/bbmap/) in these steps of filtering. To calculate k-mer similarity, we also use trim-low-abund~\citep{Zhang2015}, which applies k-mer abundance trimming to remove k-mers with abundance lower than 10 for \textit{E. coli} and Yeast datasets, and 3 for the human genome.

In k-mer similarity calculations, Jaccard similarity provides how a set of k-mers of both Illumina reads and an assembly are similar to each other. We compare the filtered Illumina reads with both polished and original (i.e., unpolished) assemblies of the small genomes (i.e., Yeast and \textit{E. coli}) and the large genomes (i.e., human); the results are in Supplementary Tables S4, S7, S10, S13, and S15. We show the percentage of both the k-mers of Illumina reads present in the assembly and the k-mers of the assembly present in Illumina reads. The latter helps us to identify \emph{how accurate} the assembly is whereas the former shows the \emph{completeness} of the assembly.

Based on our experiments on small genomes, we make three key observations. First, the tool with the highest \emph{assembly accuracy}, estimated with k-mer similarity (shown in Supplementary Tables S4, S7, S10, S13), typically provides the highest \emph{polishing score} in its category (shown in Supplementary Tables S3, S6, S9, S12), respectively. Second, Quiver usually produces more accurate assemblies than the assemblies generated by other polishing tools. Third, all polishing algorithms we evaluate dramatically increase the accuracy of the unpolished assembly generated by Miniasm. We conclude that the k-mer similarity results correlate with our findings in Section~\ref{sec:accuracy} and support our claims regarding how polished assemblies compare with the ground truth.

Based on the k-mer similarity results between the Illumina reads and the human genome assemblies, we make five key observations. First, we observe a reduction in the accuracy when polishing algorithms use raw PacBio reads as the finished assembly was generated using \emph{corrected} PacBio reads and \emph{already polished} by Quiver. Second, the polishing algorithms produce more accurate assemblies than the finished assembly \emph{only when} they use short reads to polish an assembly. This is because 1) Illumina reads are more accurate than raw PacBio reads and 2) Illumina reads have not been used when polishing the HG002 assembly, which leaves room to improve the accuracy. Third, Apollo performs better than Racon in terms of \emph{both} the completeness and the accuracy of the polished assemblies and better than Quiver in terms of accuracy (based on 51-mer results). Fourth, Apollo performs better than Pilon when it polishes the assembly using short reads. Fifth, using a low coverage readset to polish a human genome assembly dramatically reduces both the completeness of the assembly and the accuracy of the assembly. We conclude that 1) Apollo outperforms Pilon on Illumina data, and 2) it is not advisable to use raw PacBio reads to polish the large genome assemblies that have already been polished using more accurate reads than the raw PacBio reads (e.g., corrected PacBio reads).

We use QUAST~\citep{Gurevich2013}, a quality assessment tool for genome assemblies, to provide a different reference-independent assessment of the assemblies. QUAST takes paired-end \emph{filtered} Illumina reads to generate several metrics such as percentage of 1) mapped reads, 2) properly paired reads, 3) average depth of coverage, and 4) bases with at least 10$\times$ coverage. It also calculates the GC content (i.e., the ratio of bases that are either G or C) of the assembly. Based on the quality assessment results that we show in Supplementary Tables S5, S8, S11, S14, and S16, we make two key observations. First, for human genome assemblies, Apollo performs better than Racon and comparable to Pilon in terms of the percentage of the mapped reads, properly paired reads, and the bases with at least 10$\times$ read coverage. Second, for small genomes (i.e., Yeast and \textit{E. coli}), Quiver usually performs best in all of the metrics. We conclude that Apollo provides better performance when polishing large genomes than Racon, and Quiver usually performs better than any other polishing algorithm for small genomes.

\vspace{-3.2mm} \subsection{Computational Resources} \label{sec:resources}

We report the runtimes and the maximum memory requirements of both assemblers and assembly polishing algorithms in Supplementary Tables S1, S2, S3, S6, S9, and S12. Based on the runtimes of \emph{only} assembly polishing algorithms (i.e., Apollo, Nanopolish, Pilon, Quiver, and Racon), we make three observations. First, the machine learning-based assembly polishing tools, Apollo and Nanopolish, are the most time-consuming algorithms due to their computationally expensive calculations. For example, Racon is ${\sim}75\times$ and ${\sim}15\times$ faster than Apollo when polishing Miniasm-generated assemblies using PacBio and ONT read sets, respectively. Second, Racon becomes more memory-bound as the overall number of long reads in a read set increases (shown in Table~\ref{tab:son}). This shows that Racon's memory requirements are directly proportional to the size of the read set (i.e., the overall number of base pairs in a read set). Third, Quiver always requires the least amount of memory for \textit{E. coli} and Yeast genomes compared to the competing algorithms.

In Supplementary Tables S1 and S2, we evaluate the overall runtime and memory requirements of 1) polishing an assembly within a single run by using a hybrid set of reads with Apollo and 2) polishing an assembly multiple times. We observe that the overall runtime of running polishing tools multiple times is still lower at least by an order of magnitude than running Apollo once with a hybrid set of reads. However, Apollo can provide a more accurate assembly for a species when a Canu-generated assembly is polished, as discussed in Section~\ref{sec:accuracy}.

We report the runtimes, maximum memory requirements, and the parameters of the aligners we evaluated in Supplementary Tables S17 and S22, respectively, to observe how the aligner affects the overall runtime of both the aligner the assembly polishing tool. Based on the runtimes of aligners, we make two observations. First, pbalign is the most time-consuming and memory-demanding alignment tool. Overall, this makes Quiver require more time and memory than Racon, since Quiver can only work with BLASR, a part of pbalign tool. Second, all evaluated polishing tools except Quiver allow using any aligner; therefore, we only compare the runtime of the polishing tools, rather than comparing runtime of the full pipeline (i.e., aligner plus polishing tool) for the non-human genome datasets. We conclude that Quiver is the only algorithm whose runtime must be considered in conjunction with the aligner, as it can only use one aligner, pbalign, which we show in Table~\ref{tab:son}.

\vspace{-3.5mm} \subsection{Discussion} \label{sec:discussion}

We show that there is a dramatic difference between non-machine learning-based algorithms and the machine learning-based ones in terms of runtime. Apollo and Nanopolish usually require several hours to complete the polishing. Racon, Quiver, and Pilon usually require less than an hour (Supplementary Tables S1, S2, S3, S6, S9, and S12), which may suggest that Racon and Pilon can use a hybrid set of reads to polish an assembly in multiple runs instead of using Apollo in a single run. Indeed, we confirm that running Racon, Pilon, or Quiver multiple times still takes a much shorter time than running Apollo once using a hybrid set of reads within a single run. However, assembly polishing is a one-time task performed for an assembly that is usually used \emph{many} times and even made publicly available to the community. Therefore, we believe that long runtimes could still be acceptable given that genomic data produced by Apollo will probably be used \emph{many} times after it is generated. Hence, Apollo's runtime cost is paid only once but benefits are reaped many times. Note that this observation is not restricted to Apollo and applies to any polishing tool that has a long runtime. In addition, it is possible to accelerate the calculation of the Forward-Backward algorithm and the Viterbi algorithm using Tensor cores, SIMD and GPUs~\citep{Murakami2017, Eddy2011, Liu2009, Yu2015}, which we leave to future work.

Despite these slower runtimes of Apollo compared to other polishing tools, Apollo is new, unique, and useful because it provides two major functionalities that are not possible with prior tools. First, Apollo is the \emph{only} algorithm that can scale itself well to polish a large genome assembly using a readset with moderate coverage (e.g., up to ${\sim}35\times$) set of reads. Therefore, it is possible to polish a large genome with a relatively small amount of memory (i.e., less than 110GB) only with Apollo. Second, Apollo can construct more reliable Canu-generated assemblies compared to running other polishing tools multiple times when both PacBio and Illumina reads are used (i.e., a hybrid set of reads). These two advantages are \emph{only} possible if Apollo is used for assembly polishing.

\vspace{-5mm} \section{Conclusion} \label{sec:conclusion}

In this paper, we present a universal, sequencing-technology-independent assembly polishing algorithm, Apollo. Apollo uses all available reads to polish an assembly and removes the dependency of the polishing tool on sequencing technology. Apollo is the first polishing algorithm that scales well to use any arbitrary hybrid set of reads \emph{within a single run} to polish both large and small genomes. Apollo also removes the requirement of using assembly polishing algorithms multiple times to polish an assembly as it allows using a hybrid set of reads.

We show three key results. First, three state-of-the-art polishing algorithms, Quiver, Racon, and Pilon, cannot scale well to polish large genome assemblies without splitting the assembly into its contigs or read sets into smaller batches whereas Apollo scales well to polish large genomes. Second, using a hybrid set of reads with Apollo usually results in constructing Canu-generated assemblies more accurate than those generated when running other polishing tools multiple times. Third, Apollo usually polishes assemblies with comparable accuracy to state-of-the-art assembly polishing algorithms with a few exceptions that occur when long reads are used to polish Miniasm-generated assemblies. We conclude that Apollo is the first universal, sequencing-technology-independent assembly polishing algorithm that can use a hybrid set of reads within a single run to polish both large and small assemblies, while achieving high accuracy.

\vspace{-6mm} \section*{Funding}
This work was supported by gifts from Intel [to O.M.]; VMware [to O.M.]; and T\"{U}B\.{I}TAK [T\"{U}B\.{I}TAK-1001-215E172 to C.A.].\vspace*{-12pt}

\bibliographystyle{natbib}

\vspace{-3mm} \bibliography{main}

\begin{thebibliography}{}

\bibitem[Alkan {\em et~al.}(2011)Alkan, Sajjadian, and Eichler]{Alkan2011}
Alkan, C., Sajjadian, S., and Eichler, E.~E. (2011).
\newblock {Limitations of next-generation genome sequence assembly}.
\newblock {\em Nature Methods\/}, {\bf 8}(1), 61--65.

\bibitem[Alser {\em et~al.}(2017)Alser, Hassan, Xin, Ergin, Mutlu, and
  Alkan]{Alser2017a}
Alser, M., Hassan, H., Xin, H., Ergin, O., Mutlu, O., and Alkan, C. (2017).
\newblock {GateKeeper: a new hardware architecture for accelerating
  pre-alignment in DNA short read mapping}.
\newblock {\em Bioinformatics\/}, {\bf 33}(21), 3355--3363.

\bibitem[Alser {\em et~al.}(2019a)Alser, Hassan, Kumar, Mutlu, and
  Alkan]{Alser2019}
Alser, M., Hassan, H., Kumar, A., Mutlu, O., and Alkan, C. (2019a).
\newblock {Shouji: a fast and efficient pre-alignment filter for sequence
  alignment}.
\newblock {\em Bioinformatics\/}, {\bf 35}(21), 4255--4263.

\bibitem[Alser {\em et~al.}(2019b)Alser, Shahroodi, Gomez-Luna, Alkan, and
  Mutlu]{Alser2019b}
Alser, M., Shahroodi, T., Gomez-Luna, J., Alkan, C., and Mutlu, O. (2019b).
\newblock {SneakySnake: A Fast and Accurate Universal Genome Pre-Alignment
  Filter for CPUs, GPUs, and FPGAs}.

\bibitem[Au {\em et~al.}(2012)Au, Underwood, Lee, and Wong]{Au2012}
Au, K.~F., Underwood, J.~G., Lee, L., and Wong, W.~H. (2012).
\newblock {Improving PacBio Long Read Accuracy by Short Read Alignment}.
\newblock {\em PLoS One\/}, {\bf 7}(10), e46679.

\bibitem[Baum(1972)Baum]{Baum1972}
Baum, L.~E. (1972).
\newblock {An inequality and associated maximization technique in statistical
  estimation of probabilistic functions of a Markov process}.
\newblock {\em Inequalities\/}, {\bf 3}, 1--8.

\bibitem[Berlin {\em et~al.}(2015)Berlin, Koren, Chin, Drake, Landolin, and
  Phillippy]{Berlin2015}
Berlin, K., Koren, S., Chin, C.-S., Drake, J.~P., Landolin, J.~M., and
  Phillippy, A.~M. (2015).
\newblock {Assembling large genomes with single-molecule sequencing and
  locality-sensitive hashing}.
\newblock {\em Nature Biotechnology\/}, {\bf 33}(6), 623--630.

\bibitem[Chaisson {\em et~al.}(2004)Chaisson, Pevzner, and Tang]{Chaisson2004}
Chaisson, M., Pevzner, P., and Tang, H. (2004).
\newblock {Fragment assembly with short reads}.
\newblock {\em Bioinformatics\/}, {\bf 20}(13), 2067--2074.

\bibitem[Chaisson and Tesler(2012)Chaisson and Tesler]{Chaisson2012}
Chaisson, M.~J. and Tesler, G. (2012).
\newblock {Mapping single molecule sequencing reads using basic local alignment
  with successive refinement (BLASR): application and theory}.
\newblock {\em BMC Bioinformatics\/}, {\bf 13}(1), 238.

\bibitem[Chaisson {\em et~al.}(2015)Chaisson, Wilson, and
  Eichler]{Chaisson2015}
Chaisson, M. J.~P., Wilson, R.~K., and Eichler, E.~E. (2015).
\newblock {Genetic variation and the de novo assembly of human genomes}.
\newblock {\em Nature Reviews Genetics\/}, {\bf 16}(11), 627--640.

\bibitem[Chin {\em et~al.}(2013)Chin, Alexander, Marks, Klammer, Drake, Heiner,
  Clum, Copeland, Huddleston, Eichler, Turner, and Korlach]{Chin2013}
Chin, C.-S., Alexander, D.~H., Marks, P., Klammer, A.~A., Drake, J., Heiner,
  C., Clum, A., Copeland, A., Huddleston, J., Eichler, E.~E., Turner, S.~W.,
  and Korlach, J. (2013).
\newblock {Nonhybrid, finished microbial genome assemblies from long-read SMRT
  sequencing data}.
\newblock {\em Nature Methods\/}, {\bf 10}(6), 563--569.

\bibitem[D{\"{o}}ring {\em et~al.}(2008)D{\"{o}}ring, Weese, Rausch, and
  Reinert]{Doring2008}
D{\"{o}}ring, A., Weese, D., Rausch, T., and Reinert, K. (2008).
\newblock {SeqAn An efficient, generic C++ library for sequence analysis}.
\newblock {\em BMC Bioinformatics\/}, {\bf 9}(1), 11.

\bibitem[Eddy(2011)Eddy]{Eddy2011}
Eddy, S.~R. (2011).
\newblock {Accelerated Profile HMM Searches}.
\newblock {\em PLoS Computational Biology\/}, {\bf 7}(10), e1002195.

\bibitem[Firtina and Alkan(2016)Firtina and Alkan]{Firtina2016}
Firtina, C. and Alkan, C. (2016).
\newblock {On genomic repeats and reproducibility}.
\newblock {\em Bioinformatics\/}, {\bf 32}(15), 2243--2247.

\bibitem[Firtina {\em et~al.}(2018)Firtina, Bar-Joseph, Alkan, and
  Cicek]{Firtina2018}
Firtina, C., Bar-Joseph, Z., Alkan, C., and Cicek, A.~E. (2018).
\newblock {Hercules: a profile HMM-based hybrid error correction algorithm for
  long reads}.
\newblock {\em Nucleic Acids Research\/}, {\bf 46}(21), e125--e125.

\bibitem[Glenn(2011)Glenn]{Glenn2011}
Glenn, T.~C. (2011).
\newblock {Field guide to next-generation DNA sequencers}.
\newblock {\em Molecular Ecology Resources\/}, {\bf 11}(5), 759--769.

\bibitem[Gurevich {\em et~al.}(2013)Gurevich, Saveliev, Vyahhi, and
  Tesler]{Gurevich2013}
Gurevich, A., Saveliev, V., Vyahhi, N., and Tesler, G. (2013).
\newblock {QUAST: quality assessment tool for genome assemblies}.
\newblock {\em Bioinformatics\/}, {\bf 29}(8), 1072--1075.

\bibitem[Huddleston {\em et~al.}(2014)Huddleston, Ranade, Malig, Antonacci,
  Chaisson, Hon, Sudmant, Graves, Alkan, Dennis, Wilson, Turner, Korlach, and
  Eichler]{Huddleston2014}
Huddleston, J., Ranade, S., Malig, M., Antonacci, F., Chaisson, M., Hon, L.,
  Sudmant, P.~H., Graves, T.~A., Alkan, C., Dennis, M.~Y., Wilson, R.~K.,
  Turner, S.~W., Korlach, J., and Eichler, E.~E. (2014).
\newblock {Reconstructing complex regions of genomes using long-read sequencing
  technology}.
\newblock {\em Genome Research\/}, {\bf 24}(4), 688--696.

\bibitem[Jain {\em et~al.}(2018)Jain, Koren, Miga, Quick, Rand, Sasani, Tyson,
  Beggs, Dilthey, Fiddes, Malla, Marriott, Nieto, O'Grady, Olsen, Pedersen,
  Rhie, Richardson, Quinlan, Snutch, Tee, Paten, Phillippy, Simpson, Loman, and
  Loose]{Jain2018}
Jain, M., Koren, S., Miga, K.~H., Quick, J., Rand, A.~C., Sasani, T.~A., Tyson,
  J.~R., Beggs, A.~D., Dilthey, A.~T., Fiddes, I.~T., Malla, S., Marriott, H.,
  Nieto, T., O'Grady, J., Olsen, H.~E., Pedersen, B.~S., Rhie, A., Richardson,
  H., Quinlan, A.~R., Snutch, T.~P., Tee, L., Paten, B., Phillippy, A.~M.,
  Simpson, J.~T., Loman, N.~J., and Loose, M. (2018).
\newblock {Nanopore sequencing and assembly of a human genome with ultra-long
  reads}.
\newblock {\em Nature Biotechnology\/}, {\bf 36}(4), 338--345.

\bibitem[Kim {\em et~al.}(2018)Kim, {Senol Cali}, Xin, Lee, Ghose, Alser,
  Hassan, Ergin, Alkan, and Mutlu]{Kim2018}
Kim, J.~S., {Senol Cali}, D., Xin, H., Lee, D., Ghose, S., Alser, M., Hassan,
  H., Ergin, O., Alkan, C., and Mutlu, O. (2018).
\newblock {GRIM-Filter: Fast seed location filtering in DNA read mapping using
  processing-in-memory technologies}.
\newblock {\em BMC Genomics\/}, {\bf 19}(S2), 89.

\bibitem[Koren {\em et~al.}(2012)Koren, Schatz, Walenz, Martin, Howard,
  Ganapathy, Wang, Rasko, McCombie, Jarvis, and Phillippy]{Koren2012}
Koren, S., Schatz, M.~C., Walenz, B.~P., Martin, J., Howard, J.~T., Ganapathy,
  G., Wang, Z., Rasko, D.~A., McCombie, W.~R., Jarvis, E.~D., and Phillippy,
  A.~M. (2012).
\newblock {Hybrid error correction and de novo assembly of single-molecule
  sequencing reads}.
\newblock {\em Nature Biotechnology\/}, {\bf 30}(7), 693--700.

\bibitem[Koren {\em et~al.}(2017)Koren, Walenz, Berlin, Miller, Bergman, and
  Phillippy]{Koren2017}
Koren, S., Walenz, B.~P., Berlin, K., Miller, J.~R., Bergman, N.~H., and
  Phillippy, A.~M. (2017).
\newblock {Canu: scalable and accurate long-read assembly via adaptive k -mer
  weighting and repeat separation}.
\newblock {\em Genome Research\/}, {\bf 27}(5), 722--736.

\bibitem[Kurtz {\em et~al.}(2004)Kurtz, Phillippy, Delcher, Smoot, Shumway,
  Antonescu, and Salzberg]{Kurtz2004}
Kurtz, S., Phillippy, A., Delcher, A.~L., Smoot, M., Shumway, M., Antonescu,
  C., and Salzberg, S.~L. (2004).
\newblock {Versatile and open software for comparing large genomes}.
\newblock {\em Genome Biology\/}, {\bf 5}(2), R12.

\bibitem[Li(2016)Li]{Li2016a}
Li, H. (2016).
\newblock {Minimap and miniasm: fast mapping and de novo assembly for noisy
  long sequences}.
\newblock {\em Bioinformatics\/}, {\bf 32}(14), 2103--2110.

\bibitem[Li(2018)Li]{Li2018}
Li, H. (2018).
\newblock {Minimap2: pairwise alignment for nucleotide sequences}.
\newblock {\em Bioinformatics\/}, {\bf 34}(18), 3094--3100.

\bibitem[Li and Durbin(2009)Li and Durbin]{Li2009}
Li, H. and Durbin, R. (2009).
\newblock {Fast and accurate short read alignment with Burrows-Wheeler
  transform}.
\newblock {\em Bioinformatics\/}, {\bf 25}(14), 1754--1760.

\bibitem[Li {\em et~al.}(2009)Li, Handsaker, Wysoker, Fennell, Ruan, Homer,
  Marth, Abecasis, and Durbin]{Li2009a}
Li, H., Handsaker, B., Wysoker, A., Fennell, T., Ruan, J., Homer, N., Marth,
  G., Abecasis, G., and Durbin, R. (2009).
\newblock {The Sequence Alignment/Map format and SAMtools}.
\newblock {\em Bioinformatics\/}, {\bf 25}(16), 2078--2079.

\bibitem[Liu(2009)Liu]{Liu2009}
Liu, C. (2009).
\newblock {cuHMM: a CUDA Implementation of Hidden Markov Model Training and
  Classification}.
\newblock {\em The Chronicle of Higher Education\/}, pages 1--13.

\bibitem[Loman {\em et~al.}(2015)Loman, Quick, and Simpson]{Loman2015}
Loman, N.~J., Quick, J., and Simpson, J.~T. (2015).
\newblock {A complete bacterial genome assembled de novo using only nanopore
  sequencing data}.
\newblock {\em Nature Methods\/}, {\bf 12}(8), 733--735.

\bibitem[{Meltz Steinberg} {\em et~al.}(2017){Meltz Steinberg}, Schneider,
  Alkan, Montague, Warren, Church, and Wilson]{MeltzSteinberg2017}
{Meltz Steinberg}, K., Schneider, V.~A., Alkan, C., Montague, M.~J., Warren,
  W.~C., Church, D.~M., and Wilson, R.~K. (2017).
\newblock {Building and Improving Reference Genome Assemblies}.
\newblock {\em Proceedings of the IEEE\/}, {\bf 105}(3), 1--14.

\bibitem[Murakami(2017)Murakami]{Murakami2017}
Murakami, T. (2017).
\newblock {Expectation-Maximization Tensor Factorization for Practical Location
  Privacy Attacks}.
\newblock {\em Proceedings on Privacy Enhancing Technologies\/}, {\bf 2017}(4),
  138--155.

\bibitem[Niwattanakul {\em et~al.}(2013)Niwattanakul, Singthongchai, Naenudorn,
  and Wanapu]{Niwattanakul2013}
Niwattanakul, S., Singthongchai, J., Naenudorn, E., and Wanapu, S. (2013).
\newblock {Using of Jaccard Coefficient for Keywords Similarity}.
\newblock In {\em Proceedings of The International MultiConference of Engineers
  and Computer Scientists\/}, volume~1, pages 380--384.

\bibitem[Payne {\em et~al.}(2018)Payne, Holmes, Rakyan, and Loose]{Payne2018}
Payne, A., Holmes, N., Rakyan, V., and Loose, M. (2018).
\newblock {BulkVis: a graphical viewer for Oxford nanopore bulk FAST5 files}.
\newblock {\em Bioinformatics\/}.

\bibitem[Pearson and Lipman(1988)Pearson and Lipman]{Pearson1988}
Pearson, W.~R. and Lipman, D.~J. (1988).
\newblock {Improved tools for biological sequence comparison}.
\newblock {\em Proceedings of the National Academy of Sciences\/}, {\bf 85}(8),
  2444--2448.

\bibitem[Rhoads and Au(2015)Rhoads and Au]{Rhoads2015}
Rhoads, A. and Au, K.~F. (2015).
\newblock {PacBio Sequencing and Its Applications}.
\newblock {\em Genomics, Proteomics {\&} Bioinformatics\/}, {\bf 13}(5),
  278--289.

\bibitem[Salmela and Rivals(2014)Salmela and Rivals]{Salmela2014}
Salmela, L. and Rivals, E. (2014).
\newblock {LoRDEC: accurate and efficient long read error correction}.
\newblock {\em Bioinformatics\/}, {\bf 30}(24), 3506--3514.

\bibitem[Salmela {\em et~al.}(2016)Salmela, Walve, Rivals, and
  Ukkonen]{Salmela2016}
Salmela, L., Walve, R., Rivals, E., and Ukkonen, E. (2016).
\newblock {Accurate self-correction of errors in long reads using de Bruijn
  graphs}.
\newblock {\em Bioinformatics\/}, {\bf 33}(6), 799--806.

\bibitem[Sanger {\em et~al.}(1977)Sanger, Nicklen, and Coulson]{Sanger1977}
Sanger, F., Nicklen, S., and Coulson, A.~R. (1977).
\newblock {DNA sequencing with chain-terminating inhibitors}.
\newblock {\em Proceedings of the National Academy of Sciences\/}, {\bf
  74}(12), 5463--5467.

\bibitem[{Senol Cali} {\em et~al.}(2019){Senol Cali}, Kim, Ghose, Alkan, and
  Mutlu]{SenolCali2018}
{Senol Cali}, D., Kim, J.~S., Ghose, S., Alkan, C., and Mutlu, O. (2019).
\newblock {Nanopore sequencing technology and tools for genome assembly:
  computational analysis of the current state, bottlenecks and future
  directions}.
\newblock {\em Briefings in Bioinformatics\/}, {\bf 20}(4), 1542--1559.

\bibitem[{Titus Brown} and Irber(2016){Titus Brown} and Irber]{TitusBrown2016}
{Titus Brown}, C. and Irber, L. (2016).
\newblock {sourmash: a library for MinHash sketching of DNA}.
\newblock {\em The Journal of Open Source Software\/}, {\bf 1}(5), 27.

\bibitem[Vaser {\em et~al.}(2017)Vaser, Sovi{\'{c}}, Nagarajan, and
  {\v{S}}iki{\'{c}}]{Vaser2017}
Vaser, R., Sovi{\'{c}}, I., Nagarajan, N., and {\v{S}}iki{\'{c}}, M. (2017).
\newblock {Fast and accurate de novo genome assembly from long uncorrected
  reads}.
\newblock {\em Genome Research\/}, {\bf 27}(5), 737--746.

\bibitem[Viterbi(1967)Viterbi]{Viterbi1967}
Viterbi, A. (1967).
\newblock {Error bounds for convolutional codes and an asymptotically optimum
  decoding algorithm}.
\newblock {\em IEEE Transactions on Information Theory\/}, {\bf 13}(2),
  260--269.

\bibitem[Walker {\em et~al.}(2014)Walker, Abeel, Shea, Priest, Abouelliel,
  Sakthikumar, Cuomo, Zeng, Wortman, Young, and Earl]{Walker2014}
Walker, B.~J., Abeel, T., Shea, T., Priest, M., Abouelliel, A., Sakthikumar,
  S., Cuomo, C.~A., Zeng, Q., Wortman, J., Young, S.~K., and Earl, A.~M.
  (2014).
\newblock {Pilon: An Integrated Tool for Comprehensive Microbial Variant
  Detection and Genome Assembly Improvement}.
\newblock {\em PLoS One\/}, {\bf 9}(11), e112963.

\bibitem[Weirather {\em et~al.}(2017)Weirather, de~Cesare, Wang, Piazza,
  Sebastiano, Wang, Buck, and Au]{Weirather2017}
Weirather, J.~L., de~Cesare, M., Wang, Y., Piazza, P., Sebastiano, V., Wang,
  X.-J., Buck, D., and Au, K.~F. (2017).
\newblock {Comprehensive comparison of Pacific Biosciences and Oxford Nanopore
  Technologies and their applications to transcriptome analysis}.
\newblock {\em F1000Research\/}, {\bf 6}(100), 100.

\bibitem[Wenger {\em et~al.}(2019)Wenger, Peluso, Rowell, Chang, Hall,
  Concepcion, Ebler, Fungtammasan, Kolesnikov, Olson, T{\"{o}}pfer, Alonge,
  Mahmoud, Qian, Chin, Phillippy, Schatz, Myers, DePristo, Ruan, Marschall,
  Sedlazeck, Zook, Li, Koren, Carroll, Rank, and Hunkapiller]{Wenger2019}
Wenger, A.~M., Peluso, P., Rowell, W.~J., Chang, P.-C., Hall, R.~J.,
  Concepcion, G.~T., Ebler, J., Fungtammasan, A., Kolesnikov, A., Olson, N.~D.,
  T{\"{o}}pfer, A., Alonge, M., Mahmoud, M., Qian, Y., Chin, C.-S., Phillippy,
  A.~M., Schatz, M.~C., Myers, G., DePristo, M.~A., Ruan, J., Marschall, T.,
  Sedlazeck, F.~J., Zook, J.~M., Li, H., Koren, S., Carroll, A., Rank, D.~R.,
  and Hunkapiller, M.~W. (2019).
\newblock {Accurate circular consensus long-read sequencing improves variant
  detection and assembly of a human genome}.
\newblock {\em Nature Biotechnology\/}, {\bf 37}(10), 1155--1162.

\bibitem[Xin {\em et~al.}(2013)Xin, Lee, Hormozdiari, Yedkar, Mutlu, and
  Alkan]{Xin2013}
Xin, H., Lee, D., Hormozdiari, F., Yedkar, S., Mutlu, O., and Alkan, C. (2013).
\newblock {Accelerating read mapping with FastHASH}.
\newblock {\em BMC Genomics\/}, {\bf 14}(1), S13.

\bibitem[Yu {\em et~al.}(2014)Yu, Ukidave, and Kaeli]{Yu2015}
Yu, L., Ukidave, Y., and Kaeli, D. (2014).
\newblock {GPU-Accelerated HMM for Speech Recognition}.
\newblock In {\em 2014 43rd International Conference on Parallel Processing
  Workshops\/}, pages 395--402. IEEE.

\bibitem[Zhang {\em et~al.}(2015)Zhang, Awad, and Brown]{Zhang2015}
Zhang, Q., Awad, S., and Brown, C.~T. (2015).
\newblock {Crossing the streams: a framework for streaming analysis of short
  DNA sequencing reads}.
\newblock {\em PeerJ PrePrints\/}, {\bf 3}, e890v1.

\end{thebibliography}


\begin{thebibliography}{10}

\bibitem{Firtina2018}
Can Firtina, Ziv Bar-Joseph, Can Alkan, and A~Ercument Cicek.
\newblock {Hercules: a profile HMM-based hybrid error correction algorithm for
  long reads}.
\newblock {\em Nucleic Acids Research}, 46(21):e125--e125, August 2018.

\bibitem{Eddy1998}
Sean~R. Eddy.
\newblock {Profile hidden Markov models}.
\newblock {\em Bioinformatics}, 14(9):755--763, October 1998.

\bibitem{Baum1972}
L.~E. Baum.
\newblock {An inequality and associated maximization technique in statistical
  estimation of probabilistic functions of a Markov process}.
\newblock {\em Inequalities}, 3:1--8, 1972.

\bibitem{Knuth1992}
Donald~E. Knuth.
\newblock {Two Notes on Notation}.
\newblock {\em The American Mathematical Monthly}, 99(5):403, May 1992.

\bibitem{Viterbi1967}
A.~Viterbi.
\newblock {Error bounds for convolutional codes and an asymptotically optimum
  decoding algorithm}.
\newblock {\em IEEE Transactions on Information Theory}, 13(2):260--269, April
  1967.

\bibitem{Gurevich2013}
Alexey Gurevich, Vladislav Saveliev, Nikolay Vyahhi, and Glenn Tesler.
\newblock {QUAST: quality assessment tool for genome assemblies}.
\newblock {\em Bioinformatics}, 29(8):1072--1075, April 2013.

\bibitem{Koren2017}
Sergey Koren, Brian~P. Walenz, Konstantin Berlin, Jason~R. Miller, Nicholas~H.
  Bergman, and Adam~M. Phillippy.
\newblock {Canu: scalable and accurate long-read assembly via adaptive k -mer
  weighting and repeat separation}.
\newblock {\em Genome Research}, 27(5):722--736, May 2017.

\bibitem{Li2016a}
Heng Li.
\newblock {Minimap and miniasm: fast mapping and de novo assembly for noisy
  long sequences}.
\newblock {\em Bioinformatics}, 32(14):2103--2110, July 2016.

\bibitem{Li2018}
Heng Li.
\newblock {Minimap2: pairwise alignment for nucleotide sequences}.
\newblock {\em Bioinformatics}, 34(18):3094--3100, September 2018.

\bibitem{Li2009}
Heng Li and Richard Durbin.
\newblock {Fast and accurate short read alignment with Burrows-Wheeler
  transform}.
\newblock {\em Bioinformatics}, 25(14):1754--1760, July 2009.

\end{thebibliography}

\end{document}

\maketitle

\section{Constructing a profile hidden Markov model graph} \label{suppsec:phmm}

Apollo constructs a profile hidden Markov model graph (pHMM-graph) to represent the sequences of contig as well as the errors that a contig may have. A pHMM-graph includes states and directed transitions from a state to another. There are two types of probabilities that the graph contains: (1) emission and (2) transition probabilities. First, each state has emission probabilities for emitting certain characters where each character is associated with a probability value with the range $[0,1]$. Each emission probability reveals how likely it is to emit (e.g., consume or output) a certain character when a certain state is visited. Second, each transition is associated with a probability value with the range $[0,1]$. A transition probability shows the probability of visiting a state from a certain state. Thus, one can calculate the likelihood of emitting all the characters in a given sequence by traversing a certain path in the graph.

\begin{figure}
\centerline{\includegraphics[scale=0.4]{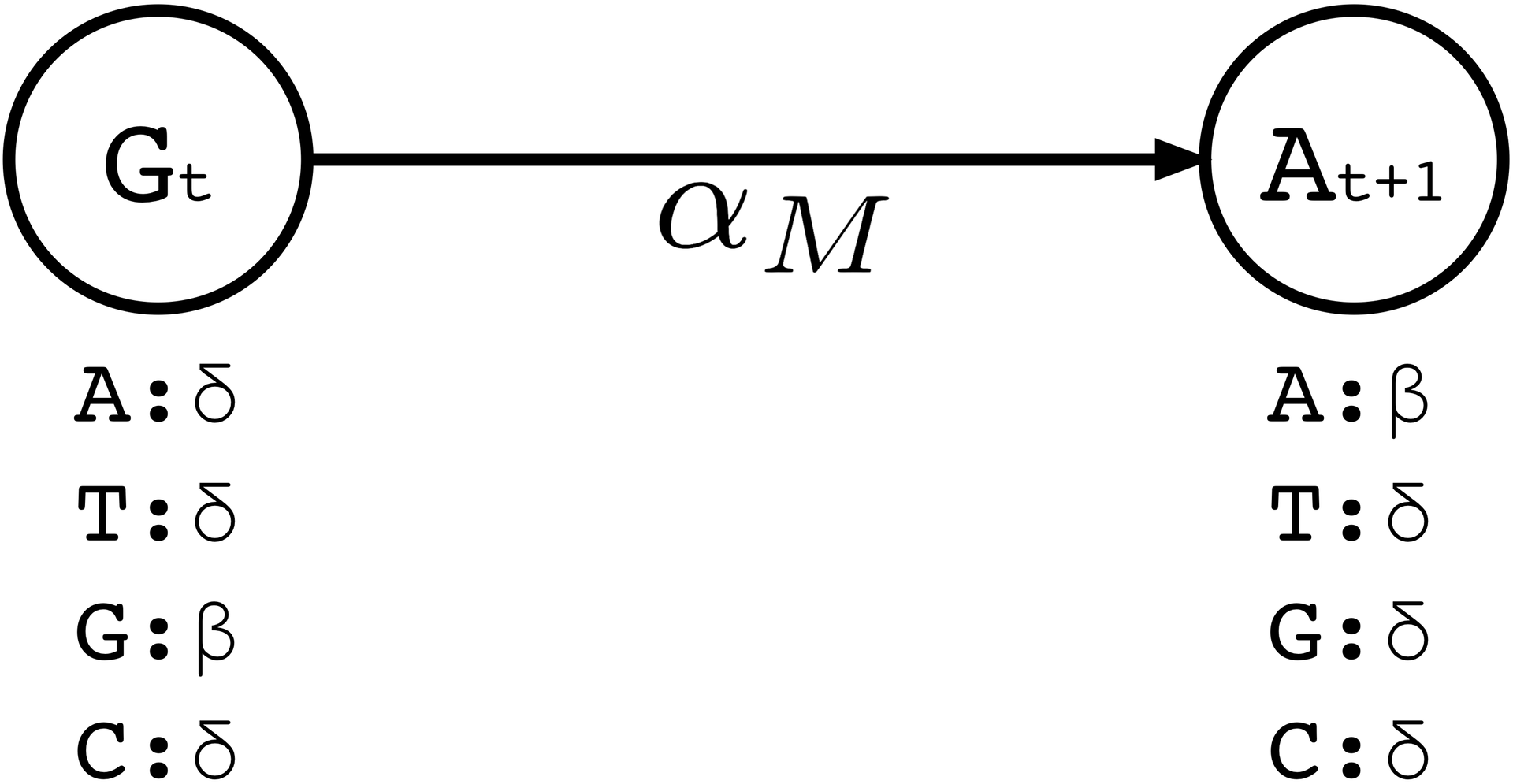}}
\caption{Two match states. Here, the contig includes the bases $G$ and $A$ at the locations $t$ and $t+1$, respectively. The corresponding match states are labeled with the bases that they correspond to (i.e., the match state $G_{t}$ represents the base $G$ at the location $t$). Each match state has a match transition to the next match state with the initial probability $\alpha_{M}$. A match state has a match emission probability, $\beta$, for the base it is labeled with. The remaining three bases have equal substitution emission probability $\delta$. The figure is taken from Hercules~\cite{Firtina2018}.}
\label{fig:match}
\end{figure}

The structure of the pHMM-graph allows us to handle insertion, deletion, and substitution errors by following certain states and transitions. Now, we will explain the structure of the graph in detail. For an assembly contig $C$, let us define the pHMM-graph that represents the contig $C$ as $G(V,E)$. Let us also define the length of the contig $C$ as $n = |C|$. A base $C[t]$  has one of the letters in the alphabet set $\Sigma=\{A,C,G,T\}$. Thus, a state emits one of the characters in $\Sigma$ with a certain probability. For a state $i$, We denote the emission probability of a base $c \in \Sigma$ as $e_{i}(c) \in [0,1]$ where $\sum\limits_{c \in \Sigma} e_{i}(c) = 1$. We denote the transition probability from a state, i, to another state, j, as $\alpha_{ij} \in [0,1]$. For the set of the states that the state $i$ has an outgoing transition to, $V_{i}$, we have $\sum\limits_{j \in V_{i}} \alpha_{ij} = 1$. Now let us define in four steps how Apollo constructs the states and the transitions of the graph $G(V,E)$:

First, Apollo constructs a start state, $v_{start} \in V$, and an end state $v_{end} \in V$. Second, for each base $C[t]$ where $1 \leq t \leq n$, Apollo constructs a match state as follows (Figure~\ref{fig:match}):
\begin{itemize}
    \item A \textit{match state} that we denote as $M_{t}$ for the base $C[t]$ where $M = C[t]$ s.t. $C[t] \in \Sigma$ and $M_{t} \in V$ (i.e., if the $t^{th}$ base of the contig $C$ is $G$, then the corresponding match state is $G_t$). For the following steps, let us assume $i = M_{t}$
    \item A \textit{match emission} with the probability $\beta$, for the base $C[t]$ s.t. $e_i(C[t]) = \beta$. $\beta$ is a parameter to Apollo.
    \item A \textit{substitution emission} with the probability $\delta$, for each base $c \in \Sigma$ and $c \neq C[t]$ s.t. $e_i(c) = \delta$ (Note that $\beta + 3\delta = 1$). $\delta$ is a parameter to Apollo.
    \item A \textit{match transition} with the probability $\alpha_{M}$, from the match state $M_t = i$ to the next match state $M_{t+1} = j$ s.t. $\alpha_{ij} = \alpha_{M}$. $\alpha_{M}$ is a parameter to Apollo.
\end{itemize}

\begin{figure}
\centerline{\includegraphics[scale=0.4]{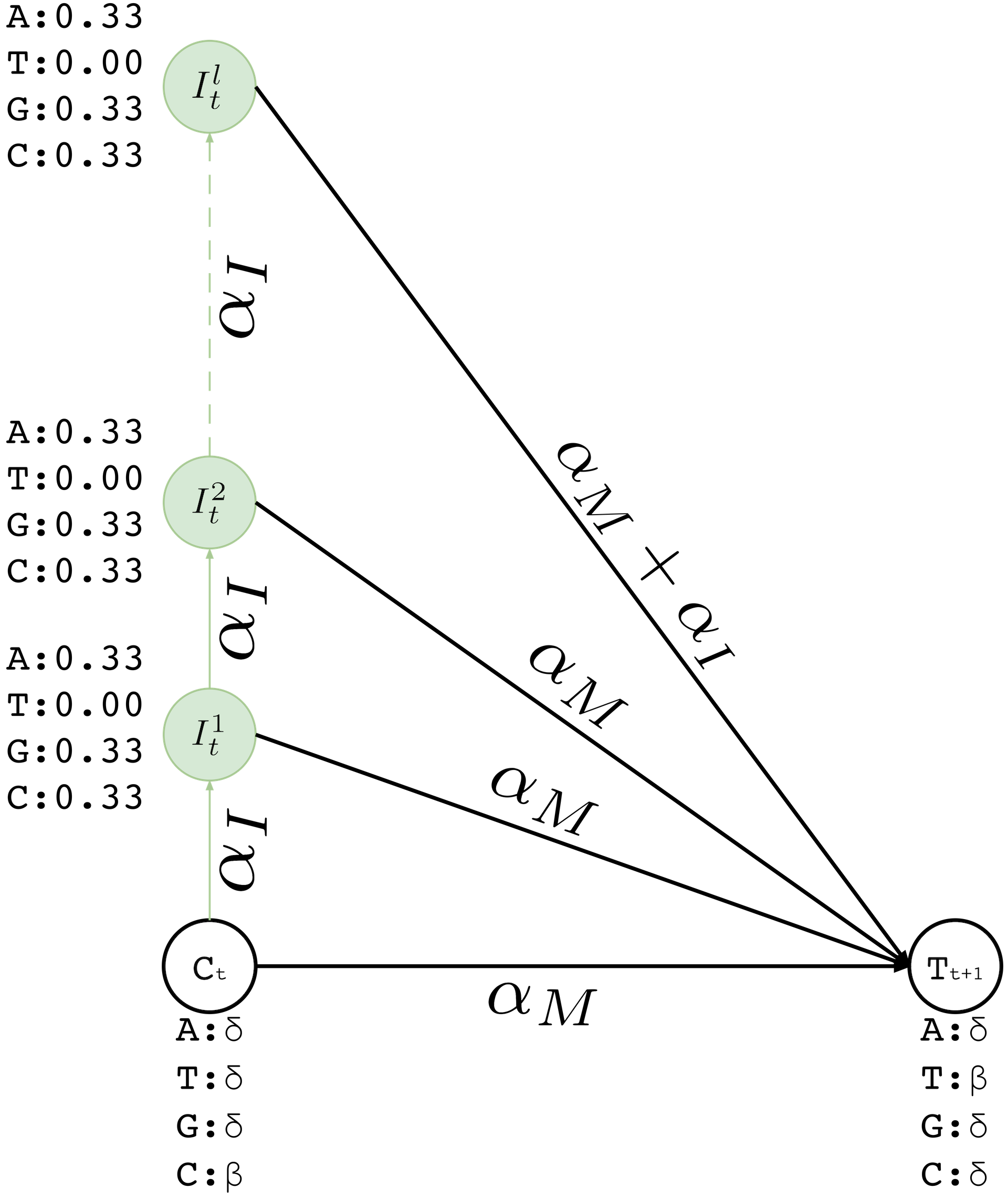}}
\caption{$l$ many insertion states for the base at location $t$. Here, the contig includes the bases $C$ and $T$ at the locations $t$ and $t+1$, respectively. The corresponding match states are labeled with the bases that they correspond to. Each insertion state has an insertion transition to the next insertion state with the initial probability $\alpha_{I}$ and a match transition to the next match state at the location $t+1$ with the initial probability $\alpha_{M}$. However, the last insertion state, $I_{t}^{l}$, does not have a transition to the next insertion state as it is the last one. Instead, it has a match transition to the next match state $T_{t+1}$ with the probability $\alpha_{M} + \alpha_{I}$. The emission probability of the base $T$ is $0$ as it appears in the next position ($t+1$) of the contig. The figure is taken from Hercules~\cite{Firtina2018}.}
\label{fig:insertion}
\end{figure}

Third, for each base $C[t]$ where $1 \leq t \leq n$, Apollo constructs the insertion states as follows (Figure~\ref{fig:insertion}):
\begin{itemize}
    \item There are $l$ many \textit{insertion states}, $I_{t}^{1}$, $I_{t}^{2}$, \dots, $I_{t}^{l}$, where $I_{t}^{i} \in V$, $1 \leq i \leq l$ and $l$ is a parameter to Apollo
    \item The match state, $M_{t} = i$, has an \textit{insertion transition} to $I_{t}^{1} = j$, with the probability $\alpha_{I}$ s.t. $\alpha_{ij} = \alpha_{I}$
    \item For each $i$ where $1 \leq i < l$, the insertion state $I_{t}^{i} = k$ has an insertion transition to the next insertion state $I_{t}^{i+1} = j$ with the probability $\alpha_{I}$ s.t. $\alpha_{kj} = \alpha_{I}$
    \item For each $i$ where $1 \leq i < l$, the insertion state $I_{t}^{i} = k$ has a match transition to the match state of the next base $M_{t+1} = j$ with the probability $\alpha_{M}$ s.t. $\alpha_{kj} = \alpha_{M}$
    \item The last insertion state, $I_{t}^{l}$, has no further insertion transitions. Instead, it has a transition to the match state of the next base $M_{t+1} = j$ with the probability $\alpha_{M} + \alpha_{I}$ s.t. $\alpha_{kj} = \alpha_{M} + \alpha_{I}$
    \item For each $i$ where $1 \leq i \leq l$, each base $c \in \Sigma$ and $c \neq C[t+1]$ has an \textit{insertion emission} probability $1/3 \approx 0.33$ for the insertion state $I_{t}^{i} = k$ s.t. $e_k(c) = 0.33$ and $e_k(C[t+1]) = 0$. Note that $\sum\limits_{c \in \Sigma} e_{k}(c) = 1$. (i.e., if the base at the location $t+1$ is T, then $e_{k}(A) = 0.33$, $e_{k}(T) = 0$, $e_{k}(G) = 0.33$, and $e_{k}(C) = 0.33$).
\end{itemize}

\begin{figure}
\centerline{\includegraphics[scale=0.5]{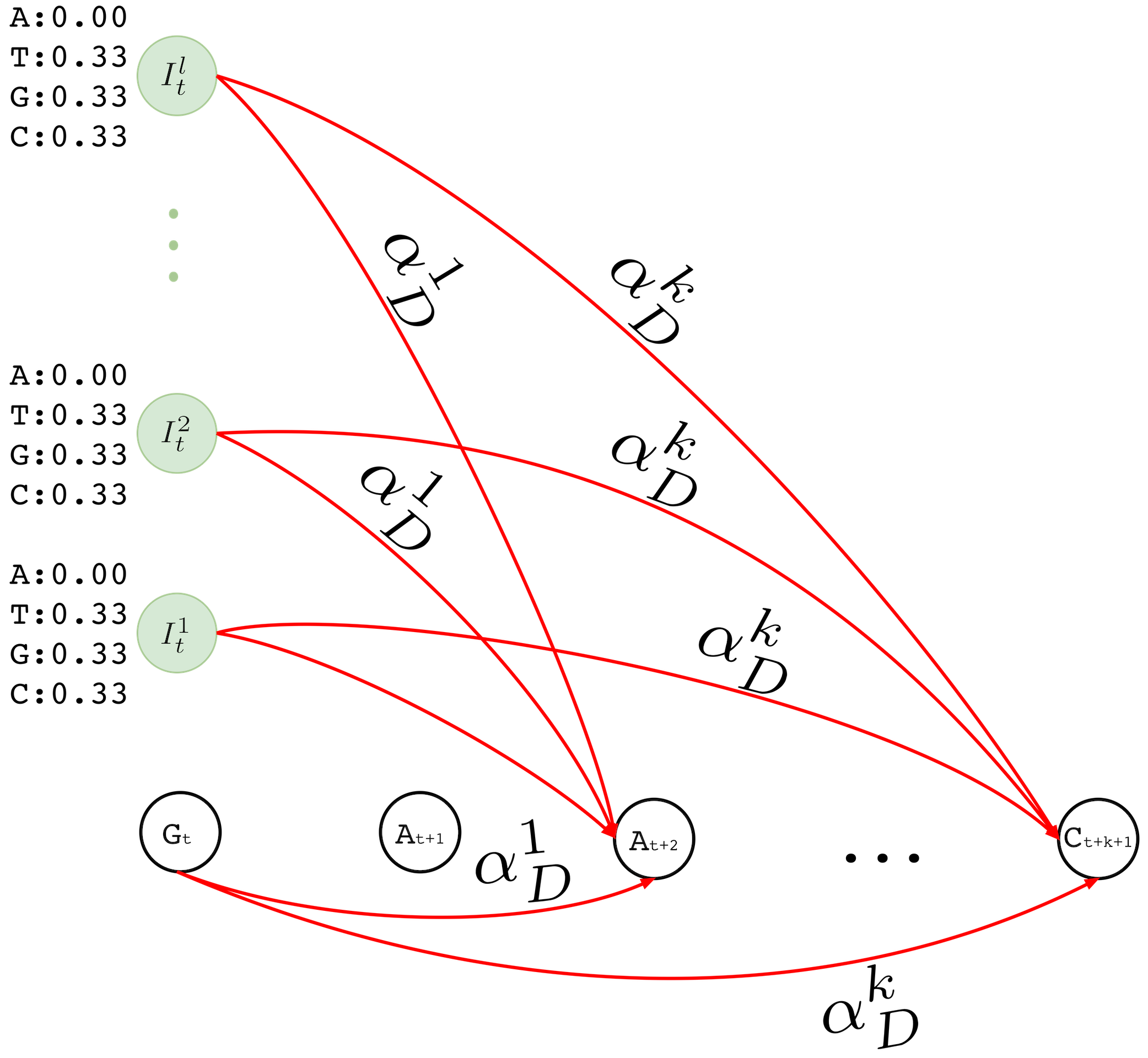}}
\caption{Deletion transitions of the match and each insertion states at location $t$. For the match and insertion states at location $t$, we show only the deletion transitions (red). Note that a deletion transition from the position $t$ to the match state of the position $t+x+1$ removes $x$ many bases with the probability $\alpha_{D}^{x}$ as it skips $x$ many match states where $1 \leq x \leq k$. The figure is taken from Hercules~\cite{Firtina2018}.}
\label{fig:deletion}
\end{figure}

Fourth step for finalizing the complete structure of the pHMM graph, for each \textit{state} $i \in V$, Apollo constructs the deletion transitions as follows (Figure~\ref{fig:deletion}):

\begin{itemize}
    \item Let us define $\alpha_{del} = 1 - (\alpha_{M} - \alpha_{I})$, which is the overall deletion transition probability.
    \item There are $k$ many deletion transitions from the state $i$, to the further \textit{match} states. $k$ is a parameter to Apollo.
    \item We assume that a transition deletes the bases if it skips the corresponding match states of the bases. We denote the transition probability of a deletion transition as $\alpha_{D}^{x}$ s.t. $1 \leq x \leq k$, if it deletes $x$ many bases in a row in one transition. Apollo calculates the deletion transition probability $\alpha_{D}^{x}$ using the normalized version of a polynomial distribution where $f \in [0, \infty)$ is a factor value for the equation:

    \begin{equation}
        \alpha_{D}^{x} = \dfrac{f^{k-x}\alpha_{del}}{\sum\limits_{j=0}^{k-1} f^{j}}\quad 1 \leq x     \leq k \tag{S1} \label{eq:deletion}
    \end{equation}

    \item If the $f$ value is set to $1$, then the each deletion transition is equally likely (i.e., $\alpha_{D}^{1} = \alpha_{D}^{10}$, if $k \geq 10$). As the $f$ value increases, the probability of deleting more bases in one transition decreases accordingly (i.e., $\alpha_{D}^{1} \gg \alpha_{D}^{10}$, if $k \geq 10$). $f$ is a parameter to Apollo.
\end{itemize}

We note that the start state $v_{start}$ also has a match transition to $M_{1}$ and deletion transitions as defined previously. There are al $l$ many insertion states, $I_{0}^{1}$, $I_{0}^{2}$, \dots, $I_{0}^{l}$, between the start state and the first match state $M_{1}$. The transitions of these insertion states are also identical to what we described before. We would also like to note that the end state $v_{end}$ has no outgoing transition. The prior states consider $v_{end}$ as a match state and connect to it accordingly. The start and end states have no emission probabilities.

Note that the design of pHMM-graph described here and proposed in Hercules~\cite{Firtina2018} is different from the conventional pHMM-graphs~\cite{Eddy1998}. One significant difference is that the conventional pHMM-graphs have deletion states for each match state whereas the pHMM-graph model of Apollo uses deletion \emph{transitions} instead of states. In the conventional model, visiting deletion states does not consume (i.e., emit) a character from a given sequence (i.e., observation). Therefore, this requires storing extra "position" information that tells which character should be consumed given a state at iteration $i$ (i.e., in each transition from a state to another). We want to make sure that each state consumes only one character (and no more) when visited to prevent storing the extra position information. In Apollo's design, iteration number $i$ equals the position of a character that is being consumed Apollo's states consume exactly one character. This allows us to remove an entire dimension, the iteration number $i$, which greatly helps us to reduce both memory requirements and runtime while calculating the Forward-Backward values.

\section{The Forward-Backward and Baum-Welch Algorithms}\label{suppsec:fwbw}

Apollo uses the region of a pHMM-graph (i.e., sub-graph) that a read (i.e., observation or a sequence) is aligned to in order to calculate the likelihood of each state emitting a certain base at position $t$ in the aligned read. However, this does not mean that position $t$ is known since we need to consider the fact that an unknown number of insertion and deletion errors may have occurred when $k$ number of transitions is followed from the start state to a certain state. Therefore, states should be measuring the likelihood of emitting a character at position $t$ where $t$ is a number in range $[1...k]$ where $k$ is the number of transitions that was taken so far. In the no error case, we have $k=t$. Apollo uses reads as observations for the Forward-Backward algorithm~\cite{Baum1972} in order to calculate the likelihoods per state. These likelihoods are calculated based on initial transition and emission probabilities of a pHMM-graph and the read itself. Apollo uses these likelihoods to make the contig similar to the aligned read. Apollo, then, trains the pHMM-graph of a contig per each read that aligns to the contig using the Baum-Welch algorithm~\cite{Baum1972}. We describe the details of both the Forward-Backward and the Baum-Welch algorithms in the following paragraphs.

For each read aligning to a contig, Apollo uses the alignment location and the sequence of the read in order to train the pHMM-graph. First, per each aligned read sequence $r$, Apollo extracts the sub-graph $G_s(V_s,E_s)$ that corresponds to the aligned region of the contig where we have $v_{start}$, $v_{end}$, match and insertion states, and the transitions as described in the Supplementary Section~\ref{suppsec:phmm}. Each transition from state $i \in V_{s}$ to state $j \in V_{s}$, $E_{ij} \in E_s$, is associated with a transition probability $\alpha_{ij}$. For every pair of states, $i \in V_s$ and $j \in V_s$, the transition probability $\alpha_{ij} = 0$ if $E_{ij} \not\in E_s$. Let us define the length of the aligned read, $r$, as $m = |r|$. Second, it calculates the forward and backward probabilities of each state based on the aligned read, $r$.

Let us assume that the forward probability of a state $j$ that observes the $t^{th}$ base of the aligned read, $r[t]$, is $F_{t}(j)$. For the forward probability, observing the $t^{th}$ base at the state $j$ means that all the previous bases ($r[1] \dots r[t-1]$ and $1 < t \leq m$) have been observed by following a path starting from the start state to the state $j$ and $j$ observes the next base, $r[t]$. All possible transitions that lead to state $j$ to observe the base $r[t]$ contribute to the probability with (1) the forward probability of the origin state $i$ calculated with the $(t-1)^{th}$ base of $r$, $F_{t-1}(i)$, (2) multiplied by the probability of the transition from $i$ to $j$, $\alpha_{ij}$, (3) multiplied by the probability of emitting the base $r[t]$ at state $j$, $e_j(r[t])$.

Let us denote the start state $v_{start}$ with the index value of $0$ (i.e., $v_{start} = 0$). For each state $j \in V_{s}$, we calculate the forward probability, $F_{t}(j)$, as follows where $F_{1}(j)$ is the initialization step:

\begin{equation}
F_{1}(j) = \alpha_{0j}e_j(r[1]) \quad s.t. \quad j \in V_{s}, \quad E_{0j} \in E_s \tag{S2.1} \label{suppeq:initforward}
\end{equation}

\begin{equation}
F_{t}(j) = \sum_{i \in V_{s}} F_{t-1}(i)\alpha_{ij}e_j(r[t]) \quad j \in V_{s}, \quad 1 < t \leq m \tag{S2.2} \label{suppeq:forward}
\end{equation}

Let us assume that the backward probability of a state $i$ that observes $t^{th}$ base of the aligned read, $r[t]$, is $B_{t}(i)$. For the backward probability, observing the $t^{th}$ base at the state $i$ means that all the further bases ($r[t+1] \dots r[m]$ and $1 \leq t < m$) have been observed by following a path starting from the end state to the state $i$ (backwards) and $i$ observes the previous base, $r[t]$. All possible transitions that lead to state $i$ to observe the base $r[t]$ contribute to the probability with (1) the backward probability of the next state $j$ calculated with the $(t+1)^{th}$ base of $r$, $B_{t+1}(j)$, (2) multiplied by the probability of the transition from $i$ to $j$, $\alpha_{ij}$, (3) multiplied by the probability of emitting the base $r[t+1]$ at state $j$, $e_j(r[t+1])$.

Let us denote the end state $v_{end}$ with the index value of $m+1$ (i.e., $v_{end} = m+1$). For each state $j \in V_{s}$, we calculate the backward probability, $B_{t}(i)$, as follows where $B_{m}(i)$ is the initialization step:

\begin{equation}
B_{m}(i) = \alpha_{i(m+1)}\quad i \in V_{s}, \quad E_{i(m+1)} \in E_s\tag{S3.1} \label{suppeq:initbackward}
\end{equation}
\begin{equation}
B_{t}(i) = \sum_{j \in V_{s}} \alpha_{ij}e_j(r[t+1])B_{t+1}(j) \quad j \in V_{s}, ~1 \leq t < m \tag{S3.2} \label{suppeq:backward}
\end{equation}

The calculations of forward and backward probabilities are referred as the Forward-Backward algorithm. After calculation of the forward and backward probabilities, Apollo uses the Baum-Welch algorithm to train the pHMM-graph by calculating the posterior transition and the emission probabilities of the sub-graph, $G_s$, as shown in equations~\ref{suppeq:emissionupd} and~\ref{suppeq:transitionupd}, respectively. In equation~\ref{suppeq:emissionupd}, we use the Iversonian brackets~\cite{Knuth1992} to denote that $[r[t] = X]$ is $1$ if the $t^{th}$ character of $r$ is the same character as $X$. Otherwise, $[r[t] = X]$ is $0$. This structure helps us to perform the summation in the numerator \emph{only when} the character at a position equals to the character given in function $e^{*}_i(X)$ (i.e., $X$). We, then, normalize this summation to make sure the sum of the emission probabilities that state $i$ can have is equal to 1.\\

\begin{equation}
e^{*}_i(X) = \dfrac{\sum\limits_{t=1}^m F_{t}(i)B_t(i)[r[t] = X]}
             {\sum\limits_{t=1}^m F_{t}(i)B_t(i)} \quad \forall X \in \{A,C,G,T\}, \forall i \in V_{s} \tag{S4} \label{suppeq:emissionupd}
\end{equation}

\begin{equation}
\alpha^{*}_{ij} = \dfrac{\sum\limits_{t=1}^{m-1} \alpha_{ij}e_{j}(r[t+1])F_{t}(i)B_{t+1}(j)}
     {\sum\limits_{t=1}^{m-1}\sum\limits_{x \in V_{s}} \alpha_{ix}e_{x}(r[t+1])F_{t}(i)B_{t+1}(x)} \quad \forall E_{ij} \in E_{s}
\tag{S5} \label{suppeq:transitionupd}
\end{equation}

\section{Joining Posterior Probabilities}\label{suppsec:join}
As we explain in the Supplementary Section~\ref{suppsec:fwbw}, for each read that aligns to the contig, Apollo extracts a sub-graph $G_s$ and uses the Forward-Backward algorithm to train the sub-graph. It is highly possible that there can be overlaps between two or many sub-graphs such that the sub-graphs can include the same states and the transitions when using high coverage reads. However, the updates on the overlapping states and the transitions are exclusive between the sub-graphs such that no two update in separate graphs affect each other while calculating the Forward or the Backward probabilities. Each sub-graph uses the initial probabilities to calculate the posterior probabilities. In order to handle training of the overlapping states and the transitions, Apollo takes the average of the posterior probabilities and reports the average probability as the final posterior probability for the entire pHMM-graph.

Let us assume that the set of sub-graphs $S$ includes the same state $i \in V$. For each $G_{s}$ in $S$, we obtain a $e^{*}_i(X)$, where $\forall X \in \Sigma$, which denotes the posterior emission probability as we explain in the Supplementary Section~\ref{suppsec:fwbw}. We denote $e^*_i(X)$ that belongs to $G_{s}$ as $e^{*,G_{s}}_i(X)$. Then, Apollo finds the final emission value $\hat{e}_i(X)$ as follows:

\begin{equation}
\hat{e}_i(X) = \dfrac{\sum\limits_{G_{s} \in S} e^{*,G_{s}}_{i}(X)}{\mid S \mid} \quad \forall X \in \Sigma
\tag{S6}\label{suppeq:emissioncapupd}
\end{equation}

Similarly, let us assume that the set of sub-graphs $S$ includes the same transition edge $E_{ij} \in E$.  For each $G_{s}$ in $S$, we obtain an $\alpha^{*}_{ij}$ that denotes the posterior transition value. We define $\alpha^{*}_{ij}$ that belongs to $G_{s}$ as $\alpha^{*,G_{s}}_{ij}$. Apollo finds the final transition value $\hat{\alpha}_{ij}$ as follows:

\begin{equation}
\hat{\alpha}_{ij} = \dfrac{\sum\limits_{G_{s} \in S} \alpha^{*,G_{s}}_{ij}}{\mid S \mid}
\tag{S7} \label{suppeq:transitioncapupd}
\end{equation}

If a state in $V$ or an edge in $E$ is not covered by a read then Apollo retains the initial emission and transition probabilities and uses as posterior probabilities, respectively.

We would like to note that the Baum-Welch algorithm is also used to train conventional hidden Markov models (HMMs). In each observation, the Baum-Welch algorithm updates the transition and emission probabilities of an HMM accordingly. The initial probabilities of such HMMs may even be assigned randomly. This means that the order of the observations (i.e., training data), and the initial probabilities used to train an HMM also affect the overall accuracy as the following observations usually use the HMM that is trained based on earlier observations. Therefore, after using all the training data, an HMM may still have room to converge to a local optimal point due to the biases caused by the initial probabilities and the order of the training data. The usual approach to mitigate such biases is to train HMMs \emph{multiple} times until the overall accuracy of an HMM converges to a certain point. We do \emph{not} follow this strategy because of three reasons. First, Apollo does not set the initial transition and emission probabilities randomly. Instead, the probabilities are usually set according to the error profile of an assembly. Second, we use the initial probabilities each time a read is used to train the pHMM-graph so the order of the training data does not matter. Third, Apollo is a very time consuming tool and taking multiple iterations until convergence would significantly increase the overall runtime, which we want to avoid.

\section{Decoding with the Viterbi Algorithm}\label{suppsec:decoding}

Apollo uses the Viterbi algorithm~\cite{Viterbi1967} to reveal the polished assembly by finding the most likely path starting from the start state, $v_{start}$, of the \emph{trained} graph $G$ to the end state, $v_{end}$. For each state $j$, the Viterbi algorithm calculates $v_{t}(j)$, which is the maximum marginal forward probability $j$ obtained from following a path starting from the start state when decoding the $t^{th}$ base of the \emph{polished} contig. Let $X_{j} \in \Sigma$ be the base that has the greatest emission probability for the state $j$, i.e., $\hat{e}_{j}(X_{j}) \geq \hat{e}_{j}(x)$, $\forall x \in \Sigma$. Then, the value of $v_{t}(j)$ depends on 1) the transition probability from state $i$ to the state $j$, $\hat{\alpha}_{ij}$, 2) the Viterbi value of the state $i$ when decoding the $(t-1)^{th}$ base of the polished contig, $v_{t-1}(i)$, and 3) the emission probability of the base $X_{j}$, $\hat{e}_{j}(X_{j})$. The Viterbi algorithm also keeps a back pointer, $b_{t}(j)$, which keeps track of the predecessor state $i$ that yields the $v_{t}(j)$ value.

Let $T$ be the length of the decoded sequence, which is initially unknown. The algorithm recursively calculates $v$ values for each position $t$ of a decoded sequence as described in the equations~\ref{suppeq:vitinitial} and~\ref{suppeq:vitrec}. The algorithm stops at iteration $T^{*}$ such that for the last $iter$ iterations, the maximum value we have observed for $v(end)$ cannot be improved and $iter$ is set to 50 by default (empirically chosen). $T$ is then set to $t^{*}$ such that $v_{t^{*}}(end)$ is the maximum among all iterations $1 \leq t \leq T^{*}$.

\begin{enumerate}
\item Initialization
\begin{equation}
v_{1}(j) = \hat{\alpha}_{start-j}\hat{e}_{j}(X_{j}) \quad \forall j \in V \tag{S8.1} \label{suppeq:vitinitial} \\
\end{equation}

\begin{equation}
b_{1}(j) = start \tag{S8.2}
\quad \forall j \in V \label{suppeq:backtraceinitial}
\end{equation}

\item Recursion
\begin{equation}
v_{t}(j) = \max_{i \in V} v_{t-1}(i)\hat{\alpha}_{ij}\hat{e}_{j}(X_{j}) \quad \forall j \in V, 1 < t \leq T \tag{S8.3} \label{suppeq:vitrec}
\end{equation}

\begin{equation}
b_{t}(j) = \argmax_{i \in V} v_{t-1}(i)\hat{\alpha}_{ij}\hat{e}_{j}(X_{j}) \quad \forall j \in V, 1 < t \leq T  \tag{S8.4} \label{suppeq:backtracerec}
\end{equation}

\item Termination
\begin{equation}
v_{T}(end) = \max_{i \in V} v_{T}(i)\hat{\alpha}_{i-end} \tag{S8.5} \label{suppeq:vitterm}
\end{equation}

\begin{equation}
b_{T}(end) = \argmax_{i \in V} v_{T}(i)\hat{\alpha}_{i-end} \tag{S8.6} \label{suppeq:backtraceterm}
\end{equation}
\end{enumerate}

The polished contig is generated by recursively following states from the end state, $v_{end}$, at time $T$ until the back pointer points back to the start state, $v_{start}$, at time $t=1$ for the state $j$ as follows:

\begin{algorithm}
\caption{Calculate $contig$}
\begin{algorithmic}
\STATE $t \leftarrow T-1$
\STATE $j \leftarrow end$

\WHILE{$t \neq 0$}
\STATE $j \leftarrow b_{t+1}(j)$
\STATE $contig[t] \leftarrow X_{j}$
\STATE $t \leftarrow t-1$
\ENDWHILE

\PRINT $contig$
\end{algorithmic}
\end{algorithm}

\clearpage

\section{Performance of the Assembly Polishing Algorithms}

In Tables~\ref{supptab:hybridcanu},~\ref{supptab:hybridminiasm},~\ref{supptab:ecolio157},~\ref{supptab:ecolio157h7}~\ref{supptab:ecolik12}, and~\ref{supptab:yeast}, we compare the assembly polishing performance of Apollo to the competing algorithms based on the difference between the assemblies and their reference genomes (i.e., ground truth). In Tables~\ref{supptab:kmerdistanceecolio157},~\ref{supptab:kmerdistanceecolio157h7},~\ref{supptab:kmerdistanceecolik12},~\ref{supptab:kmerdistanceyeast}, and~\ref{supptab:kmerdistancehum}, we show the k-mer similarities between Illumina reads and the assemblies to provide an alignment-free comparison between the tools. We also use QUAST~\cite{Gurevich2013} to make a more detailed quality assessment of the assemblies in Tables~\ref{supptab:qualassessecolio157},~\ref{supptab:qualassessecolio157h7},~\ref{supptab:qualassessecolik12},~\ref{supptab:qualassessyeast}, and~\ref{supptab:qualassesshum}.

\begin{table}[htb]
\begin{center}
\caption{Comparison between using a hybrid set of reads with Apollo and running other polishing tools multiple times to polish a Canu-generated assembly}
\label{supptab:hybridcanu}
\resizebox{\textwidth}{!}{
\begin{tabular}{|l|l|l|rrr||rr|}
\hline
\tcentb{Dataset} & \tcent{First Run} & \tcent{Second Run} & \tcentno{Aligned}     & \tcentno{Accuracy} & \tcentd{Polishing} & \tcentno{Runtime} & \tcent{Memory} \\
                 &                   &                    & \tcentno{Bases (\%)}  &                    & \tcentd{Score}     &                   & \tcent{(GB)}   \\\hhline{|=|=|=|===|==|}
\textit{E. coli} O157    & \tdash           & \tdash           & 99.94  & 0.9998 & 0.9992 & 43m 53s & 3.79 \\
\textit{E. coli} O157    & Apollo (Hybrid)  & \tdash           & 99.94  & 0.9999 & \textbf{0.9993}  & 8h 16m 08s & 13.85 \\
\textit{E. coli} O157    & Racon (PacBio)   & Racon (Illumina) & 99.94  & 0.9994 & 0.9988 & 21m 44s & 22.65 \\
\textit{E. coli} O157    & Racon (PacBio)   & Racon (PacBio)   & 99.94  & 0.9984 & 0.9978 & 4m 58s  & 2.43  \\
\textit{E. coli} O157    & Pilon (Illumina) & Pilon (Illumina) & 99.94  & 0.9999 & \textbf{0.9993}  & \textbf{4m 10s}   & 11.40 \\
\textit{E. coli} O157    & Pilon (Illumina) & Racon (PacBio)   & 99.94  & 0.9986 & 0.9980 & 4m 58s  & 11.40 \\
\textit{E. coli} O157    & Quiver (PacBio)  & Quiver (Pacbio)  & 99.94  & 0.9998 & 0.9992 & 13m 06s & \textbf{1.98}  \\
\textit{E. coli} O157    & Quiver (PacBio)  & Pilon (Illumina) & 99.94  & 0.9998 & 0.9992 & 5m 01s  & 7.50  \\
\textit{E. coli} O157    & Quiver (PacBio)  & Racon (PacBio)   & 99.94  & 0.9986 & 0.9980 & 5m 13s  & 2.48  \\\hhline{|=|=|=|===|==|}
\textit{E. coli} O157:H7 & \tdash           & \tdash           & 100 & 0.9998 & 0.9998 & 43m 19s & 3.39  \\
\textit{E. coli} O157:H7 & Apollo (Hybrid)  & \tdash           & 100 & 0.9999 & \textbf{0.9999}  & 5h 58m 05s & 8.86  \\
\textit{E. coli} O157:H7 & Racon (PacBio)   & Racon (Illumina) & 100 & 0.9995 & 0.9995 & 9m 43s  & 6.56  \\
\textit{E. coli} O157:H7 & Racon (PacBio)   & Racon (PacBio)   & 100 & 0.9970 & 0.9970 & \textbf{5m 36s} &  \textbf{2.24}  \\
\textit{E. coli} O157:H7 & Pilon (Illumina) & Pilon (Illumina) & 100 & 0.9998 & 0.9998 & 35m 12s & 10.79 \\
\textit{E. coli} O157:H7 & Pilon (Illumina) & Racon (PacBio)   & 100 & 0.9996 & 0.9996 & 6m 04s  & 10.75 \\\hhline{|=|=|=|===|==|}
\textit{E. coli} K-12    & \tdash           & \tdash           & 99.98  & 0.9794 & 0.9792 & 34h 21m 46s & 5.06  \\
\textit{E. coli} K-12    & Apollo (Hybrid)  & \tdash           & 99.99  & 0.9953 & 0.9952 & 9h 09m 50s & 9.35  \\
\textit{E. coli} K-12    & Racon (ONT)      & Racon (Illumina) & 100 & 0.9996 & \textbf{0.9996} & \textbf{11m 05s} & 5.10  \\
\textit{E. coli} K-12    & Racon (ONT)      & Racon (ONT)      & 100 & 0.9851 & 0.9851 & 14m 45s & \textbf{4.20}  \\
\textit{E. coli} K-12    & Pilon (Illumina) & Pilon (Illumina) & 99.99  & 0.9993 & 0.9992 & 18m 55s & 8.84 \\
\textit{E. coli} K-12    & Pilon (Illumina) & Racon (ONT)      & 99.99  & 0.9997 & \textbf{0.9996} & 15m 51s & 8.84 \\
\textit{E. coli} K-12    & Nanopolish (ONT) & Nanopolish (ONT) & 99.98  & 0.9929 & 0.9927 & 25h 39m 17s & 4.84 \\
\textit{E. coli} K-12    & Nanopolish (ONT) & Pilon (Illumina) & 99.99  & 0.9992 & 0.9991 & 9h 45m 01s & 18.10 \\
\textit{E. coli} K-12    & Nanopolish (ONT) & Racon (ONT)      & 100 & 0.9866 & 0.9866 & 9h 42m 24s & 4.54 \\\hhline{|=|=|=|===|==|}
Yeast S288C              & \tdash           & \tdash           & 99.89  & 0.9998 & 0.9987 & 1h 20m 39s & 6.24 \\
Yeast S288C              & Apollo (Hybrid)  & \tdash           & 99.89  & 0.9998 & \textbf{0.9987}  & 11h 08m 41s & 6.38 \\
Yeast S288C              & Racon (PacBio)   & Racon (Illumina) & 99.89  & 0.9994 & 0.9983 & 38m 21s & 6.93  \\
Yeast S288C              & Racon (PacBio)   & Racon (PacBio)   & 99.89  & 0.9949 & 0.9938 & 49m 52s & 6.93  \\
Yeast S288C              & Pilon (Illumina) & Pilon (Illumina) & 99.89  & 0.9998 & \textbf{0.9987}  & \textbf{1m 10s}  & 11.85 \\
Yeast S288C              & Pilon (Illumina) & Racon (PacBio)   & 99.89  & 0.9960 & 0.9949 & 21m 42s & 11.85 \\
Yeast S288C              & Quiver (PacBio)  & Quiver (Pacbio)  & 98.95  & 0.9998 & 0.9893 & 23m 23s & \textbf{2.96}  \\
Yeast S288C              & Quiver (PacBio)  & Pilon (Illumina) & 98.95  & 0.9998 & 0.9893 & 12m 47s & 13.28 \\
Yeast S288C              & Quiver (PacBio)  & Racon (PacBio)   & 98.93  & 0.9968 & 0.9861 & 40m 04s & 6.69  \\\hline
\end{tabular}}
\end{center}
{\footnotesize We use the long reads of \textit{E. coli} O157, \textit{E. coli} O157:H7, \textit{E. coli} K-12, and Yeast S288C datasets to generate their assemblies with \textbf{Canu}. Here, the polishing tools specified under \emph{First Run} and \emph{Second Run} polish the assembly using the set of reads specified in parentheses. The set of reads used in the second run is aligned to the assembly polished in the first run using Minimap2. PacBio and Illumina set of reads together constitute the hybrid set of reads (i.e., \emph{Hybrid}). We report the performance of the polishing tools in terms of the percentage of bases of an assembly that aligns to its reference (i.e., \emph{Aligned Bases}), the fraction of identical portions between the aligned bases of an assembly and the reference (i.e., \emph{Accuracy}) as calculated by dnadiff, and \emph{Polishing Score} value that is the product of \emph{Accuracy} and \emph{Aligned Bases} (as a fraction). We report the runtime and the memory requirements of the assembly polishing tools. We show the best result among \emph{assembly polishing algorithms} for each performance metric in \textbf{bold} text.}
\end{table}

\begin{table}[htb]
\begin{center}
\caption{Comparison between using a hybrid set of reads with Apollo and running other polishing tools multiple times to polish a Miniasm-generated assembly}
\label{supptab:hybridminiasm}
\resizebox{\textwidth}{!}{
\begin{tabular}{|l|l|l|rrr||rr|}
\hline
\tcentb{Dataset} & \tcent{First Run} & \tcent{Second Run} & \tcentno{Aligned}     & \tcentno{Accuracy} & \tcentd{Polishing} & \tcentno{Runtime} & \tcent{Memory} \\
                 &                   &                    & \tcentno{Bases (\%)}  &                    & \tcentd{Score}     &                   & \tcent{(GB)}   \\\hhline{|=|=|=|===|==|}
\textit{E. coli} O157		 & \tdash         	& \tdash  				 & 94.93 & 0.9000 & 0.8544 & 1m 48s  & 10.03 \\
\textit{E. coli} O157		 & Apollo (Hybrid) 	& \tdash  				 & 98.70 & 0.9866 & 0.9738 & 3h 51m 51s & 12.08 \\
\textit{E. coli} O157		 & Racon (PacBio)   & Racon (Illumina) & 99.37 & 0.9992 & 0.9929 & 21m 19s & 22.66 \\
\textit{E. coli} O157		 & Racon (PacBio)   & Racon (PacBio)	 & 99.51 & 0.9980 & 0.9931 & \textbf{5m 00s}& \textbf{2.46}  \\
\textit{E. coli} O157		 & Pilon (Illumina) & Pilon (Illumina) & 96.88 & 0.9872 & 0.9564 & 34m 53s & 18.60 \\
\textit{E. coli} O157		 & Pilon (Illumina) & Racon (PacBio)	 & 98.87 & 0.9970 & 0.9857 & 35m 26s & 18.60 \\
\textit{E. coli} O157    & Quiver (PacBio)  & Quiver (PacBio)  & 99.85 & 0.9994 & \textbf{0.9979} & 13m 45s & 5.05  \\
\textit{E. coli} O157		 & Quiver (PacBio)  & Pilon (Illumina) & 99.80 & 0.9994 & 0.9974 & 9m 42s & 4.76  \\
\textit{E. coli} O157		 & Quiver (PacBio)  & Racon (PacBio)	 & 99.81 & 0.9984 & 0.9965 & 10m 29s & 2.49  \\\hhline{|=|=|=|===|==|}
\textit{E. coli} O157:H7 & \tdash           & \tdash					 & 88.56 & 0.8798 & 0.7792 & 2m 57s  & 6.27  \\
\textit{E. coli} O157:H7 & Apollo (Hybrid)  & \tdash					 & 97.53 & 0.9804 & 0.9562 & 2h 54m 55s & 8.34  \\
\textit{E. coli} O157:H7 & Racon (PacBio)   & Racon (Illumina) & 99.02 & 0.9991 & \textbf{0.9893}  & 9m 24s & 6.56  \\
\textit{E. coli} O157:H7 & Racon (PacBio)   & Racon (PacBio)	 & 99.22 & 0.9954 & 0.9876 &  \textbf{5m 31s} &  \textbf{2.24}  \\
\textit{E. coli} O157:H7 & Racon (PacBio)   & Pilon (Illumina) & 99.12 & 0.9981 & \textbf{0.9893}  & 20m 37s & 12.57 \\
\textit{E. coli} O157:H7 & Pilon (Illumina) & Pilon (Illumina) & 96.32 & 0.9896 & 0.9532 & 35m 12s & 15.84 \\\hhline{|=|=|=|===|==|}
\textit{E. coli} K-12    & \tdash           & \tdash           & 86.68  & 0.8503 & 0.7370 & 4m 04s & 16.47  \\
\textit{E. coli} K-12    & Apollo (Hybrid)  & \tdash           & 97.53  & 0.9419 & 0.9186 & 2h 18m 33s & 9.12 \\
\textit{E. coli} K-12    & Racon (ONT)      & Racon (Illumina) & 99.51  & 0.9992 & \textbf{0.9943} & \textbf{8m 38s}  & 5.17  \\
\textit{E. coli} K-12    & Racon (ONT)      & Racon (ONT)      & 99.78  & 0.9840 & 0.9818 & 11m 43s & \textbf{4.06}  \\
\textit{E. coli} K-12    & Pilon (Illumina) & Pilon (Illumina) & 89.61  & 0.9622 & 0.8622 & 32m 03s & 17.78 \\
\textit{E. coli} K-12    & Pilon (Illumina) & Racon (ONT)      & 99.43  & 0.9979 & 0.9922 & 25m 15s & 32.15 \\
\textit{E. coli} K-12    & Nanopolish (ONT) & Nanopolish (ONT) & 97.35  & 0.9488 & 0.9236 & 241h 56m 10s & 8.49  \\
\textit{E. coli} K-12    & Nanopolish (ONT) & Pilon (Illumina) & 96.48  & 0.9769 & 0.9425 & 117h 29m 47s & 32.15  \\
\textit{E. coli} K-12    & Nanopolish (ONT) & Racon (ONT)      & 99.62  & 0.9814 & 0.9776 & 117h 08m 16s & 8.49 \\\hhline{|=|=|=|===|==|}
Yeast S288C		           & \tdash         	& \tdash					 & 95.05  & 0.8923 & 0.8481 & 2m 20s  & 16.59 \\
Yeast S288C		           & Apollo (Hybrid) 	& \tdash					 & 98.49  & 0.9709 & 0.9562 & 6h 37m 46s & 5.96 \\
Yeast S288C		           & Racon (PacBio)   & Racon (Illumina) & 99.26  & 0.9986 & 0.9912 & 23m 51s & 6.75  \\
Yeast S288C		           & Racon (PacBio)   & Racon (PacBio)	 & 99.33  & 0.9937 & 0.9879 & 43m 00s & 6.75  \\
Yeast S288C		           & Racon (PacBio)   & Pilon (Illumina) & 99.23  & 0.9977 & 0.9900 & 22m 07s & 14.86 \\
Yeast S288C		           & Pilon (Illumina) & Pilon (Illumina) & 95.80  & 0.9595 & 0.9192 & \textbf{2m 35s}  & 15.31 \\
Yeast S288C              & Quiver (PacBio)  & Quiver (PacBio)  & 99.42  & 0.9997 & 0.9939 & 24m 49s & \textbf{4.14}  \\
Yeast S288C              & Quiver (PacBio)  & Pilon (Illumina) & 99.45  & 0.9996 & \textbf{0.9941} & 12m 23s & 13.40 \\
Yeast S288C              & Quiver (PacBio)  & Racon (PacBio)   & 99.50  & 0.9965 & 0.9915 & 29m 31s & 6.39  \\\hline
\end{tabular}}
\end{center}
{\footnotesize We use the long reads of \textit{E. coli} O157, \textit{E. coli} O157:H7, \textit{E. coli} K-12, and Yeast S288C datasets to generate their assemblies with \textbf{Miniasm}. The polishing tools specified under \emph{First Run} and \emph{Second Run} polish the assembly using the set of reads specified in parentheses. The set of reads used in the second run is aligned to the assembly polished in the first run using Minimap2. PacBio and Illumina set of reads together constitute the hybrid set of reads (i.e., \emph{Hybrid}). We report the performance of the polishing tools in terms of the percentage of bases of an assembly that aligns to its reference (i.e., \emph{Aligned Bases}), the fraction of identical portions between the aligned bases of an assembly and the reference (i.e., \emph{Accuracy}) as calculated by dnadiff, and \emph{Polishing Score} value that is the product of \emph{Accuracy} and \emph{Aligned Bases} (as a fraction). We report the runtime and the memory requirements of the assembly polishing tools. We show the best result among \emph{assembly polishing algorithms} for each performance metric in \textbf{bold} text.}
\end{table}

\begin{table}[htb]
\begin{center}
\caption{Assembly polishing performance of the tools for the \textit{E. coli} O157 dataset}
\label{supptab:ecolio157}
\setlength{\tabcolsep}{0.4em}
\resizebox{\textwidth}{!}{
\begin{tabular}{|l|l|l|l|l|rrr||rr|}
\hline
\tcentb{Dataset} & \tcent{Assembler} & \tcent{Aligner} & \tcent{Sequencing Tech.} & \tcent{Polishing} & \tcentno{Aligned}    & \tcentno{Accuracy} & \tcentd{Polishing} & \tcentno{Runtime} & \tcent{Memory}\\
                 & 			            & 		            & \tcent{of the Reads} 	   & \tcent{Algorithm} & \tcentno{Bases (\%)} & 		               & \tcentd{Score}		  & 				         & \tcent{(GB)}\\\hhline{|=|=|=|=|=|===|==|}
PacBio 		            & Miniasm    & \tdash   & \tdash 			        & \tdash 		& 94.93	   & 0.9000	  & 0.8544 			& 1m 48s 		   & 10.03		   \\
PacBio 		            & Miniasm    & Minimap2 & PacBio 		          & Apollo 	  & 98.49	   & 0.9798   & 0.9650 			& 2h 27m 49s 	   & 7.07 		   \\
PacBio 		            & Miniasm    & Minimap2 & PacBio 		          & Pilon 	  & 96.43	   & 0.9528   & 0.9188 			& 1h 31m 32s 	   & 17.68 		   \\
PacBio 		            & Miniasm    & Minimap2 & PacBio		          & Racon 	  & 99.35	   & 0.9951   & \textbf{0.9886} & \textbf{2m 13s}  & \textbf{2.44} \\\hhline{|=|=|=|=|=|===|==|}
PacBio 		            & Miniasm    & pbalign  & PacBio 		          & Quiver	  & 99.80	   & 0.9993   & \textbf{0.9973} & \textbf{7m 31s}  & \textbf{0.51} \\\hhline{|=|=|=|=|=|===|==|}
PacBio 		            & Miniasm    & Minimap2 & Illumina 		        & Apollo 	  & 97.61	   & 0.9816   & \textbf{0.9581} & 4h 25m 17s 	   & \textbf{9.22} \\
PacBio 		            & Miniasm    & Minimap2 & Illumina 			      & Pilon 	  & 96.52	   & 0.9775   & 0.9435 			& 32m 48s 		   & 18.60		   \\
PacBio 		            & Miniasm    & Minimap2 & Illumina 		        & Racon 	  & 96.45	   & 0.9876   & 0.9525 			& \textbf{14m 09s} & 21.57 		   \\\hhline{|=|=|=|=|=|===|==|}
PacBio 		            & Miniasm    & BWA-MEM  & Illumina 			      & Apollo 	  & 96.62	   & 0.9738   & 0.9409 			& 3h 32m 45s 	   & \textbf{9.21} \\
PacBio 		            & Miniasm    & BWA-MEM  & Illumina 			      & Pilon 	  & 96.13	   & 0.9693   & 0.9318 			& 31m 21s 		   & 18.45		   \\
PacBio 		            & Miniasm    & BWA-MEM  & Illumina 			      & Racon 	  & 96.90	   & 0.9813   & \textbf{0.9509} & \textbf{12m 05s} & 20.85 		   \\\hhline{|=|=|=|=|=|===|==|}
PacBio 		            & Canu		   & \tdash   & \tdash 				      & \tdash 		& 99.94	   & 0.9998   & 0.9992      & 43m 53s		   & 3.79		   \\
PacBio 		            & Canu		   & Minimap2 & PacBio 		   	      & Apollo 	  & 99.94	   & 0.9997   & \textbf{0.9991} 			& 3h 42m 03s 	   & 8.82 		   \\
PacBio 		            & Canu		   & Minimap2 & PacBio 		          & Racon 	  & 99.94	   & 0.9986   & 0.9980 			& \textbf{2m 17s}  & \textbf{2.34} \\\hhline{|=|=|=|=|=|===|==|}
PacBio                & Canu		   & pbalign  & PacBio 		          & Quiver	  & 99.94	   & 0.9998   & \textbf{0.9992} & \textbf{7m 06s}  & \textbf{0.20} \\\hhline{|=|=|=|=|=|===|==|}
PacBio 		            & Canu		   & BWA-MEM  & Illumina 			      & Apollo 	  & 99.94	   & 0.9999   & \textbf{0.9993} & 4h 49m 15s 	   & \textbf{11.05}\\
PacBio 		            & Canu		   & BWA-MEM  & Illumina 			      & Pilon 	  & 99.94	   & 0.9998   & 0.9992 			& \textbf{2m 05s}  & 11.40 		   \\
PacBio 		            & Canu		   & BWA-MEM  & Illumina 			      & Racon 	  & 99.94	   & 0.9999   & \textbf{0.9993} & 14m 58s	 	   & 21.04 		   \\\hhline{|=|=|=|=|=|===|==|}
PacBio (30$\times$)   & Miniasm$^*$& \tdash   & \tdash 				      & \tdash 		& \tdashno & \tdashno & \tdashd 	  	& \tdashno 		& \tdash \\
PacBio (30$\times$)   & Canu		   & \tdash   & \tdash 				      & \tdash 		& 99.98	   & 0.9981   & 0.9979 			& 21m 03s 		   & 3.70		   \\
PacBio (30$\times$)   & Canu		   & Minimap2 & PacBio (30$\times$) & Apollo 	  & 99.98	   & 0.9982   & \textbf{0.9980} & 43m 32s 	 	   & 8.00 		   \\
PacBio (30$\times$)   & Canu		   & Minimap2 & PacBio (30$\times$) & Racon 	  & 99.98	   & 0.9980   & 0.9978 			& \textbf{15s} 		       & \textbf{0.59}\\\hhline{|=|=|=|=|=|===|==|}
PacBio (30$\times$)   & Canu		   & Minimap2 & PacBio (30$\times$, Corr.) & Apollo 	  & 99.97	   & 0.9976   & 0.9973 			& 46m 10s 	 	   & 7.99 		   \\
PacBio (30$\times$)   & Canu		   & Minimap2 & PacBio (30$\times$, Corr.) & Racon 	    & 99.98	   & 0.9983   & \textbf{0.9981} & \textbf{7s}      & \textbf{0.37} \\\hhline{|=|=|=|=|=|===|==|}
PacBio (30$\times$)   & Canu		   & BWA-MEM  & Illumina 			      & Apollo 	  & 99.98	   & 0.9997   & 0.9995 			& 4h 48m 31s       & 10.35 		   \\
PacBio (30$\times$)   & Canu		   & BWA-MEM  & Illumina 			      & Pilon 	  & 99.98	   & 0.9998   & \textbf{0.9996} & \textbf{3m 03s}  & \textbf{8.52} \\
PacBio (30$\times$)   & Canu		   & BWA-MEM  & Illumina 			      & Racon 	  & 99.98	   & 0.9997   & 0.9995 			& 14m 42s 	       & 21.04 		   \\\hline
\end{tabular}}
\end{center}
{\footnotesize We polish the PacBio assemblies of \textit{E. coli} O157 for different combinations of sequencing technology, assembler, aligner, and polishing algorithm. Canu-corrected long reads are labeled as "Corr.". We report the performance of the tools in terms of percentage of bases of an assembly that aligns to its reference (i.e., \emph{Aligned Bases}), the fraction of identical portions between the aligned bases of an assembly and the reference (i.e., \emph{Accuracy}) as calculated by dnadiff, and a \emph{Polishing Score} value that is the product of \emph{Accuracy} and \emph{Aligned Bases} (as a fraction). We report the runtime and the memory requirements of the assembly polishing tools. For the rows that do not specify assembly polishing algorithms, we only report the runtime and the memory requirements of the assemblers as well as accuracy of the unpolished assembly that they construct. We show the best result among \emph{assembly polishing algorithms} for each performance metric in \textbf{bold} text. $^*$ denotes that Miniasm cannot produce an assembly given the specified set of reads.}
\vspace{-5mm}
\end{table}

\begin{table}[htb]
\begin{center}
\caption{K-mer similarity between Illumina reads and the \textit{E. coli} O157 assemblies}
\label{supptab:kmerdistanceecolio157}
\resizebox{0.95\textwidth}{!}{
\begin{tabular}{|l|l|l|l|l|l|l|l|l|}
\hline
\tcentb{Dataset} & \tcent{Assembler} & \tcent{Aligner} & \tcent{Sequencing Tech.}   & \tcent{Polishing} & \tcent{11-mer}    & \tcent{21-mer}    & \tcent{31-mer}    & \tcent{51-mer}\\
                  &                   &                 & \tcent{of the Reads} & \tcent{Algorithm} & \tcent{Sim. (\%)} & \tcent{Sim. (\%)} & \tcent{Sim. (\%)} & \tcent{Sim. (\%)}\\\hhline{|=|=|=|=|=|=|=|=|=|}
PacBio       & Reference & \tdash   & \tdash              & \tdash & 100 / 100 & 99.89 / 99.98 & 99.92 / 99.96 & 99.66 / 99.96\\
PacBio       & Miniasm   & \tdash   & \tdash              & \tdash & 90.67 / 83.48 & 14.31 / 13.53 & 5.61 / 5.21 & 1.12 / 1.04\\
PacBio       & Miniasm   & Minimap2 & PacBio              & Apollo & 96.19 / 94.94 & 76.20 / 74.70 & 66.76 / 64.01 & 54.77 / 52.38\\
PacBio       & Miniasm   & Minimap2 & PacBio              & Pilon  & 93.63 / 89.91 & 46.18 / 44.24 & 31.07 / 28.92 & 14.57 / 13.70\\
PacBio       & Miniasm   & Minimap2 & PacBio              & Racon  & 99.47 / 98.70 & 94.89 / 94.11 & 91.11 / 89.05 & 85.22 / 84.67\\
PacBio       & Miniasm   & pbalign  & PacBio              & Quiver & 100 / 99.61 & 99.81 / 99.06 & 99.65 / 98.41 & 99.16 / 98.31\\\hline
PacBio       & Miniasm   & Minimap2 & Illumina            & Apollo & 97.11 / 95.42 & 83.33 / 82.33 & 78.23 / 76.56 & 71.05 / 69.02\\
PacBio       & Miniasm   & Minimap2 & Illumina            & Pilon  & 96.52 / 93.93 & 83.74 / 80.15 & 82.25 / 77.44 & 79.02 / 74.49\\
PacBio       & Miniasm   & Minimap2 & Illumina            & Racon  & 97.31 / 96.42 & 90.35 / 90.02 & 88.61 / 87.88 & 87.98 / 87.34\\\hline
PacBio       & Miniasm   & BWA-MEM  & Illumina            & Apollo & 96.98 / 94.19 & 80.06 / 77.20 & 75.18 / 72.08 & 67.71 / 64.42\\
PacBio       & Miniasm   & BWA-MEM  & Illumina            & Pilon  & 96.32 / 93.20 & 79.65 / 75.30 & 76.75 / 72.32 & 72.92 / 67.16\\
PacBio       & Miniasm   & BWA-MEM  & Illumina            & Racon  & 96.91 / 95.10 & 85.89 / 85.27 & 84.00 / 83.88 & 82.36 / 81.06\\\hline
PacBio       & Canu      & \tdash   & \tdash              & \tdash & 100 / 99.93 & 99.63 / 99.78 & 99.46 / 99.42 & 98.93 / 99.00\\
PacBio       & Canu      & Minimap2 & PacBio              & Apollo & 100 / 99.93 & 99.50 / 99.74 & 99.17 / 99.50 & 98.50 / 99.11\\
PacBio       & Canu      & Minimap2 & PacBio              & Racon  & 99.87 / 99.74 & 98.44 / 98.52 & 97.37 / 97.39 & 95.63 / 95.78\\
PacBio       & Canu      & pbalign  & PacBio              & Quiver & 100 / 100 & 99.80 / 99.72 & 99.67 / 99.44 & 99.40 / 99.25\\\hline
PacBio       & Canu      & BWA-MEM  & Illumina            & Apollo & 100 / 100 & 99.83 / 99.91 & 99.73 / 99.77 & 99.59 / 99.61\\
PacBio       & Canu      & BWA-MEM  & Illumina            & Pilon  & 100 / 100 & 99.83 / 99.93 & 99.73 / 99.77 & 99.59 / 99.62\\
PacBio       & Canu      & BWA-MEM  & Illumina            & Racon  & 100 / 100 & 99.81 / 99.91 & 99.71 / 99.75 & 99.57 / 99.53\\\hline
PacBio (30$\times$) & Canu      & \tdash   & \tdash              & \tdash & 99.47 / 99.41 & 96.74 / 96.88 & 95.20 / 94.92 & 92.31 / 91.39\\
PacBio (30$\times$) & Canu      & Minimap2 & PacBio (30$\times$)        & Apollo & 99.61 / 99.41 & 97.04 / 97.40 & 95.41 / 95.63 & 92.67 / 92.48\\
PacBio (30$\times$) & Canu      & Minimap2 & PacBio (30$\times$)        & Racon  & 99.80 / 99.61 & 97.00 / 97.34 & 95.12 / 95.16 & 92.63 / 92.98\\\hline
PacBio (30$\times$) & Canu      & Minimap2 & PacBio (30$\times$, Corr.) & Apollo & 99.41 / 99.41 & 97.00 / 97.47 & 95.31 / 95.72 & 92.44 / 92.93\\
PacBio (30$\times$) & Canu      & Minimap2 & PacBio (30$\times$, Corr.) & Racon  & 99.67 / 99.48 & 97.48 / 98.19 & 96.00 / 96.56 & 93.12 / 94.07\\\hline
PacBio (30$\times$) & Canu      & BWA-MEM  & Illumina            & Apollo & 100 / 99.93 & 99.83 / 99.54 & 99.69 / 99.52 & 99.55 / 99.23\\
PacBio (30$\times$) & Canu      & BWA-MEM  & Illumina            & Pilon  & 100 / 99.93 & 99.83 / 99.70 & 99.69 / 99.58 & 99.51 / 99.31\\
PacBio (30$\times$) & Canu      & BWA-MEM  & Illumina            & Racon  & 100 / 99.93 & 99.89 / 99.63 & 99.73 / 99.62 & 99.55 / 99.29\\\hline
\end{tabular}}
\end{center}
{\footnotesize We report the pairs of the percentage of 1) k-mers of Illumina reads present in the assembly and 2) k-mers of the assembly present in the Illumina reads (separated by ``/") in \emph{k-mer Sim.} using a fixed k-mer size (i.e., $k \in \{11,21,31,51\}$). We generate the assemblies for the \emph{Dataset}s using the reads sequenced from PacBio. We use Canu and Miniasm assemblers as specified in \emph{Assembler}. The reads specified under \emph{Sequencing Tech. of the Reads} are sequenced by the specified sequencing technology and are aligned to the assembly using the \emph{Aligner}. For the rows that do not specify assembly polishing algorithms, we only report the k-mer similarity between Illumina set of reads and either the unpolished assembly or the reference.}
\end{table}

\begin{table}[htb]
\begin{center}
\caption{Quality assessment of the \textit{E. coli} O157 assemblies}
\label{supptab:qualassessecolio157}
\resizebox{0.95\textwidth}{!}{
\begin{tabular}{|l|l|l|l|l|r|r|r|r|r|}
\hline
\tcentb{Dataset} & \tcent{Assembler} & \tcent{Aligner} & \tcent{Sequencing Tech.}   & \tcent{Polishing} & \tcent{GC}   & \tcent{Mapped}     & \tcent{Properly}    & \tcent{Avg.}     & \tcent{Coverage} \\
                 &                   &                 & \tcent{of the Reads} & \tcent{Algorithm} & \tcent{(\%)} & \tcent{Reads (\%)} & \tcent{Paired (\%)} & \tcent{Coverage} & \tcent{$\geq$ 10$\times$ (\%)} \\\hhline{|=|=|=|=|=|=|=|=|=|=|}
PacBio       & Reference & \tdash   & \tdash              & \tdash & 50.48 & 99.92 & 99.49 & 564 & 99.94\\
PacBio       & Miniasm   & \tdash   & \tdash              & \tdash & 49.88 & 92.08 & 87.50 & 434 & 87.90\\
PacBio       & Miniasm   & Minimap2 & PacBio              & Apollo & 50.28 & 98.74 & 97.43 & 531 & 96.19\\
PacBio       & Miniasm   & Minimap2 & PacBio              & Pilon  & 50.14 & 99.17 & 97.20 & 526 & 93.78\\
PacBio       & Miniasm   & Minimap2 & PacBio              & Racon  & 50.52 & 99.63 & 99.03 & 542 & 98.35\\
PacBio       & Miniasm   & pbalign  & PacBio              & Quiver & 50.56 & 99.83 & 99.40 & 545 & 98.56\\\hline
PacBio       & Miniasm   & Minimap2 & Illumina            & Apollo & 50.37 & 96.49 & 94.60 & 513 & 93.74\\
PacBio       & Miniasm   & Minimap2 & Illumina            & Pilon  & 50.36 & 95.58 & 92.04 & 499 & 89.57\\
PacBio       & Miniasm   & Minimap2 & Illumina            & Racon  & 50.45 & 96.48 & 94.73 & 514 & 94.11\\\hline
PacBio       & Miniasm   & BWA-MEM  & Illumina            & Apollo & 50.30 & 95.55 & 92.22 & 498 & 89.58\\
PacBio       & Miniasm   & BWA-MEM  & Illumina            & Pilon  & 50.30 & 94.48 & 89.64 & 478 & 86.54\\\hline
PacBio       & Miniasm   & BWA-MEM  & Illumina            & Racon  & 50.37 & 94.63 & 90.69 & 508 & 90.76\\\hline
PacBio       & Canu      & \tdash   & \tdash              & \tdash & 50.36 & 99.90 & 99.46 & 547 & 99.73\\
PacBio       & Canu      & Minimap2 & PacBio              & Apollo & 50.36 & 99.90 & 99.46 & 547 & 99.92\\
PacBio       & Canu      & Minimap2 & PacBio              & Racon  & 50.35 & 99.89 & 99.44 & 547 & 99.89\\
PacBio       & Canu      & pbalign  & PacBio              & Quiver & 50.36 & 99.90 & 99.46 & 547 & 99.38\\\hline
PacBio       & Canu      & BWA-MEM  & Illumina            & Apollo & 50.36 & 99.90 & 99.46 & 547 & 99.73\\
PacBio       & Canu      & BWA-MEM  & Illumina            & Pilon  & 50.36 & 99.90 & 99.46 & 547 & 99.73\\
PacBio       & Canu      & BWA-MEM  & Illumina            & Racon  & 50.36 & 99.90 & 99.46 & 547 & 99.73\\\hline
PacBio (30$\times$) & Canu      & \tdash   & \tdash              & \tdash & 50.44 & 99.89 & 99.42 & 560 & 99.61\\
PacBio (30$\times$) & Canu      & Minimap2 & PacBio (30$\times$)        & Apollo & 50.46 & 99.89 & 99.44 & 560 & 99.91\\
PacBio (30$\times$) & Canu      & Minimap2 & PacBio (30$\times$)        & Racon  & 50.44 & 99.89 & 99.43 & 560 & 99.94\\\hline
PacBio (30$\times$) & Canu      & Minimap2 & PacBio (30$\times$, Corr.) & Apollo & 50.46 & 99.89 & 99.42 & 560 & 99.92\\
PacBio (30$\times$) & Canu      & Minimap2 & PacBio (30$\times$, Corr.) & Racon  & 50.46 & 99.89 & 99.42 & 560 & 99.97\\\hline
PacBio (30$\times$) & Canu      & BWA-MEM  & Illumina            & Apollo & 50.47 & 99.89 & 99.44 & 560 & 99.70\\
PacBio (30$\times$) & Canu      & BWA-MEM  & Illumina            & Pilon  & 50.47 & 99.89 & 99.44 & 560 & 99.71\\
PacBio (30$\times$) & Canu      & BWA-MEM  & Illumina            & Racon  & 50.47 & 99.89 & 99.43 & 560 & 99.69\\\hline
\end{tabular}}
\end{center}
{\footnotesize We report the quality assessment of the assemblies as reported by QUAST~\cite{Gurevich2013}. QUAST reports the \emph{GC} content and uses the filtered Illumina reads to measure 1) percentage of the short reads that mapped to the assembly (\emph{Mapped Reads}), 2) percentage of \emph{Properly Paired} reads that mapped within the expected range of each other to the assembly, 3) average depth of coverage (\emph{Avg. Coverage}), and 4) percentage of the bases with at least 10$\times$ coverage (\emph{Coverage $\geq$ 10$\times$}). We generate the assemblies for the \emph{Dataset}s using the reads sequenced from PacBio. We use Canu and Miniasm assemblers as specified in \emph{Assembler}. The reads specified under \emph{Sequencing Tech. of the Reads} are sequenced by the specified sequencing technology and are aligned to the assembly using the \emph{Aligner} to polish the assembly. For the rows that do not specify assembly polishing algorithms, we only report the quality assessment of either the unpolished assembly or the reference.}
\end{table}

\begin{table}[htb]
\begin{center}
\caption{Assembly polishing performance of the tools for the \textit{E. coli} O157:H7 dataset}
\label{supptab:ecolio157h7}
\resizebox{\textwidth}{!}{
\begin{tabular}{|l|l|l|l|l|rrr||rr|}
\hline
\tcentb{Dataset} & \tcent{Assembler} & \tcent{Aligner} & \tcent{Sequencing Tech.} & \tcent{Polishing} & \tcentno{Aligned}    & \tcentno{Accuracy} & \tcentd{Polishing} & \tcentno{Runtime} & \tcent{Memory}\\
                 &                   &                 & \tcent{of the Reads}     & \tcent{Algorithm} & \tcentno{Bases (\%)} &                    & \tcentd{Score}     &                  & \tcent{(GB)}\\\hhline{|=|=|=|=|=|===|==|}
PacBio & Miniasm & \tdash   & \tdash   & \tdash & 88.56  & 0.8798 & 0.7792 & 2m 57s     & 6.27\\
PacBio & Miniasm & Minimap2 & PacBio   & Apollo & 96.99  & 0.9636 & 0.9346 & 1h 10m 23s & 7.07\\
PacBio & Miniasm & Minimap2 & PacBio   & Racon  & 98.94  & 0.9899 & \textbf{0.9794} & \textbf{2m 24s}     & \textbf{2.14}\\\hline
PacBio & Miniasm & Minimap2 & Illumina & Apollo & 96.06  & 0.9781 & 0.9396 & 2h 17m 28s & \textbf{5.66}\\
PacBio & Miniasm & Minimap2 & Illumina & Pilon  & 95.09  & 0.9791 & 0.9310 & 28m 54s    & 15.84\\
PacBio & Miniasm & Minimap2 & Illumina & Racon  & 96.17  & 0.9883 & \textbf{0.9504} & \textbf{4m 39s}     & 6.29\\\hline
PacBio & Canu    & \tdash   & \tdash   & \tdash & 100 & 0.9998 & 0.9998 & 43m 19s    & 3.39\\
PacBio & Canu    & Minimap2 & PacBio   & Apollo & 100 & 0.9997 & \textbf{0.9997} & 2h 57m 18s & 7.58\\
PacBio & Canu    & Minimap2 & PacBio   & Racon  & 100 & 0.9975 & 0.9975 & \textbf{2m 50s}     & \textbf{2.23}\\\hline
PacBio & Canu    & Minimap2 & Illumina & Apollo & 100 & 0.9997 & 0.9997 & 3h 10m 16s & \textbf{6.18}\\
PacBio & Canu    & Minimap2 & Illumina & Pilon  & 100 & 0.9999 & \textbf{0.9999} & \textbf{1m 27s}     & 10.75\\
PacBio & Canu    & Minimap2 & Illumina & Racon  & 100 & 0.9996 & 0.9996 & 7m 14s     & 6.53\\\hline
\end{tabular}}
\end{center}
{\footnotesize We polish the PacBio assemblies of \textit{E. coli} O157:H7 for different combinations of sequencing technology, assembler, aligner, and polishing algorithm. Canu-corrected long reads are labeled as "Corr.". We report the performance of the tools in terms of percentage of bases of an assembly that aligns to its reference (i.e., \emph{Aligned Bases}), the fraction of identical portions between the aligned bases of an assembly and the reference (i.e., \emph{Accuracy}) as calculated by dnadiff, and a \emph{Polishing Score} value that is the product of \emph{Accuracy} and \emph{Aligned Bases} (as a fraction). We report the runtime and the memory requirements of the assembly polishing tools. For the rows that do not specify assembly polishing algorithms, we only report the runtime and the memory requirements of the assemblers as well as accuracy of the unpolished assembly that they construct. We show the best result among \emph{assembly polishing algorithms} for each performance metric in \textbf{bold} text. $^*$ denotes that Miniasm cannot produce an assembly given the specified set of reads.}
\end{table}

\begin{table}[htb]
\begin{center}
\caption{K-mer similarity between Illumina reads and the \textit{E. coli} O157:H7 assemblies}
\label{supptab:kmerdistanceecolio157h7}
\resizebox{0.95\textwidth}{!}{
\begin{tabular}{|l|l|l|l|l|l|l|l|l|}
\hline
\tcentb{Dataset} & \tcent{Assembler} & \tcent{Aligner} & \tcent{Sequencing Tech.}   & \tcent{Polishing} & \tcent{11-mer}    & \tcent{21-mer}    & \tcent{31-mer}    & \tcent{51-mer}\\
                  &                   &                 & \tcent{of the Reads} & \tcent{Algorithm} & \tcent{Sim. (\%)} & \tcent{Sim. (\%)} & \tcent{Sim. (\%)} & \tcent{Sim. (\%)}\\\hhline{|=|=|=|=|=|=|=|=|=|}
\textit{E. coli} O157:H7 & Reference & \tdash   & \tdash   & \tdash & 99.93 / 100 & 99.78 / 99.94 & 99.73 / 99.96 & 99.70 / 99.92\\
\textit{E. coli} O157:H7 & Miniasm   & \tdash   & \tdash   & \tdash & 91.14 / 81.04 & 9.01 / 7.94 & 3.25 / 2.74 & 0.37 / 0.33\\
\textit{E. coli} O157:H7 & Miniasm   & Minimap2 & PacBio   & Apollo & 96.46 / 91.36 & 61.52 / 57.92 & 52.73 / 48.27 & 35.22 / 32.38\\
\textit{E. coli} O157:H7 & Miniasm   & Minimap2 & PacBio   & Racon  & 98.10 / 96.95 & 88.45 / 85.70 & 84.37 / 80.22 & 74.61 / 70.87\\\hline
\textit{E. coli} O157:H7 & Miniasm   & Minimap2 & Illumina & Apollo & 97.97 / 93.43 & 81.92 / 78.79 & 77.05 / 72.69 & 66.69 / 63.21\\
\textit{E. coli} O157:H7 & Miniasm   & Minimap2 & Illumina & Pilon  & 97.64 / 92.25 & 85.57 / 79.87 & 84.74 / 78.02 & 80.92 / 75.85\\
\textit{E. coli} O157:H7 & Miniasm   & Minimap2 & Illumina & Racon  & 98.36 / 94.57 & 91.28 / 89.04 & 90.77 / 87.49 & 88.78 / 87.23\\\hline
\textit{E. coli} O157:H7 & Canu      & \tdash   & \tdash   & \tdash & 99.80 / 99.93 & 99.41 / 99.57 & 99.13 / 99.46 & 99.06 / 98.99\\
\textit{E. coli} O157:H7 & Canu      & Minimap2 & PacBio   & Apollo & 99.80 / 99.93 & 99.35 / 99.57 & 99.08 / 99.44 & 98.82 / 98.88\\
\textit{E. coli} O157:H7 & Canu      & Minimap2 & PacBio   & Racon  & 99.54 / 99.61 & 96.81 / 96.67 & 95.27 / 95.22 & 91.84 / 91.72\\\hline
\textit{E. coli} O157:H7 & Canu      & Minimap2 & Illumina & Apollo & 99.93 / 99.93 & 99.48 / 99.85 & 99.17 / 99.71 & 98.95 / 99.70\\
\textit{E. coli} O157:H7 & Canu      & Minimap2 & Illumina & Pilon  & 99.87 / 100 & 99.78 / 99.91 & 99.73 / 99.88 & 99.70 / 99.79\\
\textit{E. coli} O157:H7 & Canu      & Minimap2 & Illumina & Racon  & 99.80 / 100 & 99.31 / 99.83 & 99.00 / 99.88 & 98.63 / 99.47\\\hline
\end{tabular}}
\end{center}
{\footnotesize  We report the pairs of the percentage of 1) k-mers of Illumina reads present in the assembly and 2) k-mers of the assembly present in the Illumina reads (separated by ``/") in \emph{k-mer Sim.} using a fixed k-mer size (i.e., $k \in \{11,21,31,51\}$). We generate the assemblies for the \emph{Dataset}s using the reads sequenced from PacBio. We use Canu and Miniasm assemblers as specified in \emph{Assembler}. The reads specified under \emph{Sequencing Tech. of the Reads} are sequenced by the specified sequencing technology and are aligned to the assembly using the \emph{Aligner}. For the rows that do not specify assembly polishing algorithms, we only report the k-mer similarity between Illumina set of reads and either the unpolished assembly or the reference.}
\end{table}

\begin{table}[htb]
\begin{center}
\caption{Quality assessment of the \textit{E. coli} O157:H7 assemblies}
\label{supptab:qualassessecolio157h7}
\resizebox{0.95\textwidth}{!}{
\begin{tabular}{|l|l|l|l|l|r|r|r|r|r|}
\hline
\tcentb{Dataset} & \tcent{Assembler} & \tcent{Aligner} & \tcent{Sequencing Tech.}   & \tcent{Polishing} & \tcent{GC}   & \tcent{Mapped}     & \tcent{Properly}    & \tcent{Avg.}     & \tcent{Coverage} \\
                 &                   &                 & \tcent{of the Reads} & \tcent{Algorithm} & \tcent{(\%)} & \tcent{Reads (\%)} & \tcent{Paired (\%)} & \tcent{Coverage} & \tcent{$\geq$ 10$\times$ (\%)} \\\hhline{|=|=|=|=|=|=|=|=|=|=|}
\textit{E. coli} O157:H7 & Reference & \tdash   & \tdash   & \tdash & 50.43 & 97.42 & 94.3 & 183 & 99.93\\
\textit{E. coli} O157:H7 & Miniasm   & \tdash   & \tdash   & \tdash & 49.61 & 80.51 & 68.24 & 108 & 76.01\\
\textit{E. coli} O157:H7 & Miniasm   & Minimap2 & PacBio   & Apollo & 50.09 & 95.0  & 88.69 & 163 & 91.74\\
\textit{E. coli} O157:H7 & Miniasm   & Minimap2 & PacBio   & Racon  & 50.55 & 97.03 & 93.06 & 173 & 96.59\\\hline
\textit{E. coli} O157:H7 & Miniasm   & Minimap2 & Illumina & Apollo & 50.39 & 93.6  & 87.69 & 162 & 90.65\\
\textit{E. coli} O157:H7 & Miniasm   & Minimap2 & Illumina & Pilon  & 50.36 & 93.01 & 85.66 & 159 & 86.75\\
\textit{E. coli} O157:H7 & Miniasm   & Minimap2 & Illumina & Racon  & 50.48 & 93.84 & 88.52 & 163 & 91.67\\\hline
\textit{E. coli} O157:H7 & Canu      & \tdash   & \tdash   & \tdash & 50.43 & 97.42 & 94.32 & 182 & 99.71\\
\textit{E. coli} O157:H7 & Canu      & Minimap2 & PacBio   & Apollo & 50.44 & 97.42 & 94.32 & 182 & 99.87\\
\textit{E. coli} O157:H7 & Canu      & Minimap2 & PacBio   & Racon  & 50.41 & 97.4  & 94.22 & 182 & 99.73\\\hline
\textit{E. coli} O157:H7 & Canu      & Minimap2 & Illumina & Apollo & 50.45 & 97.42 & 94.31 & 182 & 99.95\\
\textit{E. coli} O157:H7 & Canu      & Minimap2 & Illumina & Pilon  & 50.44 & 97.42 & 94.33 & 182 & 99.71\\
\textit{E. coli} O157:H7 & Canu      & Minimap2 & Illumina & Racon  & 50.45 & 97.42 & 94.29 & 182 & 99.98\\\hline
\end{tabular}}
\end{center}
{\footnotesize We report the quality assessment of the assemblies as reported by QUAST~\cite{Gurevich2013}. QUAST reports the \emph{GC} content and uses the filtered Illumina reads to measure 1) percentage of the short reads that mapped to the assembly (\emph{Mapped Reads}), 2) percentage of \emph{Properly Paired} reads that mapped within the expected range of each other to the assembly, 3) average depth of coverage (\emph{Avg. Coverage}), and 4) percentage of the bases with at least 10$\times$ coverage (\emph{Coverage $\geq$ 10$\times$}). We generate the assemblies for the \emph{Dataset}s using the reads sequenced from PacBio. We use Canu and Miniasm assemblers as specified in \emph{Assembler}. The reads specified under \emph{Sequencing Tech. of the Reads} are sequenced by the specified sequencing technology and are aligned to the assembly using the \emph{Aligner} to polish the assembly. For the rows that do not specify assembly polishing algorithms, we only report the quality assessment of either the unpolished assembly or the reference.}
\end{table}

\begin{table}[htb]
\begin{center}
\caption{Assembly polishing performance of the tools for \textit{E. coli} K-12 MG1655 dataset}
\label{supptab:ecolik12}
\resizebox{\textwidth}{!}{
\begin{tabular}{|l|l|l|l|l|rrr||rr|}
\hline
\tcentb{Dataset} & \tcent{Assembler} & \tcent{Aligner} & \tcent{Sequencing Tech.} & \tcent{Polishing} & \tcentno{Aligned}    & \tcentno{Accuracy} & \tcentd{Polishing} & \tcentno{Runtime} & \tcent{Memory}\\
                 &                   &                 & \tcent{of the Reads}     & \tcent{Algorithm} & \tcentno{Bases (\%)} &                    & \tcentd{Score}     &                  & \tcent{(GB)}\\\hhline{|=|=|=|=|=|===|==|}
ONT       & Miniasm   & \tdash   & \tdash           & \tdash     & 86.68      & 0.8503   & 0.7370          & 4m 04s      & 16.47\\
ONT       & Miniasm   & Minimap2 & ONT              & Apollo     & 97.50      & 0.9209   & 0.8979          & 1h 40m 08s  & 7.96\\
ONT       & Miniasm   & Minimap2 & ONT              & Nanopolish & 96.01      & 0.9182   & 0.8816          & 117h 02m 10s & 8.49\\
ONT       & Miniasm   & Minimap2 & ONT              & Racon      & 99.41      & 0.9769   & \textbf{0.9711}          & \textbf{4m 55s}      & \textbf{3.70}\\\hhline{|=|=|=|=|=|===|==|}
ONT       & Miniasm   & Minimap2 & Illumina         & Apollo     & 89.41      & 0.9291   & \textbf{0.8307}     & 54m 46s          & \textbf{6.20}\\
ONT       & Miniasm   & Minimap2 & Illumina         & Pilon      & 89.22      & 0.9310   & 0.8306     & \textbf{17m 28s}          & 10.58\\\hhline{|=|=|=|=|=|===|==|}
ONT       & Canu      & \tdash   & \tdash           & \tdash     & 99.98      & 0.9794   & 0.9792          & 34h 21m 46s & 5.06\\
ONT       & Canu      & Minimap2 & ONT              & Apollo     & 99.99      & 0.9803   & 0.9802          & 6h 08m 05s  & 8.09\\
ONT       & Canu      & Minimap2 & ONT              & Nanopolish & 99.98      & 0.9925   & \textbf{0.9923}          & 9h 35m 26s  & 4.54\\
ONT       & Canu      & Minimap2 & ONT              & Racon      & 100     & 0.9840   & 0.9840          & \textbf{7m 22s}      & \textbf{4.20}\\\hhline{|=|=|=|=|=|===|==|}
ONT       & Canu      & Minimap2 & Illumina         & Apollo     & 99.96      & 0.9982   & 0.9978     & 2h 09m 47s       & \textbf{6.43}\\
ONT       & Canu      & Minimap2 & Illumina         & Pilon      & 99.99      & 0.9987   & \textbf{0.9986}     & \textbf{11m 59s}          & 8.84\\\hhline{|=|=|=|=|=|===|==|}
ONT (30$\times$) & Miniasm$^*$& \tdash  & \tdash           & \tdash    & \tdashno & \tdashno & \tdashd       & \tdashno    & \tdash \\
ONT (30$\times$) & Canu      & \tdash   & \tdash           & \tdash     & 99.98      & 0.9744   & 0.9742          & 3h 17m 47s  & 4.54\\
ONT (30$\times$) & Canu      & Minimap2 & ONT (30$\times$)        & Apollo     & 99.98      & 0.9752   & 0.9750          & 40m 37s     & 7.74\\
ONT (30$\times$) & Canu      & Minimap2 & ONT (30$\times$)        & Nanopolish & 99.99      & 0.9857   & \textbf{0.9856}          & 4h 07m 06s  & 2.15\\
ONT (30$\times$) & Canu      & Minimap2 & ONT (30$\times$)        & Racon      & 100     & 0.9825   & 0.9825          & \textbf{20s}         & \textbf{0.59}\\\hhline{|=|=|=|=|=|===|==|}
ONT (30$\times$) & Canu      & Minimap2 & ONT (30$\times$, Corr)  & Apollo     & 99.96      & 0.9755   & 0.9751          & 46m 40s     & 7.75\\
ONT (30$\times$) & Canu      & Minimap2 & ONT (30$\times$, Corr)  & Racon      & 100     & 0.9799   & \textbf{0.9799}          & \textbf{9s}      & \textbf{0.42}\\\hline
\end{tabular}}
\end{center}
{\footnotesize We polish the ONT assemblies of \textit{E. coli} K-12 MG1655 for different combinations of assembler and polishing algorithm. Canu-corrected long reads are labeled as "Corr.". We report the performance of the tools in terms of percentage of bases of an assembly that aligns to its reference (i.e., \emph{Aligned Bases}), the fraction of identical portions between the aligned bases of an assembly and the reference (i.e., \emph{Accuracy}) as calculated by dnadiff, and a \emph{Polishing Score} value that is the product of \emph{Accuracy} and \emph{Aligned Bases} (as a fraction). We report the runtime and the memory requirements of the assembly polishing tools. For the rows that do not specify assembly polishing algorithms, we only report the runtime and the memory requirements of the assemblers as well as accuracy of the unpolished assembly that they construct. We show the best result among \emph{assembly polishing algorithms} for each performance metric in \textbf{bold} text. $^*$ denotes that Miniasm cannot produce an assembly given the specified set of reads.}
\end{table}

\begin{table}[htb]
\begin{center}
\caption{K-mer similarity between Illumina reads and the \textit{E. coli} K-12 assemblies}
\label{supptab:kmerdistanceecolik12}
\resizebox{0.95\textwidth}{!}{
\begin{tabular}{|l|l|l|l|l|l|l|l|l|}
\hline
\tcentb{Dataset} & \tcent{Assembler} & \tcent{Aligner} & \tcent{Sequencing Tech.}   & \tcent{Polishing} & \tcent{11-mer}    & \tcent{21-mer}    & \tcent{31-mer}    & \tcent{51-mer}\\
                  &                   &                 & \tcent{of the Reads} & \tcent{Algorithm} & \tcent{Sim. (\%)} & \tcent{Sim. (\%)} & \tcent{Sim. (\%)} & \tcent{Sim. (\%)}\\\hhline{|=|=|=|=|=|=|=|=|=|}
ONT       & Reference & \tdash   & \tdash           & \tdash     & 99.79 / 100 & 99.37 / 99.70 & 99.35 / 99.51 & 99.22 / 99.65\\
ONT       & Miniasm   & \tdash   & \tdash           & \tdash     & 82.92 / 80.97 & 13.49 / 14.57 & 5.40 / 5.59 & 1.22 / 1.29\\
ONT       & Miniasm   & Minimap2 & ONT              & Apollo     & 88.09 / 87.01 & 39.46 / 41.06 & 26.10 / 27.06 & 12.11 / 12.20\\
ONT       & Miniasm   & Minimap2 & ONT              & Nanopolish & 89.67 / 87.09 & 47.47 / 48.79 & 38.19 / 37.81 & 25.04 / 25.30\\
ONT       & Miniasm   & Minimap2 & ONT              & Racon      & 93.25 / 95.02 & 75.24 / 74.16 & 63.69 / 63.36 & 48.72 / 47.87\\\hline
ONT       & Miniasm   & Minimap2 & Illumina         & Apollo     & 91.25 / 87.17 & 50.96 / 53.20 & 44.37 / 44.28 & 32.54 / 32.75\\
ONT       & Miniasm   & Minimap2 & Illumina         & Pilon      & 89.60 / 86.45 & 56.27 / 58.38 & 51.30 / 52.20 & 44.28 / 45.29\\\hline
ONT       & Canu      & \tdash   & \tdash           & \tdash     & 92.08 / 95.91 & 76.08 / 76.15 & 66.05 / 66.09 & 49.94 / 49.87\\
ONT       & Canu      & Minimap2 & ONT              & Apollo     & 92.15 / 95.91 & 76.93 / 77.04 & 67.52 / 67.37 & 51.35 / 50.94\\
ONT       & Canu      & Minimap2 & ONT              & Nanopolish & 97.04 / 98.60 & 90.74 / 91.32 & 86.95 / 86.20 & 79.33 / 78.49\\
ONT       & Canu      & Minimap2 & ONT              & Racon      & 94.49 / 96.89 & 80.33 / 80.42 & 72.03 / 71.58 & 57.39 / 56.79\\\hline
ONT       & Canu      & Minimap2 & Illumina         & Apollo     & 99.24 / 99.65 & 97.72 / 97.88 & 97.35 / 96.94 & 96.26 / 95.82\\
ONT       & Canu      & Minimap2 & Illumina         & Pilon      & 99.59 / 99.59 & 98.10 / 98.39 & 98.37 / 97.70 & 97.06 / 96.46\\\hline
ONT (30$\times$) & Canu      & \tdash   & \tdash           & \tdash     & 90.36 / 94.87 & 71.60 / 72.15 & 59.89 / 59.83 & 41.94 / 42.42\\
ONT (30$\times$) & Canu      & Minimap2 & ONT (30$\times$)        & Apollo     & 91.05 / 94.84 & 72.62 / 73.06 & 60.96 / 61.17 & 43.50 / 43.84\\
ONT (30$\times$) & Canu      & Minimap2 & ONT (30$\times$)        & Nanopolish & 95.94 / 96.80 & 83.00 / 82.30 & 75.37 / 73.76 & 61.85 / 60.96\\
ONT (30$\times$) & Canu      & Minimap2 & ONT (30$\times$)        & Racon      & 93.73 / 96.46 & 79.13 / 78.91 & 68.55 / 68.62 & 53.17 / 53.00\\\hline
ONT (30$\times$) & Canu      & Minimap2 & ONT (30$\times$, Corr)  & Apollo     & 91.05 / 94.97 & 72.64 / 73.56 & 61.21 / 62.11 & 42.89 / 43.54\\
ONT (30$\times$) & Canu      & Minimap2 & ONT (30$\times$, Corr)  & Racon      & 92.08 / 95.91 & 74.79 / 76.08 & 64.47 / 65.09 & 47.06 / 47.38\\\hline
\end{tabular}}
\end{center}
{\footnotesize We report the pairs of the percentage of 1) k-mers of Illumina reads present in the assembly and 2) k-mers of the assembly present in the Illumina reads (separated by ``/") in \emph{k-mer Sim.} using a fixed k-mer size (i.e., $k \in \{11,21,31,51\}$). We generate the assemblies for the \emph{Dataset}s using the reads sequenced from PacBio. We use Canu and Miniasm assemblers as specified in \emph{Assembler}. The reads specified under \emph{Sequencing Tech. of the Reads} are sequenced by the specified sequencing technology and are aligned to the assembly using the \emph{Aligner}. For the rows that do not specify assembly polishing algorithms, we only report the k-mer similarity between Illumina set of reads and either the unpolished assembly or the reference.}
\end{table}

\begin{table}[htb]
\begin{center}
\caption{Quality assessment of the \textit{E. coli} K-12 assemblies}
\label{supptab:qualassessecolik12}
\resizebox{0.95\textwidth}{!}{
\begin{tabular}{|l|l|l|l|l|r|r|r|r|r|}
\hline
\tcentb{Dataset} & \tcent{Assembler} & \tcent{Aligner} & \tcent{Sequencing Tech.}   & \tcent{Polishing} & \tcent{GC}   & \tcent{Mapped}     & \tcent{Properly}    & \tcent{Avg.}     & \tcent{Coverage} \\
                 &                   &                 & \tcent{of the Reads} & \tcent{Algorithm} & \tcent{(\%)} & \tcent{Reads (\%)} & \tcent{Paired (\%)} & \tcent{Coverage} & \tcent{$\geq$ 10$\times$ (\%)} \\\hhline{|=|=|=|=|=|=|=|=|=|=|}
ONT       & Reference & \tdash   & \tdash           & \tdash     & 50.79 & 99.70 & 98.96 & 237 & 99.55\\
ONT       & Miniasm   & \tdash   & \tdash           & \tdash     & 52.62 & 90.85 & 82.50 & 147 & 75.72\\
ONT       & Miniasm   & Minimap2 & ONT              & Apollo     & 52.23 & 97.44 & 94.28 & 216 & 94.84\\
ONT       & Miniasm   & Minimap2 & ONT              & Nanopolish & 52.10 & 96.97 & 90.32 & 200 & 90.35\\
ONT       & Miniasm   & Minimap2 & ONT              & Racon      & 51.12 & 99.09 & 97.71 & 234 & 98.51\\\hline
ONT       & Miniasm   & Minimap2 & Illumina         & Apollo     & 51.89 & 92.90 & 86.52 & 181 & 80.33\\
ONT       & Miniasm   & Minimap2 & Illumina         & Pilon      & 52.11 & 92.59 & 85.77 & 175 & 78.64\\\hline
ONT       & Canu      & \tdash   & \tdash           & \tdash     & 51.05 & 99.61 & 98.71 & 233 & 98.75\\
ONT       & Canu      & Minimap2 & ONT              & Apollo     & 50.90 & 99.67 & 98.57 & 234 & 98.31\\
ONT       & Canu      & Minimap2 & ONT              & Nanopolish & 51.04 & 99.66 & 98.83 & 234 & 98.77\\
ONT       & Canu      & Minimap2 & ONT              & Racon      & 51.01 & 99.65 & 98.75 & 234 & 99.24\\\hline
ONT       & Canu      & Minimap2 & Illumina         & Apollo     & 50.81 & 99.68 & 98.80 & 235 & 98.58\\
ONT       & Canu      & Minimap2 & Illumina         & Pilon      & 50.80 & 99.68 & 98.77 & 235 & 98.76\\\hline
ONT (30$\times$) & Canu      & \tdash   & \tdash           & \tdash     & 51.11 & 99.60 & 98.57 & 234 & 99.04\\
ONT (30$\times$) & Canu      & Minimap2 & ONT (30$\times$)        & Apollo     & 51.14 & 99.60 & 98.59 & 234 & 99.19\\
ONT (30$\times$) & Canu      & Minimap2 & ONT (30$\times$)        & Nanopolish & 51.12 & 99.65 & 98.72 & 235 & 98.92\\
ONT (30$\times$) & Canu      & Minimap2 & ONT (30$\times$)        & Racon      & 51.05 & 99.64 & 98.78 & 234 & 99.35\\\hline
ONT (30$\times$) & Canu      & Minimap2 & ONT (30$\times$, Corr)  & Apollo     & 51.14 & 99.60 & 98.65 & 234 & 99.28\\
ONT (30$\times$) & Canu      & Minimap2 & ONT (30$\times$, Corr)  & Racon      & 51.08 & 99.63 & 98.80 & 234 & 99.40\\\hline
\end{tabular}}
\end{center}
{\footnotesize We report the quality assessment of the assemblies as reported by QUAST~\cite{Gurevich2013}. QUAST reports the \emph{GC} content and uses the filtered Illumina reads to measure 1) percentage of the short reads that mapped to the assembly (\emph{Mapped Reads}), 2) percentage of \emph{Properly Paired} reads that mapped within the expected range of each other to the assembly, 3) average depth of coverage (\emph{Avg. Coverage}), and 4) percentage of the bases with at least 10$\times$ coverage (\emph{Coverage $\geq$ 10$\times$}). We generate the assemblies for the \emph{Dataset}s using the reads sequenced from PacBio. We use Canu and Miniasm assemblers as specified in \emph{Assembler}. The reads specified under \emph{Sequencing Tech. of the Reads} are sequenced by the specified sequencing technology and are aligned to the assembly using the \emph{Aligner} to polish the assembly. For the rows that do not specify assembly polishing algorithms, we only report the quality assessment of either the unpolished assembly or the reference.}
\end{table}

\begin{table}[htb]
\begin{center}
\caption{Assembly polishing performance of the tools for the Yeast S288C dataset}
\label{supptab:yeast}
\resizebox{\textwidth}{!}{
\begin{tabular}{|l|l|l|l|l|rrr||rr|}
\hline
\tcentb{Dataset} & \tcent{Assembler} & \tcent{Aligner} & \tcent{Sequencing Tech.} & \tcent{Polishing} & \tcentno{Aligned}    & \tcentno{Accuracy} & \tcentd{Polishing} & \tcentno{Runtime} & \tcent{Memory}\\
                 &                   &                 & \tcent{of the Reads}     & \tcent{Algorithm} & \tcentno{Bases (\%)} &                    & \tcentd{Score}     &                  & \tcent{(GB)}\\\hhline{|=|=|=|=|=|===|==|}
PacBio & Miniasm & \tdash   & \tdash   & \tdash & 95.05 & 0.8923 & 0.8481 & 2m 23s      & 16.59\\
PacBio & Miniasm & Minimap2 & PacBio   & Apollo & 98.44 & 0.9706 & 0.9555 & 6h 53m 51s  & 4.62\\
PacBio & Miniasm & Minimap2 & PacBio   & Racon  & 99.15 & 0.9895 & 0.9811 & 18m 55s     & 6.63\\
PacBio & Miniasm & Minimap2 & PacBio   & Quiver & 99.44 & 0.9995 & \textbf{0.9939} & \textbf{16m 11s}     & \textbf{0.26}\\\hline
PacBio & Miniasm & Minimap2 & Illumina & Apollo & 97.26 & 0.9733 & 0.9466 & 2h 05m 58s  & \textbf{2.83}\\
PacBio & Miniasm & Minimap2 & Illumina & Pilon  & 97.06 & 0.9761 & 0.9474 & \textbf{4m 00s}      & 26.64\\
PacBio & Miniasm & Minimap2 & Illumina & Racon  & 97.27 & 0.9835 & \textbf{0.9567} & 5m 00s      & 7.34\\\hline
PacBio & Canu    & \tdash   & \tdash   & \tdash & 99.89 & 0.9998 & 0.9987 & 1h 20m 39s  & 6.24\\
PacBio & Canu    & Minimap2 & PacBio   & Apollo & 98.95 & 0.9997 & 0.9892 & 10h 59m 10s & 5.05\\
PacBio & Canu    & Minimap2 & PacBio   & Racon  & 98.93 & 0.9964 & 0.9857 & 19m 16s     & 6.82\\
PacBio & Canu    & Minimap2 & PacBio   & Quiver & 98.95 & 0.9998 & \textbf{0.9893} & \textbf{12m 02s}     & \textbf{0.29}\\\hline
PacBio & Canu    & Minimap2 & Illumina & Apollo & 98.95 & 0.9998 & \textbf{0.9893} & 1h 22m 24s  & \textbf{3.25}\\
PacBio & Canu    & Minimap2 & Illumina & Pilon  & 98.95 & 0.9998 & \textbf{0.9893} & \textbf{43s}         & 13.83\\
PacBio & Canu    & Minimap2 & Illumina & Racon  & 98.95 & 0.9998 & \textbf{0.9893} & 2m 55s      & 5.15\\\hline
\end{tabular}}
\end{center}
{\footnotesize We polish the PacBio assemblies of Yeast S288C for different combinations of sequencing technology, assembler, aligner, and polishing algorithm. Canu-corrected long reads are labeled as "Corr.". We report the performance of the tools in terms of percentage of bases of an assembly that aligns to its reference (i.e., \emph{Aligned Bases}), the fraction of identical portions between the aligned bases of an assembly and the reference (i.e., \emph{Accuracy}) as calculated by dnadiff, and a \emph{Polishing Score} value that is the product of \emph{Accuracy} and \emph{Aligned Bases} (as a fraction). We report the runtime and the memory requirements of the assembly polishing tools. For the rows that do not specify assembly polishing algorithms, we only report the runtime and the memory requirements of the assemblers as well as accuracy of the unpolished assembly that they construct. We show the best result among \emph{assembly polishing algorithms} for each performance metric in \textbf{bold} text. $^*$ denotes that Miniasm cannot produce an assembly given the specified set of reads.}
\end{table}

\begin{table}[htb]
\begin{center}
\caption{K-mer similarity between Illumina reads and the Yeast S288C assemblies}
\label{supptab:kmerdistanceyeast}
\resizebox{0.95\textwidth}{!}{
\begin{tabular}{|l|l|l|l|l|l|l|l|l|}
\hline
\tcentb{Dataset} & \tcent{Assembler} & \tcent{Aligner} & \tcent{Sequencing Tech.}   & \tcent{Polishing} & \tcent{11-mer}    & \tcent{21-mer}    & \tcent{31-mer}    & \tcent{51-mer}\\
                 &                   &                 & \tcent{of the Reads} & \tcent{Algorithm} & \tcent{Sim. (\%)} & \tcent{Sim. (\%)} & \tcent{Sim. (\%)} & \tcent{Sim. (\%)}\\\hhline{|=|=|=|=|=|=|=|=|=|}
Yeast S288C           & Reference    & \tdash   & \tdash         & \tdash & 100 / 100 & 99.96 / 99.87 & 99.87 / 99.71 & 99.73 / 99.59\\
Yeast S288C           & Miniasm      & \tdash   & \tdash         & \tdash & 95.49 / 91.36 & 12.06 / 10.85 & 4.38 / 3.84 & 0.62 / 0.55\\
Yeast S288C           & Miniasm      & Minimap2 & PacBio         & Apollo & 98.79 / 96.71 & 65.93 / 62.88 & 53.80 / 50.13 & 35.83 / 33.02\\
Yeast S288C           & Miniasm      & Minimap2 & PacBio         & Racon  & 99.39 / 98.63 & 88.15 / 86.21 & 82.35 / 79.89 & 72.60 / 69.48\\
Yeast S288C           & Miniasm      & Minimap2 & PacBio         & Quiver & 99.89 / 99.34 & 99.38 / 98.42 & 99.07 / 98.19 & 98.98 / 97.63\\\hline
Yeast S288C           & Miniasm      & Minimap2 & Illumina       & Apollo & 98.35 / 96.65 & 77.96 / 74.13 & 69.85 / 66.35 & 59.06 / 55.89\\
Yeast S288C           & Miniasm      & Minimap2 & Illumina       & Pilon  & 98.84 / 96.25 & 84.87 / 79.60 & 82.25 / 77.24 & 80.12 / 74.60\\
Yeast S288C           & Miniasm      & Minimap2 & Illumina       & Racon  & 98.51 / 97.18 & 89.53 / 87.02 & 87.49 / 84.96 & 87.02 / 83.89\\\hline
Yeast S288C           & Canu         & \tdash   & \tdash         & \tdash & 100 / 99.45 & 99.91 / 99.09 & 99.86 / 98.97 & 99.60 / 98.56\\
Yeast S288C           & Canu         & Minimap2 & PacBio         & Apollo & 99.94 / 99.45 & 99.87 / 99.11 & 99.74 / 98.95 & 99.46 / 98.58\\
Yeast S288C           & Canu         & Minimap2 & PacBio         & Racon  & 99.94 / 99.40 & 96.37 / 94.96 & 94.20 / 92.48 & 89.17 / 87.70\\
Yeast S288C           & Canu         & Minimap2 & PacBio         & Quiver & 100 / 99.62 & 99.93 / 99.19 & 99.89 / 98.95 & 99.76 / 98.69\\\hline
Yeast S288C           & Canu         & Minimap2 & Illumina       & Apollo & 100 / 99.45 & 99.92 / 99.10 & 99.88 / 98.93 & 99.68 / 98.58\\
Yeast S288C           & Canu         & Minimap2 & Illumina       & Pilon  & 100 / 99.45 & 99.94 / 99.13 & 99.89 / 98.95 & 99.74 / 98.69\\
Yeast S288C           & Canu         & Minimap2 & Illumina       & Racon  & 100 / 99.45 & 99.94 / 99.15 & 99.89 / 98.95 & 99.75 / 98.67\\\hline
\end{tabular}}
\end{center}
{\footnotesize  We report the pairs of the percentage of 1) k-mers of Illumina reads present in the assembly and 2) k-mers of the assembly present in the Illumina reads (separated by ``/") in \emph{k-mer Sim.} using a fixed k-mer size (i.e., $k \in \{11,21,31,51\}$). We generate the assemblies for the \emph{Dataset}s using the reads sequenced from PacBio. We use Canu and Miniasm assemblers as specified in \emph{Assembler}. The reads specified under \emph{Sequencing Tech. of the Reads} are sequenced by the specified sequencing technology and are aligned to the assembly using the \emph{Aligner}. For the rows that do not specify assembly polishing algorithms, we only report the k-mer similarity between Illumina set of reads and either the unpolished assembly or the reference.}
\end{table}

\begin{table}[htb]
\begin{center}
\caption{Quality assessment of the Yeast S288C assemblies}
\label{supptab:qualassessyeast}
\resizebox{0.95\textwidth}{!}{
\begin{tabular}{|l|l|l|l|l|r|r|r|r|r|}
\hline
\tcentb{Dataset} & \tcent{Assembler} & \tcent{Aligner} & \tcent{Sequencing Tech.}   & \tcent{Polishing} & \tcent{GC}   & \tcent{Mapped}     & \tcent{Properly}    & \tcent{Avg.}     & \tcent{Coverage} \\
                 &                   &                 & \tcent{of the Reads} & \tcent{Algorithm} & \tcent{(\%)} & \tcent{Reads (\%)} & \tcent{Paired (\%)} & \tcent{Coverage} & \tcent{$\geq$ 10$\times$ (\%)} \\\hhline{|=|=|=|=|=|=|=|=|=|=|}
Yeast S288C           & Reference    & \tdash   & \tdash         & \tdash & 38.30 & 99.94 & 99.71 & 73  & 99.95\\
Yeast S288C           & Miniasm      & \tdash   & \tdash         & \tdash & 38.42 & 93.88 & 83.94 & 57  & 82.63\\
Yeast S288C           & Miniasm      & Minimap2 & PacBio         & Apollo & 38.00 & 99.11 & 97.45 & 69  & 94.38\\
Yeast S288C           & Miniasm      & Minimap2 & PacBio         & Racon  & 38.26 & 99.51 & 98.64 & 70  & 96.34\\
Yeast S288C           & Miniasm      & Minimap2 & PacBio         & Quiver & 38.39 & 99.61 & 99.29 & 71  & 98.04\\\hline
Yeast S288C           & Miniasm      & Minimap2 & Illumina       & Apollo & 38.22 & 97.10 & 94.98 & 66  & 90.53\\
Yeast S288C           & Miniasm      & Minimap2 & Illumina       & Pilon  & 38.41 & 96.86 & 88.65 & 66  & 87.78\\
Yeast S288C           & Miniasm      & Minimap2 & Illumina       & Racon  & 38.42 & 97.03 & 95.33 & 66  & 91.35\\\hline
Yeast S288C           & Canu         & \tdash   & \tdash         & \tdash & 38.17 & 99.94 & 99.73 & 71  & 98.81\\
Yeast S288C           & Canu         & Minimap2 & PacBio         & Apollo & 38.17 & 99.94 & 99.73 & 71  & 98.83\\
Yeast S288C           & Canu         & Minimap2 & PacBio         & Racon  & 38.09 & 99.94 & 99.23 & 71  & 98.21\\
Yeast S288C           & Canu         & Minimap2 & PacBio         & Quiver & 38.17 & 99.94 & 99.74 & 71  & 98.74\\\hline
Yeast S288C           & Canu         & Minimap2 & Illumina       & Apollo & 38.17 & 99.94 & 99.73 & 71  & 98.81\\
Yeast S288C           & Canu         & Minimap2 & Illumina       & Pilon  & 38.17 & 99.94 & 99.74 & 71  & 98.81\\
Yeast S288C           & Canu         & Minimap2 & Illumina       & Racon  & 38.17 & 99.94 & 99.73 & 71  & 98.81\\\hline
\end{tabular}}
\end{center}
{\footnotesize We report the quality assessment of the assemblies as reported by QUAST~\cite{Gurevich2013}. QUAST reports the \emph{GC} content and uses the filtered Illumina reads to measure 1) percentage of the short reads that mapped to the assembly (\emph{Mapped Reads}), 2) percentage of \emph{Properly Paired} reads that mapped within the expected range of each other to the assembly, 3) average depth of coverage (\emph{Avg. Coverage}), and 4) percentage of the bases with at least 10$\times$ coverage (\emph{Coverage $\geq$ 10$\times$}). We generate the assemblies for the \emph{Dataset}s using the reads sequenced from PacBio. We use Canu and Miniasm assemblers as specified in \emph{Assembler}. The reads specified under \emph{Sequencing Tech. of the Reads} are sequenced by the specified sequencing technology and are aligned to the assembly using the \emph{Aligner} to polish the assembly. For the rows that do not specify assembly polishing algorithms, we only report the quality assessment of either the unpolished assembly or the reference.}
\end{table}

\begin{table}[htb]
\begin{center}
\caption{K-mer similarity between Illumina reads and the human genome assemblies}
\label{supptab:kmerdistancehum}
\resizebox{0.95\textwidth}{!}{
\begin{tabular}{|l|l|l|l|r|r|r|}
\hline
\tcentb{Dataset} & \tcent{Assembler} & \tcent{Aligner} & \tcent{Polishing} & \tcent{21-mer}    & \tcent{31-mer}    & \tcent{51-mer}\\
                 &                   &                 & \tcent{Algorithm} & \tcent{Sim. (\%)} & \tcent{Sim. (\%)} & \tcent{Sim. (\%)}\\\hhline{|=|=|=|=|=|=|=|}
Human HG002       & Reference        & \tdash      & \tdash       & 98.05 / 87.02 & 96.98 / 84.73 & 93.56 / 80.14\\
Human HG002       & Minimap2         & PacBio      & Apollo       & 93.74 / 82.62 & 91.05 / 79.18 & 85.26 / 73.11\\
Human HG002       & Minimap2         & PacBio      & Quiver$^*$   & 94.55 / 83.49 & 91.50 / 79.47 & 84.95 / 72.36\\
Human HG002       & Minimap2         & PacBio      & Racon$^*$    & 85.96 / 74.53 & 79.07 / 67.58 & 67.19 / 56.73\\\hline
Human HG002       & Minimap2         & Illumina    & Apollo       & 98.33 / 87.22 & 97.41 / 85.05 & 94.26 / 80.64\\
Human HG002       & BWA-MEM          & Illumina    & Apollo       & 98.32 / 87.17 & 97.39 / 84.98 & 94.23 / 80.57\\
Human HG002       & BWA-MEM          & Illumina    & Pilon$^*$    & 98.19 / 87.14 & 97.23 / 84.95 & 93.99 / 80.49\\\hline
Human HG002       & Minimap2         & PacBio (9$\times$) & Apollo       & 54.00 / 43.72 & 45.59 / 36.91 & 36.82 / 30.24\\
Human HG002       & BWA-MEM          & PacBio (9$\times$) & Apollo       & 53.97 / 42.76 & 45.61 / 36.10 & 36.95 / 29.66\\
Human HG002       & Minimap2         & PacBio (9$\times$) & Racon        & 48.93 / 37.77 & 39.97 / 31.08 & 31.04 / 24.62\\
Human HG002       & BWA-MEM          & PacBio (9$\times$) & Racon        & 46.83 / 34.91 & 37.69 / 28.35 & 28.67 / 22.07\\\hline
\end{tabular}}
\end{center}
{\footnotesize We report the pairs of the percentage of 1) k-mers of Illumina reads present in the assembly and 2) k-mers of the assembly present in the Illumina reads (separated by ``/") in \emph{k-mer Sim.} using a fixed k-mer size (i.e., $k \in \{21,31,51\}$). We polish the human genome assembly in \emph{Dataset} using PacBio or Illumina reads. The reads specified under \emph{Sequencing Tech. of the Reads} are sequenced by the specified sequencing technology and are aligned to the assembly using the \emph{Aligner}. For the row that does not specify any assembly polishing algorithm, we only report the k-mer similarity between Illumina set of reads and the unpolished assembly that is already constructed and we use as reference. $^*$ denotes that we polish the assembly contig by contig in these runs and collect the results once all of the contigs are polished separately.}
\end{table}

\begin{table}[htb]
\begin{center}
\caption{Quality assessment of the human genome assemblies}
\label{supptab:qualassesshum}
\resizebox{0.95\textwidth}{!}{
\begin{tabular}{|l|l|l|l|r|r|r|r|r|}
\hline
\tcentb{Dataset} & \tcent{Aligner} & \tcent{Sequencing Tech.}   & \tcent{Polishing} & \tcent{GC}   & \tcent{Mapped}     & \tcent{Properly}    & \tcent{Avg.}     & \tcent{Coverage} \\
                 &                 & \tcent{of the Reads} & \tcent{Algorithm} & \tcent{(\%)} & \tcent{Reads (\%)} & \tcent{Paired (\%)} & \tcent{Coverage} & \tcent{$\geq$ 10$\times$ (\%)} \\\hhline{|=|=|=|=|=|=|=|=|=|}
Human HG002       & \tdash           & \tdash      & \tdash       & 40.86 & 99.92 & 98.35 & 10 & 44.82\\
Human HG002       & Minimap2         & PacBio      & Apollo       & 40.81 & 99.91 & 97.75 & 10 & 44.81\\
Human HG002       & Minimap2         & PacBio      & Quiver       & 40.84 & 99.92 & 98.21 & 10 & 44.55\\
Human HG002       & Minimap2         & PacBio      & Racon$^*$    & 40.74 & 99.89 & 97.34 & 10 & 44.30\\\hline
Human HG002       & Minimap2         & Illumina    & Apollo       & 40.86 & 99.92 & 98.21 & 10 & 44.93\\
Human HG002       & BWA-MEM          & Illumina    & Apollo       & 40.86 & 99.92 & 98.19 & 10 & 44.90\\
Human HG002       & BWA-MEM          & Illumina    & Pilon$^*$    & 40.86 & 99.92 & 98.22 & 10 & 44.86\\\hline
Human HG002       & Minimap2         & PacBio (9$\times$) & Apollo       & 40.62 & 99.36 & 83.34 & 10 & 37.17\\
Human HG002       & BWA-MEM          & PacBio (9$\times$) & Apollo       & 40.62 & 99.29 & 82.54 & 10 & 36.04\\
Human HG002       & Minimap2         & PacBio (9$\times$) & Racon        & 40.95 & 98.00 & 78.70 & 9  & 33.82\\
Human HG002       & BWA-MEM          & PacBio (9$\times$) & Racon        & 40.94 & 97.27 & 76.30 & 9  & 32.07\\\hline
\end{tabular}}
\end{center}
{\footnotesize We report the quality assessment of the assemblies as reported by QUAST~\cite{Gurevich2013}. QUAST reports the \emph{GC} content and uses the filtered Illumina reads to measure 1) percentage of the short reads that mapped to the assembly (\emph{Mapped Reads}), 2) percentage of \emph{Properly Paired} reads that mapped within the expected range of each other to the assembly, 3) average depth of coverage (\emph{Avg. Coverage}), and 4) percentage of the bases with at least 10$\times$ coverage (\emph{Coverage $\geq$ 10$\times$}). We polish the human genome assembly in \emph{Dataset} using PacBio or Illumina reads. The reads specified under \emph{Sequencing Tech. of the Reads} are sequenced by the specified sequencing technology and are aligned to the assembly using the \emph{Aligner}. For the rows that do not specify assembly polishing algorithms, we only report the quality assessment of the reference. $^*$ denotes that we polish the assembly contig by contig in these runs and collect the results once all of the contigs are polished separately.}
\end{table}

\clearpage
\section{Performance of the Aligners}

Here in Table~\ref{supptab:aligner}, we show the performances of the aligners in terms of number of alignments that the aligners generate given the assembly and the reads to align, runtime (wall clock), and the memory requirement.

\begin{table}[htb]
\begin{center}
\caption{Performance of the aligners}
\label{supptab:aligner}
\resizebox{\textwidth}{!}{
\begin{tabular}{|l|l|l|l|rrr|}
\hline \textbf{Dataset for} & \textbf{Assembler} & \textbf{Aligner} & \textbf{Platform of the} & \textbf{Number of} & \textbf{Runtime} & \textbf{Memory}\\
\textbf{the Assembly} & & & \textbf{Aligned Reads} & \textbf{Alignments} & & \textbf{(GB)} \\\hhline{|=|=|=|=|===|}
\textit{E. coli} K-12 - ONT          & Miniasm & Minimap2 & ONT                & 8,095,856  & 3m 30s  & 4.88 \\
\textit{E. coli} K-12 - ONT          & Canu    & Minimap2 & ONT                & 1,662,306  & 39s     & 2.10 \\
\textit{E. coli} K-12 - ONT (30$\times$)    & Canu    & Minimap2 & ONT (30$\times$)          & 170,910    & 6s      & 0.60 \\\hhline{|=|=|=|=|===|}
\textit{E. coli} O157 - PacBio       & Miniasm & Minimap2 & PacBio             & 732,397    & 25s     & 1.79 \\
\textit{E. coli} O157 - PacBio       & Miniasm & Minimap2 & Illumina           & 21,933,051 & 1m 35s  & 3.16 \\
\textit{E. coli} O157 - PacBio       & Canu    & Minimap2 & PacBio             & 741,343    & 22s     & 1.80 \\
\textit{E. coli} O157 - PacBio (30$\times$) & Canu    & Minimap2 & PacBio (30$\times$)       & 148,241    & 5s      & 0.67 \\
\textit{E. coli} O157 - PacBio (30$\times$) & Canu    & Minimap2 & PacBio (30$\times$, Corr) & 137,620    & 3s      & 0.47 \\
\textit{E. coli} O157 - PacBio       & Miniasm & BWA-MEM  & Illumina           & 19,799,002 & 2m 34s  & 3.17 \\
\textit{E. coli} O157 - PacBio       & Canu    & BWA-MEM  & Illumina           & 23,328,379 & 1m 16s  & 2.89 \\
\textit{E. coli} O157 - PacBio (30$\times$) & Canu    & BWA-MEM  & Illumina           & 23,326,202 & 1m 20s  & 2.96 \\
\textit{E. coli} O157 - PacBio       & Miniasm & pbalign  & PacBio             & 49,561     & 12m 55s & 6.36 \\
\textit{E. coli} O157 - PacBio       & Canu    & pbalign  & PacBio             & 51,994     & 11m 29s & 6.28 \\\hline
\end{tabular}}
\end{center}
{\footnotesize We generate the assembly using the reads specified under \textit{Dataset for the Assembly}. We use Canu~\cite{Koren2017} and Miniasm~\cite{Li2016a} assemblers as specified in \textit{Assembler}. The reads specified under \textit{Platform of the Aligned Reads} are aligned to the assembly using the \textit{Aligner}. We use Minimap2~\cite{Li2018} aligner for aligning both long and short reads to the assembly and BWA-MEM~\cite{Li2009} aligner to align the short reads to the assembly. We report the performance of the aligners in terms of the number of the aligners (\textit{Number of Alignments}), the runtime (\textit{Runtime}), and the maximum memory requirement \textit{Memory}.}
\end{table}

\clearpage
\section{Robustness of Apollo}

Here in Tables~\ref{supptab:chunk},~\ref{supptab:deletion},~\ref{supptab:insertion},~\ref{supptab:transition}, we show the robustness of Apollo based on the parameters that has a direct affect on the machine learning algorithm. In each of the tables we show that Apollo is robust to different set of parameters.

\begin{table}[htb]
\begin{center}
\caption{Apollo's robustness based on the chunk size of the long read and the contig}
\label{supptab:chunk}
\begin{tabular}{|l|l|rrr|}
\hline \textbf{Long Read}  & \textbf{Contig Chunk} & \textbf{Aligned}   & \textbf{Aligned}    & \textbf{Accuracy}\\
       \textbf{Chunk Size} & \textbf{Size}         & \textbf{Bases}     & \textbf{Bases (\%)} &         \\\hhline{|=|=|===|}
1000              & Original     & 5,708,747 & 98.49      & 0.9798 \\
1000  & 25000     & 5,487,736 & 94.46 & 0.9733 \\
1000  & 50000     & 5,689,120 & 97.95 & 0.9728 \\
1000  & 100000    & 5,493,663 & 94.52 & 0.9727 \\
5000  & 25000     & 5,430,700 & 93.06 & 0.8974 \\
5000  & 50000     & 5,411,163 & 92.68 & 0.8971 \\
5000  & 100000    & 5,516,599 & 94.49 & 0.8970 \\
10000 & 25000     & 5,415,333 & 92.65 & 0.8918 \\
10000 & 50000     & 5,423,340 & 92.75 & 0.8914 \\
10000 & 100000    & 5,474,159 & 93.61 & 0.8914 \\\hline
\end{tabular}
\end{center}
{\footnotesize Here we divide the long reads and the assembly into smaller chunks. We use \textit{E. coli} O157 dataset, assembled with Miniasm. We divide long reads into smaller reads with lengths 1000, 5000, and 10000. Similarly, we divide the assembly contigs into smaller contigs with lengths 25000, 50000, and 100000. We align each chunked read to each chunked contig. We report the performance of Apollo given the chunked assembly and chunked reads.}
\end{table}

\begin{table}[htb]
\begin{center}
\caption{Apollo's robustness based on the maximum deletion and filter size parameters}
\label{supptab:deletion}
\begin{tabular}{|l|l|rrr|}
\hline \textbf{Max}           & \textbf{Filter}    & \textbf{Aligned} & \textbf{Aligned}    & \textbf{Accuracy} \\
       \textbf{Deletion (-d)} & \textbf{Size (-f)} & \textbf{Bases}   & \textbf{Bases (\%)} &          \\\hhline{|=|=|===|}
3  & 100 & 5,699,182 & 97.91 & 0.9739 \\
5  & 100 & 5,696,138 & 97.93 & 0.9735 \\
15 & 100 & 5,678,838 & 97.90 & 0.9731 \\
3  & 200 & 5,705,130 & 98.12 & 0.9751 \\
5  & 200 & 5,704,582 & 98.12 & 0.9750 \\
15 & 200 & 5,702,478 & 98.14 & 0.9751 \\\hline
\end{tabular}
\end{center}
{\footnotesize Performance of Apollo with respect to the parameter that defines the maximum number of deletion in one transition ($d=3$, $d=5$, $d=15$). We also adjust the filter size ($f=100$, $f=200$)}
\end{table}

\begin{table}[htb]
\begin{center}
\caption{Apollo's robustness based on the maximum insertion and filter size parameters}
\label{supptab:insertion}
\begin{tabular}{|l|l|rrr|}
\hline \textbf{Max} & \textbf{Filter} & \textbf{Aligned} & \textbf{Aligned} & \textbf{Accuracy} \\
\textbf{Insertion (-i)} & \textbf{Size (-f)} & \textbf{Bases} & \textbf{Bases (\%)} & \\\hhline{|=|=|===|}
1  & 100 & 5,685,635 & 97.89 & 0.9660 \\
5  & 100 & 5,638,585 & 97.62 & 0.9696 \\
10 & 100 & 5,365,978 & 95.54 & 0.9531 \\
1  & 200 & 5,685,040 & 98.02 & 0.9668 \\
5  & 200 & 5,692,813 & 98.07 & 0.9740 \\
10 & 200 & 5,623,736 & 97.62 & 0.9701 \\\hline
\end{tabular}
\end{center}
{\footnotesize Performance of Apollo with respect to the parameter that defines the maximum number of insertion states for each base ($i=1$, $i=5$, $i=10$). We also adjust the filter size ($f=100$, $f=200$)}
\end{table}

\begin{table}[htb]
\begin{center}
\caption{Apollo's robustness based on the match transition, insertion transition probabilities, and the filter size parameters}
\label{supptab:transition}
\begin{tabular}{|l|l|l|rrr|}
\hline \textbf{Match Transition} & \textbf{Insertion Transition} & \textbf{Filter} & \textbf{Aligned} & \textbf{Aligned} & \textbf{Accuracy} \\
\textbf{Probability (-tm)} & \textbf{Probability (-ti)} & \textbf{Size (-f)} & \textbf{Bases} & \textbf{Bases (\%)} & \\\hhline{|=|=|=|===|}
0.60 & 0.25 & 100 & 5,670,852 & 97.95 & 0.9625 \\
0.60 & 0.30 & 100 & 5,660,957 & 97.90 & 0.9596 \\
0.80 & 0.10 & 100 & 5,699,660 & 98.02 & 0.9788 \\
0.90 & 0.05 & 100 & 5,685,770 & 97.89 & 0.9774 \\
0.60 & 0.25 & 200 & 5,682,512 & 98.10 & 0.9644 \\
0.60 & 0.30 & 200 & 5,681,993 & 98.13 & 0.9618 \\
0.80 & 0.10 & 200 & 5,707,293 & 98.16 & 0.9803 \\
0.90 & 0.05 & 200 & 5,695,902 & 98.05 & 0.9789 \\\hline
\end{tabular}
\end{center}
{\footnotesize Performance of Apollo with respect to the parameters that define the match and insertion transition probabilities ($tm=0.60$ \& $ti=0.25$, $tm=0.60$ \& $ti=0.30$, $tm=0.80$ \& $ti=0.10$, $tm=0.90$ \& $ti=0.05$). We also adjust the filter size ($f=100$, $f=200$)}
\end{table}

\clearpage

\section{Parameters}

We show the parameter settings of the aligners that we used to align the reads to the assembly in Table~\ref{supptab:aligner-params}.

\begin{table}[htb]
\begin{center}
\caption{List of the parameters that are used to align the reads to the assemblies}
\label{supptab:aligner-params}
\begin{tabular}{|l|c|}
\hline \textbf{Aligner} & \textbf{Parameters} \\\hhline{|=|=|}
BWA-MEM & -t 45\\\hline
Minimap2 (for PacBio) & -x map-pb -a -t 45\\\hline
Minimap2 (for ONT) & -x map-ont -a -t 45\\\hline
Minimap2 (for Illumina) &  -a -x sr -t 45\\\hline
pbalign & --nproc 45 \\\hline
\end{tabular}
\end{center}
\end{table}

\clearpage

\bibliographystyle{unsrt}
\bibliography{main}